\documentclass[reqno]{amsart}

\usepackage{booktabs}
\usepackage{IEEEtrantools}
\usepackage{amssymb,latexsym,amsfonts,amsmath}
\usepackage{graphicx}       
\usepackage{mathrsfs}
\usepackage{hyperref}
\usepackage{subfigure}
\usepackage{dsfont}

\usepackage{algorithm,algpseudocode}
\usepackage[noadjust]{cite}

\usepackage{extarrows}
\usepackage{caption}

\usepackage{graphicx}
\usepackage{paralist}
\usepackage{capt-of}
\usepackage{xcolor}
\usepackage{diagbox}
\usepackage{multirow}

\newcommand{\R}{{\mathbb{R}}}

\newcommand{\N}{{\mathbb{N}}}

\newcommand{\ie}{{\it i.e.}}

\newcommand{\argmax}{\textrm{arg}\max}

\newcommand{\ul}{\underline}

\definecolor{myco}{rgb}{0.55, 0.0, 0.63}

\newtheorem{theorem}{Theorem}[section]
\newtheorem{lemma}[theorem]{Lemma}
\newtheorem{problem}[theorem]{Problem}
\newtheorem{proposition}[theorem]{Proposition}
\newtheorem{corollary}[theorem]{Corollary}

\newtheorem{definition}[theorem]{Definition}

\newtheorem{remark}[theorem]{Remark}
\newtheorem{assumption}[theorem]{Assumption}
\numberwithin{equation}{section}

\begin{document}
	
\begin{abstract}
In this work, we propose an abstraction and refinement methodology for the controller synthesis of discrete-time stochastic systems to enforce complex logical properties expressed by deterministic finite automata (\emph{a.k.a.} DFA). 
Our proposed scheme is based on a notion of so-called $(\epsilon,\delta)$-approximate probabilistic relations, allowing one to quantify the similarity between stochastic systems modeled by discrete-time stochastic games and their corresponding finite abstractions. 
Leveraging this type of relations, the lower bound for the probability of satisfying the desired specifications can be well ensured by refining controllers synthesized over abstract systems to the original games. 
Moreover, we propose an algorithmic procedure to construct such a relation for a particular class of nonlinear stochastic systems with slope restrictions on the nonlinearity. 
The proposed methods are demonstrated on a quadrotor example, and the results indicate that the desired lower bound for the probability of satisfaction is guaranteed.
\end{abstract}

\title[Automata-based Controller Synthesis for Stochastic Systems: A Game Framework]{Automata-based Controller Synthesis for Stochastic Systems: \\A Game Framework via Approximate Probabilistic Relations}

\author{Bingzhuo Zhong$^{1}$}
\author{Abolfazl Lavaei$^{2}$}
\author{Majid Zamani$^{3,4}$}
\author{Marco Caccamo$^{1}$}
\address{$^1$TUM School of Engineering and Design, Technical University of Munich, Germany}
\email{\{bingzhuo.zhong,mcaccamo\}@tum.de}
\address{$^2$School of Computing, Newcastle University, United Kingdom}
\email{abolfazl.lavaei@newcastle.ac.uk}
\address{$^3$Department of Computer Science, University of Colorado Boulder, USA}
\address{$^4$Department of Computer Science, LMU Munich, Germany}
\email{majid.zamani@colorado.edu}
\maketitle

\section{Introduction}\label{sec:intro}
{\bf Motivation.} 
Formal synthesis of controllers for continuous-space stochastic systems have gained significant attention in the past two decades due to the increasing demand for synthesizing correct-by-construction controllers in real-life safety-critical applications, including self-driving cars, power grids, etc., to name a few.
In particular, these problems are more challenging when controllers are required to enforce high-level logic properties, \emph{e.g.,} those expressed by automata~\cite{Baier2008Principles}.
Since closed-form solutions of synthesized policies for continuous-space stochastic systems are not available, a promising approach is to approximate these models by simpler ones with finite state sets. 
A challenging step during this approximation phase is to provide formal guarantees such that the controller synthesized over (simpler) finite models can be refined back to original complex ones.

{\bf Related Works} 
There have been many results on the controller synthesis for discrete-time stochastic systems in the past few years.
Results in~\cite{Asselborn2015Probabilistic,Cannon2012Stochastic} focus on enforcing invariance properties for stochastic linear systems.
As for nonlinear stochastic systems with continuous state and input sets, an abstraction-based approximation approach is initially proposed in \cite{Abate2008Probabilistic}.
This result is later improved in~\cite{Soudjani2014Formal} regarding the scalability issue and extended in~\cite{Tkachev2013Quantitative,Kamgarpour2013Control,Kamgarpour2017Control} for abstraction-based policy synthesis enforcing temporal logic properties characterized by deterministic finite automata. 
An ($\epsilon, \delta$)-approximate probabilistic relation is introduced in~\cite{Haesaert2017Verification} to characterize the probabilistic dependency between an original model and its finite abstraction.
With this type of relation, one can synthesize controllers enforcing the desired properties with less conservative lower bounds on probability of satisfaction.
Later, results in~\cite{Haesaert2018Temporal,Haesaert2020Robust} propose Bellman operators for synthesizing controllers enforcing co-safe LTL$_F$ properties~\cite{Faruq2018Simultaneous} based on this relation.
In the context of constructing finite abstractions for large-scale stochastic systems, compositional abstraction-based techniques have been introduced to alleviate the scalability issue due to discretizing the state sets; see for example~\cite{Soudjani2015Dynamic,lavaei2018ADHSJ,lavaei2018CDCJ,AmyJournal2020,lavaei2019HSCC_J,lavaei2019NAHS,AmyIFAC2020,Lavaei2021Compositional,lavaei2019Thesis,Lavaei2021Automated}.

Note that the above-mentioned works mainly focus on stochastic systems that are only affected by control inputs and noises. 
In some safety-critical real-life applications, systems are also affected by (rational) adversarial inputs, whose objectives are opposed to that of control inputs.
	In these scenarios, synthesis approaches for non-cooperative stochastic games~\cite{Zhu2015Game} are required.
Results in~\cite{Svorenova2017Temporal} handle synthesis problem for linear stochastic games based on iterative abstraction-refinement~\cite{Kattenbelt2010game}.
Results in~\cite{Kamgarpour2011Discrete,Ding2013stochastic} utilize a grid-based approximation framework~\cite{Abate2010Approximate} to synthesize controllers for nonlinear stochastic games. 
Within the same framework, those results which are initially proposed for stochastic games with a finite or countably infinite number of states (\emph{e.g.,} stochastic games with generalized mean-payoff objectives~\cite{Chatterjee2016Perfect,Chatterjee2015Qualitative}, 
	reachability objectives~\cite{Kwiatkowska2016Model,Chatterjee2020Stochastic,Frederiksen2013Monomial,Henzinger2006Strategy},
	multiple objectives~\cite{Basset2018Compositional,Wiltsche2015Assume,Kwiatkowska2019Verification})
	can also be employed to synthesize controllers for stochastic games with continuous state sets.
However, the guarantee provided under this framework is sometimes very conservative (this is shown with an example in Section~\ref{sec:comp1}).
There have also been some results to synthesize controllers for non-cooperative stochastic games (see e.g.~\cite{Hou2013game,Moon2016Discrete,Mukaidani2017Infinite,GonzalezTrejo2002Minimax,Aberkane2019solution}), but they are not applicable to enforce high-level temporal logic properties for nonlinear systems.

{\bf Contributions.} 
In this work, we focus on synthesizing controllers for discrete-time nonlinear stochastic systems with continuous state and input sets over a finite time horizon.
	Particularly, we consider those systems modeled by zero-sum stochastic games~\cite{Zhu2015Game}, which are subject to not only control inputs but also (rational) adversarial inputs whose objective is opposed to that of control inputs.
	Moreover, we are interested in a class of complex logical properties expressed by deterministic finite automata (DFA), which are powerful in specifying behaviours that occurs within finite time~\cite{Faruq2018Simultaneous}.
	Here, we propose an abstraction-based technique to synthesize controllers based on an ($\epsilon, \delta$)-approximate probabilistic relation.
	Concretely, we first construct a finite abstraction of the original game and then establish such a relation between the finite abstraction and the original one. 
Leveraging the probabilistic relation, we then propose new Bellman operators to synthesize a controller over the finite abstraction and finally refine this controller back over the original game while providing probabilistic guarantees for the satisfaction of desired properties.
Here, we summarize our contributions as follows:
\begin{enumerate}[(i)]
	\setlength{\itemsep}{0pt}
	\setlength{\parsep}{0pt}
	\setlength{\parskip}{0pt}
	\item Given the notion of approximate probabilistic relations similar to~\cite{Lavaei2021Compositional} for stochastic games,  
	we propose a new Bellman operator to synthesize controllers for \emph{nonlinear} stochastic games with continuous state and input sets.
	Leveraging the proposed operators, we are able to deal with complex logic properties modeled by deterministic finite automata, while providing less conservative probabilistic guarantees for satisfying those properties in comparison with the results in~\cite{Kamgarpour2011Discrete,Ding2013stochastic}(cf. Section~\ref{sec:comp1}).
	\item For a class of \emph{nonlinear} stochastic games, we propose a systematic algorithm to establish an approximate probabilistic relation between the original game and its abstraction.
	In comparison, results in~\cite{Haesaert2018Temporal,Haesaert2020Robust,Huijgevoort2020Similarity} only establish such a relation for \emph{linear} stochastic systems without rational adversarial inputs. 
	Although~\cite{Lavaei2021Compositional} provides the notion of approximate probabilistic relations for stochastic games, it does not provide any constructive algorithm for establishing the relation.
	\item The proposed operators in this work can also be applied to synthesis problems for stochastic systems without adversarial inputs.
	In this case, compared with the operators in~\cite{Haesaert2020Robust}, our results provide a less conservative probabilistic guarantee for satisfying the desired properties (cf. Lemma~\ref{lem:less_con}, Corollary~\ref{co:less_con}, and Section~\ref{sec:comp2}).
\end{enumerate}

{\bf Organizations.} The remainder of the paper is structured as follows.
In Section~\ref{sec:2}, we provide notations, underlying models, and a preliminary discussion on the problem that we aim to solve.
We present in Section \ref{sec:apr_general} an ($\epsilon, \delta$)-approximate probabilistic relation between two stochastic games.
This is followed by Section~\ref{sec:nonlinear_case}, in which we focus on the construction of finite abstractions together with ($\epsilon, \delta$)-approximate probabilistic relations for a particular class of nonlinear stochastic games.
In Section~\ref{sec:safetyctr}, we synthesize controllers given an ($\epsilon, \delta$)-approximate probabilistic relation between the original game and its finite abstraction.
In Section~\ref{Case_study}, we apply our results to a control problem for a quadrotor, and we also compare our results with some existing methods. 
Finally, we conclude our work in Section~\ref{sec:discussion}.

\section{Problem Formulation}\label{sec:2}
\subsection{Preliminaries}
A topological space $S$ is called a Borel space if it is homeomorphic to a Borel subset of a Polish space (\emph{i.e.,} a separable and completely metrizable space). 
One of the examples of Borel space is the Euclidean spaces $\mathbb{R}^n$.
Here, any Borel space $S$ is assumed to be endowed with a Borel $\sigma$-algebra denoted by $\mathcal{B}(S)$. 
A map $f:X\rightarrow Y$ is measurable whenever it is Borel measurable. 
Moreover, a map $f:X\rightarrow Y$ is universally measurable if the inverse image of every Borel set under $f$ is measurable w.r.t. every complete probability measure on $X$ that measures all Borel subsets of $X$.

A probability space in this work is presented by $(\hat \Omega,\mathcal F_{\hat \Omega},\mathbb{P}_{\hat \Omega})$, where $\hat \Omega$ is the sample space,
$\mathcal F_{\hat \Omega}$ is a sigma-algebra on $\hat \Omega$ which comprises subsets of $\hat\Omega$ as events, and $\mathbb{P}_{\hat \Omega}$ is a probability measure that assigns probabilities to events.
Throughout this paper, we focus on random variables, denoted by $X$, that take values from measurable spaces ($S$,$\mathcal{B}(S)$), \emph{i.e.,} random variables here are measurable functions $X:(\hat \Omega,\mathcal F_{\hat \Omega})\rightarrow(S,\mathcal{B}(S))$ such that one has $Prob\{\mathcal{Q}\} = \mathbb{P}_{\hat \Omega}\{X^{-1}(\mathcal{Q})\}$, $\forall \mathcal{Q}\in \mathcal{B}(S)$. 
	For brevity, we directly present the probability measure on ($S$,$\mathcal{B}(S)$) without explicitly mentioning the underlying probability space and the function $X$ itself. 
	Additionally, we denote by $\mathbf{P}(S,\mathcal{B}(S))$ a set of probability measures on the ($S$,$\mathcal{B}(S)$).

\subsection{Notations}
We use $\mathbb{R}$ and $\mathbb{N}$ to denote sets of real and natural numbers, respectively. These symbols are annotated by subscripts to restrict the sets in a usual way, \emph{e.g.,} $\mathbb{R}_{\geq0}$ denotes the set of non-negative real numbers.
Moreover, $\mathbb{R}^{n\times m}$ with $n,m\in \mathbb{N}_{\geq 1}$ denotes the vector space of real matrices with $n$ rows and $m$ columns.
For $a,b\in\mathbb{R}$ (resp. $a,b\in \N$) with $a\leq b$, the close, open, and half-open intervals in $\mathbb{R}$ (resp. $\N$) are denoted by $[a,b]$, $(a,b)$ ,$[a,b)$, and $(a,b]$, respectively. 
Given $N$ vectors $x_i \in \mathbb R^{n_i}$, $n_i\in \mathbb N_{\ge 1}$, and $i\in\{1,\ldots,N\}$, we use $x = [x_1;\ldots;x_N]$ to denote the corresponding column vector of the dimension $\sum_i n_i$.
We denote respectively by $\mathbf{0}_n$ and $\mathbf{1}_n$ the column vector in $\R^n$ with all elements equal to 0 and 1. We also denote the identity matrix in $\R^{n\times n}$ and zero matrix in $\R^{m\times n}$ by $I_n$ and $0_{m\times n}$, respectively. 
We denote the chi-square inverse cumulative distribution function with $n$ degrees of freedom by $\chi^{-1}_n: [0,1]\rightarrow \R$~\cite{Bernardo2009Bayesian}. 
Moreover, given a vector $x\in\R^n$, $\lVert x\rVert$ denotes the Euclidean norm of $x$, and $\lVert x\rVert_{\infty}$ denotes the infinity norm of $x$.
Given a matrix $A\in \R^{n\times n}$, $\textsl{im}(A)$ denotes the image of A, and $\lVert A\rVert$ represents the operator norm of A, which is equal to the largest singular value of A.
Given sets X and Y, a relation $\mathscr{R} \in X\times Y$ is a subset of the Cartesian product $X\times Y$ that relates $x\in X$ with $y\in Y$ if $(x,y)\in\mathscr{R}$, which is equivalently denoted by $x\mathscr{R}y$.
Given a set $\mathsf{M} =X_1\times X_2\times\ldots\times X_n$ and vector $\mathsf{m}=(x_1,x_2,\ldots,x_n)\in\mathsf{M}$ with $x_i\in X_i$, we define $\mathsf{m}_{X_i}=x_i$.
Moreover, given a set $X$, $X^{\mathbb{N}}$ denotes the Cartesian product of the countably infinite number of set $X$. 
Additionally, given functions $f:X\rightarrow Y$ and $g:Y\rightarrow Z$, we denote by $g\circ f : X\rightarrow Z$ the composition of functions $f$ and $g$.

\subsection{General Discrete-Time Stochastic Games} \label{sec:gDTSG}
In this paper, we focus on stochastic systems modeled as general discrete-time stochastic games (gDTSGs) between two non-cooperative players. 
Following standard conventions, we refer to the control input as Player~\uppercase\expandafter{\romannumeral1} and the adversary input as Player~\uppercase\expandafter{\romannumeral2}. 
This class of games, formalized in the next definition, evolves over continuous or uncountable state sets with an output set over which properties of interest are defined. 
\begin{definition}
	\label{def:gDTSG}
	A general discrete-time stochastic game (gDTSG) is a tuple
	\begin{equation}
	\label{eq:dt-SCS}
	\mathfrak{D} =(X,U,W,X_0,T,Y,h),
	\end{equation}
	where,
	\begin{itemize}
		\setlength{\itemsep}{0pt}
		\setlength{\parsep}{0pt}
		\setlength{\parskip}{0pt}
		\item $X\subseteq \mathbb R^s$ is a Borel set as the state set. We denote by $(X, \mathcal B (X))$ the measurable space with $\mathcal B (X)$  being  the Borel sigma-algebra on $X$;
		\item $U\subseteq \mathbb R^m$ is a compact Borel set as the input set of Player~\uppercase\expandafter{\romannumeral1};
		\item $W\subseteq \mathbb R^p$ is a compact Borel set as the input set of Player~\uppercase\expandafter{\romannumeral2};
		\item $X_0\subseteq X$ is the set of initial states;
		\item $T:\mathcal B(X)\times X\times U\times W\rightarrow[0,1]$ 
		is a conditional stochastic kernel that assigns to any $x \in X$, $u\in U$, and $w\in W$ a probability measure $T(\cdot | x,u,w)$ on the measurable space $(X,\mathcal B(X))$. This stochastic kernel specifies probabilities over executions $\{x(k),k\in\mathbb N\}$ of the gDTSG such that for any set $\mathcal{Q} \subseteq \mathcal B(X)$ and for any $k\in\mathbb N$, 
		\begin{align*}
		\mathbb P \big\{x(k+1)\in \mathcal{Q}\,\big|\, x(k),u(k),w(k)\big\}=\int_\mathcal{Q} T (\mathsf dx(k+1)|x(k),u(k),w(k));
		\end{align*}
		\item $Y\subseteq \mathbb R^q$ is a Borel set as the output set; 
		\item $h:X\rightarrow Y$ is a measurable function that maps a state $x\in X$ to its output $y = h(x)$.
	\end{itemize}
\end{definition}
\begin{remark}\label{rem:asy_info}
	For robustness concern, we consider an asymmetric information pattern that favors Player~\uppercase\expandafter{\romannumeral2} in this paper, \ie, Player~\uppercase\expandafter{\romannumeral2} may select its action in a rational fashion based upon the choice of Player~\uppercase\expandafter{\romannumeral1}.
	This results in a zero-sum Stackelberg game~\cite{Breton1988Sequential} with Player~\uppercase\expandafter{\romannumeral1} as the leader, which is crucial for the existence of deterministic policies (cf. Definition~\ref{def_mp}, Remarks~\ref{randomp}and~\ref{det_Mar}) in this work.
		Note that our setting here is common for robust control problems in which control inputs are selected considering that adversarial inputs are provided in a worst-case manner.
		The motivation for using such a setting is to provide formal probabilistic guarantees regardless of how adversarial inputs are chosen by Player~\uppercase\expandafter{\romannumeral2}.
		This also indicates that Player~\uppercase\expandafter{\romannumeral2} does not have to select adversarial inputs rationally in practice. 
\end{remark}

Alternatively, a gDTSG $\mathfrak{D}$ as in~\eqref{eq:dt-SCS} can be described by the following difference equations
\begin{equation}\label{eq:gMDP_f}
\mathfrak{D}\!:
\left\{\hspace{-0.15cm}\begin{array}{l}
x(k+1)=f(x(k),u(k),w(k),\varsigma(k)),\\
y(k)=h(x(k)),\quad \quad k\in\mathbb N,\end{array}\right.
\end{equation}
where $x(k)\in X$, $u(k)\in U$, $w(k)\in W$, $y(k)\in Y$, and $\varsigma:=\{\varsigma(k): \hat\Omega\rightarrow V_{\varsigma}, k\in \N\}$ is a sequence of independent and identically distributed (i.i.d.) random variables from the sample space $\hat\Omega$ to a set $V_{\varsigma}$. 
With this notion, the evolution of a gDTSG can be described by its paths and output sequences as defined below. 

\begin{definition}\label{def:path}\emph{(Path)}
	A path of a gDTSG $\mathfrak{D}$ as in~\eqref{eq:dt-SCS} is 
	\begin{align*}
	\omega\,=\,(x(0),u(0),&w(0),\ldots,x(k-1), u(k-1),w(k-1),x(k),\ldots),
	\end{align*}
	where $x(k)\!\in\!X$, $u(k)\!\in\!U$, and $w(k)\!\in\! W$ with $k\!\in\!\N$. 
	We denote by $\omega_{x}=(x(0),x(1),\ldots,x(k),\ldots)$, $\omega_{u}=(u(0),u(1),\ldots,u(k),\ldots)$, and $\omega_{w}=(w(0),w(1),\ldots,$ $w(k),\ldots)$ the subsequences of states, control inputs of Player~\uppercase\expandafter{\romannumeral1}, and adversarial inputs of Player~\uppercase\expandafter{\romannumeral2}, respectively.
	The corresponding \emph{output sequence} is denoted by 
	\begin{equation*}
	y_{\omega} = (y(0),y(1),\ldots,y(k),\ldots),
	\end{equation*}
	with $y(k)=h(x(k))$.
	In addition, we denote by $\omega_k$ the path up to time instant $k$, and by $y_{\omega k}$ its corresponding output sequence.
\end{definition}

For a better illustration of the theoretical results, we employ a running example throughout the paper as follows.

{\bf Running example.} Consider the following gDTSG
\begin{equation*}
\mathfrak{D}\!:
\left\{\hspace{-0.15cm}\begin{array}{l}
\begin{aligned}
x(k+1) = ~\!&Ax(k)\!+\!Bu(k)\!+\!E\sin(Fx(k))\!+\!Dw(k)\!+\!R\varsigma(k),
\end{aligned}\\
y(k)=Cx(k),\quad \quad k\in\mathbb N,\end{array}\right.
\end{equation*}
with 
\begin{align*}
&A \!=\!  \begin{bmatrix}
\begin{smallmatrix}0.9204\ &0.4512\ &0.9491\\0.7865\ &0.8269\ &1.074\\0.6681\ &0.3393\ &0.5110\end{smallmatrix}
\end{bmatrix}\!, ~~
B \!=\! \begin{bmatrix}
\begin{smallmatrix}9.001\ &1.611\ &3.663\\1.404\ &11.76\ &2.386\\5.568\ &4.560\ &5.156\end{smallmatrix}
\end{bmatrix}\!,
\end{align*}
$E \!=\! [0.6740;0.6367;0.7030]$, $D \!=\! [0.6;0.4;0.6]$, $R =[0.5110;$ $0.3347;0.5336]$, $F\!=\! [0.5439;0.9578;0.2493]^T$, and $C = [0.1;$ $0.1;0.1]^T$, where $x(t)\!=\![x_1(k);x_2(k);x_3(k)]$ is the state, $u(k)\in[-2.5,2.5]^3$ denotes the control input of Player~\uppercase\expandafter{\romannumeral1}, $w(k)\in[-0.5,$ $0.5]$ denotes the adversarial input of Player~\uppercase\expandafter{\romannumeral2}, $\varsigma(k)$ is a sequence of standard Gaussian random variables, and $y(k)$ is the output.

The space for all infinite paths $\Omega = (X\times U\times W)^{\N}$ along with its product $\sigma$-algebra $(\mathcal B(X)\times \mathcal B(U)\times \mathcal B(W))^{\N}$ is called a canonical sample space for the gDTSG.
Next, we define Markov policy for controlling the gDTSG.
\begin{definition}\label{def_mp}
	\emph{(Markov Policy)} Consider a gDTSG $\mathfrak{D} =(X,U,$ $W,X_0,T,Y,h)$. A \textit{Markov policy} $\rho$ defined over the time horizon $[0,H-1]\subset\N$ for Player~\uppercase\expandafter{\romannumeral1} is a sequence $\rho\!=\!(\rho_{0},\rho_{1},\ldots,\rho_{H-1})$ of universally measurable maps $\rho_{k}:X\rightarrow\mathbf{P}(U,\mathcal{B}(U))$, with
	\begin{equation*}
	\rho_{k}(U\big | x(k)) =1.
	\end{equation*} 
	Similarly, 	
	a \textit{Markov policy} $\lambda$ for Player~\uppercase\expandafter{\romannumeral2} is a sequence $\lambda\,=\,(\lambda_{0},\,\lambda_{1},\,\ldots,\lambda_{H-1})$ of universally measurable maps $\lambda_{k}:\,X\times U\,\rightarrow\,\mathbf{P}(W,\mathcal{B}(W))$, with
	\begin{equation*}
	\lambda_{k}(W\big | x(k),u(k)) =1,
	\end{equation*}
	for all $k\in[0,H-1]$.
	We use $\mathcal{P}$ and $\Lambda$ to denote the set of all Markov policies for Players~\uppercase\expandafter{\romannumeral1} and~\uppercase\expandafter{\romannumeral2}, respectively. 
	Moreover, we denote by $\mathcal{P}^H$ and $\Lambda^H$ the set of all Markov policies for Players~\uppercase\expandafter{\romannumeral1} and~\uppercase\expandafter{\romannumeral2} within time horizon $[0,H-1]$, respectively.
\end{definition}
\begin{remark}\label{randomp}
	In general, the Markov policy assigns a probability measure over $(U,$ $\mathcal{B}(U))$ (resp. $(W,\mathcal{B}(W))$).
	From practical implementations' point of view, we are interested in \emph{nonrandomized} Markov policies~\cite[Definition 8.2]{Shreve1978Stochastic}.
	In this paper, by Markov policies, we refer to~\emph{nonrandomized} ones; otherwise, we explicitly say that the Markov policies are randomized ones. 
\end{remark}
Next, we define a more general set of control strategies for Players~\uppercase\expandafter{\romannumeral1} and~\uppercase\expandafter{\romannumeral2}. 
The definition here is adapted from~\cite{Haesaert2020Robust} by allowing their output update map to be time dependent.
\begin{definition}\label{def_p}
	\emph{(Control Strategy)} A control strategy for Player~\uppercase\expandafter{\romannumeral1} or~\uppercase\expandafter{\romannumeral2} of a gDTSG $\mathfrak{D} =(X,U,W,X_0,T,Y,h)$ is a tuple
	\begin{align}
	\mathbf{C} = (\mathsf{M},\mathsf{U},\mathsf{Y},\mathsf{H},\mathsf{M}_{0},\pi_{\mathsf{M}},\pi_{\mathsf{Y}}),
	\end{align}
	where $\mathsf{M}$ is a Borel set as the \emph{memory state set}; 
	$\mathsf{U}$ is a Borel set as the \emph{observation set}; 
	$\mathsf{Y}$ is a Borel set as the \emph{output set}, which should be equal to $U$ for Player~\uppercase\expandafter{\romannumeral1}, and to $W$ for Player~\uppercase\expandafter{\romannumeral2};	
	$\mathsf{H}\subseteq \N$ is the \emph{time domain}; 
	$\mathsf{M}_{0}\subseteq\mathsf{M}$ is the set of initial memory state; 
	$\pi_{\mathsf{M}}\!: \!\mathsf{M}\times\mathsf{U}\times\mathsf{H}\!\rightarrow\! \mathbf{P}(\mathsf{M},\mathcal{B}(\mathsf{M}))$ is a \emph{memory update function}; $\pi_{\mathsf{Y}}:\mathsf{M}\times\mathsf{H}\rightarrow\mathbf{P}(\mathsf{Y},\mathcal{B}(\mathsf{Y}))$ is an \emph{output update function}.
\end{definition}
\begin{remark}
	A Markov policy $\rho\,=\,(\rho_{0},\,\rho_{1},\,\ldots,\rho_{H-1})$ for Player~\uppercase\expandafter{\romannumeral1} (resp.~$\lambda\,=\,(\lambda_{0},\,\lambda_{1},\,\ldots,\lambda_{H-1})$ for Player~\uppercase\expandafter{\romannumeral2}) can be redefined as a control strategy $\mathbf{C} = (\mathsf{M},\mathsf{U},\mathsf{Y},\mathsf{H},$ $\mathsf{M}_{0},\pi_{\mathsf{M}},\pi_{\mathsf{Y}})$ with $\mathsf{M}=\{\mathsf{m}\}$, where $\mathsf{m}$ is the sole element in $\mathsf{M}$; $\mathsf{U}=X$ (resp. $\mathsf{U}=X\times U$); $\mathsf{Y}=U$ (resp. $\mathsf{Y}=W$); $\mathsf{H}=[0,H-1]$; $\mathsf{M}_{0}=\{\mathsf{m}\}$; and $\pi_{\mathsf{Y}}:=\rho_k$ (resp. $\pi_{\mathsf{Y}}:=\lambda_k$) for all $k\in\mathsf{H}$.
\end{remark}

Given a gDTSG $\mathfrak{D}$, we denote by ($\rho$,$\lambda$)$\times \mathfrak{D}$ the controlled gDTSG when $\mathfrak{D}$ is controlled by Markov policies $\rho$ for Player~\uppercase\expandafter{\romannumeral1} and $\lambda$ for Player~\uppercase\expandafter{\romannumeral2}.
Analogously, consider a control strategy for Player~\uppercase\expandafter{\romannumeral1}, denoted by $\mathbf{C}_{\rho}$, and a control strategy for Player~\uppercase\expandafter{\romannumeral2}, denoted by $\mathbf{C}_{\lambda}$.
The controlled gDTSG is denoted by ($\mathbf{C}_{\rho}$,$\mathbf{C}_{\lambda}$)$\times \mathfrak{D}$.
With this notation, we denote by $\mathbb{P}_{(\rho,\lambda)\times \mathfrak{D}}$ (resp. $\mathbb{P}_{(\mathbf{C}_{\rho},\mathbf{C}_{\lambda})\times \mathfrak{D}}$) the probability measure over the space of output sequences of the controlled gDTSG ($\rho$,$\lambda$)$\times \mathfrak{D}$ (resp. ($\mathbf{C}_{\rho}$,$\mathbf{C}_{\lambda}$)$\times \mathfrak{D}$).
In the next subsection, we discuss the logical properties of interest. 
\subsection{Deterministic Finite Automata}\label{sec:DFA}
In this paper, deterministic finite automata (DFA) would be leveraged to model the desired properties, as introduced below.
\begin{definition} \emph{(DFA)}
	A deterministic finite automata (DFA) is a tuple $\mathcal{A}\ =\!(Q, q_0, \Pi,\tau, F)$, where $Q$ is a finite set of states, $q_0\!\in\! Q$ is the initial state, $\Pi$ is a finite set of alphabet, $\tau : Q\times\Pi \rightarrow Q$ is a transition function, and $F\!\subseteq\! Q$ is a set of accepting states.
\end{definition}
Without loss of generality~\cite[Section 4.1]{Baier2008Principles}., we focus on those DFA which is \emph{total}, i.e., given any $q\in Q$, $\forall \sigma' \in \Pi$, $\exists q'\in Q$ such that $q'\! =\! \tau(q,\sigma')$.
A finite word $\sigma = (\sigma_0, \sigma_1,\ldots,\sigma_{k-1})\in \Pi^k$ is accepted by $\mathcal{A}$ if there exists a finite state run $q\! =\!(q_0,q_1,\ldots,q_k)\!\in\!Q^{k+1}$ such that $q_{z+1}\! =\! \tau(q_z,\sigma_z)$, $\sigma_z\! \in\! \Pi$ for all $0\!\leq\! z\!<\!k$ and $q_k\!\in\! F$.
The set of words accepted by $\mathcal{A}$ is called the language of $\mathcal{A}$ and denoted by $\mathcal{L}(\mathcal{A})$.
Next, we introduce how to connect the gDTSG $\mathfrak{D}$ as in~\eqref{eq:dt-SCS} to a DFA $\mathcal{A}$ using a measurable labelling function.

\begin{definition}\label{def:sactisfaction_DFA} \emph{(Labelling Function)}
	Consider a gDTSG $\mathfrak{D}\!=\!(X,U,W,X_0,T,Y,h)$, a DFA $\mathcal{A}\!=\! (Q, q_0, \Pi,\tau,$ $ F)$, and a finite output sequence $y_{\omega (H-1)}\!=\!(y(0),y(1),\ldots,y(H-1))\!\in\! Y^H$ of $\mathfrak{D}$ with some $H\!\in\!\mathbb{N}_{>0}$.	
	The trace of $y_{\omega (H-1)}$ over $\Pi$ is $\sigma\!=\!L_H(y_{\omega (H-1)})\!=\!$ $(\sigma_0,\sigma_1,\ldots,\sigma_{H-1})$ with $\sigma_k=L(y(k))$ for all $k\in[0,H-1]$, where $L: Y\rightarrow \Pi$ is a measurable labelling function and $L_H:Y^H\rightarrow \Pi^H$ is a measurable function.
	Moreover, $y_{\omega (H-1)}$ is accepted by $\mathcal{A}$, denoted by $y_{\omega (H-1)}\models \mathcal{A}$, if $L_H(y_{\omega (H-1)})\in \mathcal{L}(\mathcal{A})$. 
\end{definition}
Throughout the paper, we denote by $(\mathcal{A},H)$ the property of interest, with $\mathcal{A}$ being a DFA and $H$ being the finite time horizon over which the property should be satisfied.
Accordingly, we evaluate the satisfaction of a gDTSG $\mathfrak{D}$ with respect to this property in terms of $\mathbb{P}_{\mathfrak{D}}\{y_{\omega (H-1)}\models \mathcal{A} \}$ within a bounded-time horizon, where $y_{\omega (H-1)}$ is the output sequences generated by $\mathfrak{D}$.  
For this purpose, we need to construct a product gDTSG based on $\mathfrak{D}$ and $\mathcal{A}$, as defined below.
\begin{definition}\label{def:product_gmdp_dfa} \emph{(Product gDTSG)}
	Consider a gDTSG $\mathfrak{D} =(X,U,W,X_0,T,Y,h)$, a DFA $\mathcal{A} = (Q, q_0, \Pi, \tau, F)$, and a labelling function $L: Y\rightarrow \Pi$ as in Definition~\ref{def:sactisfaction_DFA}.
	The product of $\mathfrak{D}$ and $\mathcal{A}$ is a gDTSG defined as
	\begin{equation*}
	\mathfrak{D}\otimes\mathcal{A} = \{\bar{X},\bar{U},\bar{W},\bar{X}_0,\bar{T},\bar{Y},\bar{h}\},
	\end{equation*}
	where
	$\bar{X}:=X\times Q$ is the state set;
	$\bar{U}:= U$ is the input set for Player~\uppercase\expandafter{\romannumeral1}; 
	$\bar{W}:= W$ is the input set for Player~\uppercase\expandafter{\romannumeral2};	
	$\bar{X}_0$ is the initial state set, with $\bar{x}_0:=(x_0,\bar{q}_0)\in\bar{X_0}$, $x_0\in X_0$ and
	\begin{equation}
	\bar{q}_0 = \tau(q_0,L\circ h(x_0));\label{eq:compute_initial_q0}
	\end{equation}
	$\bar{T}(\mathsf dx'\times\{q'\}|x,q,u,w)$ is the stochastic kernel that assigns for any $(x,q)\in\bar{X}$, $u\in\bar{U}$, and $w\in\bar{W}$ the probability $\bar{T}(\mathsf dx'\times\{q'\}|x,q,u,w)=T(\mathsf dx'|x,u,w)$ when $q' =\tau(q,L\circ h(x'))$, and $\bar{T}(\mathsf dx'\times\{q'\}|x,q,w,u)=0$, otherwise; $\bar{Y}:=Y$ is the output set and $\bar{h}(x,q):=h(x)$ is the output map.
\end{definition}

In the following, we formally define the problems of interest in this paper.
For some properties, \emph{e.g.,} co-safe-LTL$_F$~\cite{Faruq2018Simultaneous}, all infinite output sequences satisfying them have a finite good prefix~\cite[Section 2.2]{Kupferman2001Model}. 
In this case, we model such properties with DFAs that accept all good prefixes. 
Accordingly, we focus on the lower bound of satisfaction probability, which yields a \emph{problem of robust satisfaction} as follows.
\begin{problem}\label{problem1}
	\emph{(Robust Satisfaction)} Consider a gDTSG $\mathfrak{D} =(X,U,$ $W,X_0,T,Y,h)$ and the desired property $(\mathcal{A},H)$.
	The \emph{problem of robust satisfaction} is to design a control strategy $\mathbf{C}_{\rho}$ for Player~\uppercase\expandafter{\romannumeral1} such that for any control strategy $\mathbf{C}_{\lambda}$ for Player~\uppercase\expandafter{\romannumeral2}, we have
	\begin{equation}
	\mathbb{P}_{(\mathbf{C}_{\rho},\mathbf{C}_{\lambda})\times \mathfrak{D}}\Big\{\exists k\leq H, y_{\omega k}\models \mathcal{A}\Big\}\geq \mathbf{s},
	\end{equation}
	where $\mathbf{s}$ is the \emph{robust satisfaction probability} guaranteed by $\mathbf{C}_{\rho}$.
\end{problem}
Meanwhile, for some other logical properties, \emph{e.g.,} safe-LTL$_F$~\cite{Saha2014Automated}, all infinite output sequences that violate these properties have a finite bad prefix~\cite[Section 2.2]{Kupferman2001Model}. 
Thus, we model such properties with DFAs that accept all bad prefixes, and an upper bound of the violation probability is of interest.
This results in a \emph{problem of worst-case violation} as defined below.
\begin{problem}\label{problem2}
	\emph{(Worst-case Violation)} Consider a gDTSG $\mathfrak{D} =(X,U,W,X_0,T,Y,h)$ and a property $(\mathcal{A},H)$.
	The \emph{problem of worst-case violation} is to design a control strategy $\mathbf{C}_{\rho}$ for Player~\uppercase\expandafter{\romannumeral1} such that for any control strategy $\mathbf{C}_{\lambda}$ for Player~\uppercase\expandafter{\romannumeral2}, we have
	\begin{equation}
	\mathbb{P}_{(\mathbf{C}_{\rho},\mathbf{C}_{\lambda})\times \mathfrak{D}}\Big\{\exists k\leq H, y_{\omega k}\models \mathcal{A}\Big\}\leq \mathbf{v},
	\end{equation}
	with $\mathbf{v}$ being the \emph{worst-case violation probability} ensured by $\mathbf{C}_{\rho}$.
\end{problem}
{\bf Running example (continued).}
\begin{figure}
	\centering
	\includegraphics[width=8cm]{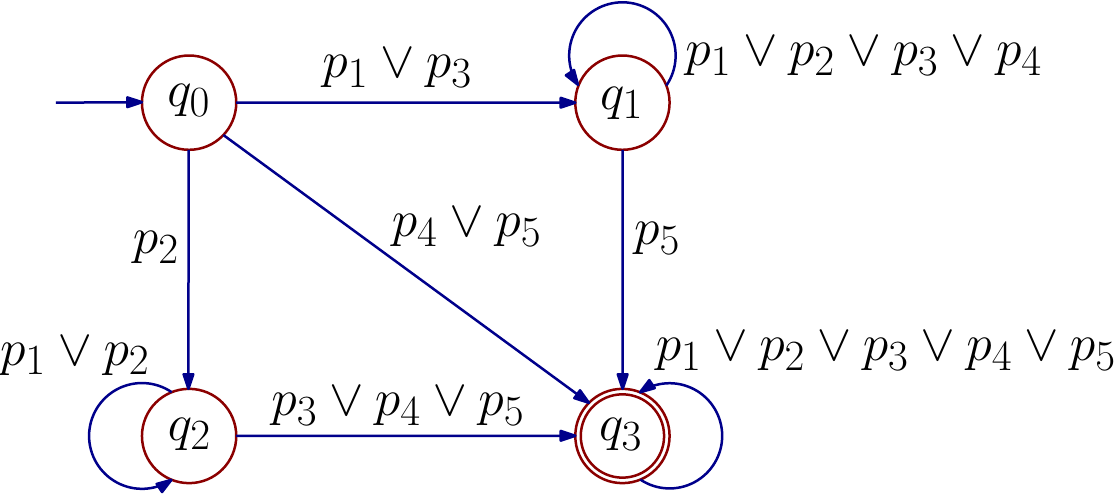}
	\caption{DFA for modeling $\psi$, with accepting state $q_3$, alphabet $\Pi=\{p_1,p_2,p_3,p_4,p_5\}$, and labelling function $L:Y \rightarrow \Pi$, where $L(y)=p_1$ when $y \in [0, 1.8]$, $L(y)=p_2$ when $y \in [-1.8, 0)$, $L(y)=p_3$ when $y \in (1.8,2]$, $L(y)=p_4$ when $y \in [-2,-1.8)$, and $L(y)=p_5$ when $y \in (-\infty, -2)\cup(2,+\infty)$.	
	} \label{fig:case}
\end{figure}
Here, we focus on the following property $\psi$: within $20$ time steps (i.e., $H=20$), if the output of the system starts from $[0,2]$, it should stay within $[-2,2]$; if it instead starts from $[-1.8,0]$, it should then stay within $[-1.8,1.8]$.
The DFA for modeling $\psi$ is shown in Figure~\ref{fig:case}.
Here, we focus on the problem of worst-case violation corresponding to this DFA.
\begin{remark}
	To construct a DFA for modeling the desired property, one can first write down the temporal logic formula corresponding to this property, e.g., safe-LTL$_F$ or co-safe-LTL$_F$ formulae. 
	Alternatively, one can also translate the desired property written in natural language into a temporal logic formula using the results in~\cite{Buzhinsky2019Formalization}. 
	Having the logical formula, one can then build the corresponding DFA using existing tools such as SPOT~\cite{DuretLutz2016Spot}.
\end{remark}

\section{Approximate Probabilistic Relations between gDTSGs}\label{sec:apr_general}
The probabilistic guarantee provided in this paper relies on an approximate probabilistic relation that captures the probabilistic dependency between the executions of two gDTSGs.
This relation is an extension of the approximate probabilistic relation between two stochastic systems without rational adversarial input~\cite{Haesaert2017Verification}.
Here, we first define $\delta$-lifted relation over general state spaces, which pave the way for defining the approximate probabilistic relation between gDTSGs afterward.
\begin{definition}\label{lifting}
	\emph{($\delta$-lifted Relation~\cite{Haesaert2017Verification})}
	Let $X, \hat X$ be two sets with associated measurable spaces $(X, \mathcal B(X))$ and $(\hat X, \mathcal B(\hat X))$.
	Consider a relation	$\mathscr{R}\subseteq X\times \hat X$ that is measurable, i.e., $\mathscr{R} \in \mathcal B(X \times \hat X)$, probability distributions $\Phi\in\mathbf{P}(X, \mathcal B(X))$, and $\Theta\in\mathbf{P}(\hat X, \mathcal B(\hat X))$.
	One has $(\Phi,\Theta)\in\mathscr{\bar R}_{\delta}$, denoted by $\Phi\mathscr{\bar R}_{\delta}\Theta$, with $\mathscr{\bar R}_{\delta}\subseteq  \mathbf{P}(X, \mathcal B(X))\times \mathbf{P}(\hat X, \mathcal B(\hat X))$ being a \emph{$\delta$-lifted relation}, if there exists a probability measure $\mathscr{L}$, referred to as a \emph{lifting}, with a probability space $(X \times \hat X, \mathcal B(X \times \hat{X}), \mathscr{L})$ such that
	\begin{itemize} 
		\setlength{\itemsep}{0pt}
		\setlength{\parsep}{0pt}
		\setlength{\parskip}{0pt}
		\item $\forall \mathcal{X} \in \mathcal B(X), ~\mathscr{L}(\mathcal{X}\times\hat X ) = \Phi (\mathcal{X})$,
		\item $\forall \mathcal{\hat X} \in \mathcal B(\hat X), ~\mathscr{L}( X\times \mathcal{\hat X} ) = \Theta (\mathcal{\hat X})$,
		\item $\mathscr{L}(\mathscr{R})\geq 1-\delta$, i.e., for the probability space $(X \times \hat{X}, \mathcal B(X \times \hat{X}), \mathscr{L})$, one has $x\mathscr{R} \hat x$ with a probability of at least $1-\delta$.
	\end{itemize}
\end{definition}
Next, inspired by~\cite[Definition 3.2]{Lavaei2021Compositional}, we define ($\epsilon, \delta$)-approximate probabilistic relations between two gDTSGs based on the $\delta$-lifted relations between their probability measures.
\begin{definition}\label{Def: apr} \emph{(($\epsilon, \delta$)-Approximate Probabilistic Relations)}
	Consider gDTSGs $\mathfrak{D} =(X,U,W,X_0,Y,h)$ and $\widehat{\mathfrak{D}} =(\hat X,\hat U,$ $ \hat W,\hat{X}_0,\hat T,Y,\hat h)$ with the same output set.
	The gDTSG $\widehat{\mathfrak{D}}$ is ($\epsilon, \delta $)-stochastically simulated by $\mathfrak{D}$, denoted by $ \widehat{\mathfrak{D}}\preceq_{\epsilon}^{\delta}\mathfrak{D} $, if there exist relations $\mathscr{R}\subseteq X \times \hat X$, $\mathscr{R}_w\subseteq W \times \hat W$ and a Borel measurable stochastic kernel $\mathscr{L}_{T}(\cdot~|~ x, \hat{x},\hat{u}, w, \hat{w})$ on $X \times \hat X$ such that 
	\begin{itemize}
		\setlength{\itemsep}{0pt}
		\setlength{\parsep}{0pt}
		\setlength{\parskip}{0pt}
		\item $\forall (x,\hat x)\! \in\! \mathscr{R}$, $\Vert y- \hat y \Vert \!\leq\! \epsilon$, with $y \!=\! h(x)$ and $\hat y \!= \!\hat h (\hat x)$;
		\item $\forall (x,\hat x) \!\in\! \mathscr{R}$, and $\forall \hat u \!\in\! \hat U$, $\exists u \!\in \!U$ such that $\forall w\!\in\! W$,  $\exists \hat w \in \hat W$ with $(w,\hat w) \in \mathscr{R}_w$ such that one has $T(\cdot~|~ x, u, w)~\mathscr{\bar R}_{\delta} ~ \hat T(\cdot| \hat x,\hat u, \hat w)$ with lifting $\mathscr{L}_{T}(\cdot| x, \hat{x}, \hat{u}, w, \hat{w})$;
		\item $\forall x_0\in X_0$, $\exists \hat{x}_0\in\hat{X}_0$ such that $x_0 \mathscr{R}\hat{x}_0 $.
	\end{itemize}
\end{definition}
The second condition of Definition~\ref{Def: apr} implies implicitly that for any $\hat u \in \hat U$, there exists an \emph{interface function}~\cite{Girard2009Hierarchical} $u=\nu(x,\hat x, \hat u)$ with $u\in U$ such that the state probability measures are in the $\delta$-lifted relation after one-step transition. 
This function can be employed for refining $\hat u$ for $\widehat{\mathfrak{D}}$ to $u$ for $\mathfrak{D}$. 
Note that unlike~\cite[Definition 3.2]{Lavaei2021Compositional}, the interface map is not a function of $w\in W$ and $\hat{w}\in \hat{W}$ due to the asymmetric information pattern of the game, as discussed in Remark~\ref{rem:asy_info}. 
Once we have $ \widehat{\mathfrak{D}}\preceq_{\epsilon}^{\delta}\mathfrak{D} $, we are able to construct a product gDTSG based on $\mathfrak{D}$ and $\widehat{\mathfrak{D}}$, as defined in the following.
\begin{definition}\label{Def:couplingmodel}
	Consider gDTSGs $\mathfrak{D} =(X,U,W,X_0,T,Y,h)$ and $\widehat{\mathfrak{D}} =(\hat X,\hat U,\hat W,\hat{X}_0,\hat T,Y,\hat h)$ with $ \widehat{\mathfrak{D}}\preceq_{\epsilon}^{\delta}\mathfrak{D}$, interface function $\nu(x,\hat x, \hat u)$, and the corresponding lifted kernel $\mathscr{L}_T$.
	The product gDTSG of $\mathfrak{D}$ and $\widehat{\mathfrak{D}}$ is a gDTSG and defined as
	\begin{align*}
	\mathfrak{D}||_{\mathscr{R}}\widehat{\mathfrak{D}}:=(X_{||},U_{||},W_{||},X_{0||},T_{||},Y_{||},h_{||}),
	\end{align*}
	where  $X_{||}\!\!:=\!\!X \!\times \!\hat X$ is the state set; $U_{||}\!\!:=\!\hat U $ is the input set of Player~\uppercase\expandafter{\romannumeral1}; $W_{||}\!\!:=\!\!W $ is the input set of Player~\uppercase\expandafter{\romannumeral2}; $X_{0||}$ the initial state set, with $x_{0||}\!:=(x_0,\hat{x}_0)\in X_{0||}$, $x_0\in X_0$, $\hat{x}_0 \in \hat{X}_0$, and $(x_0,\hat{x}_0)\in \mathscr{R}$; $T_{||}:=\mathscr{L}_{T}$ is the stochastic transition kernel; $Y_{||}:= Y$ is the output set; and $h_{||}(x,\hat x):=h(x)$ is the output map.
\end{definition}
\begin{figure}[htbp]
	\centering
	\includegraphics[width=0.4\textwidth]{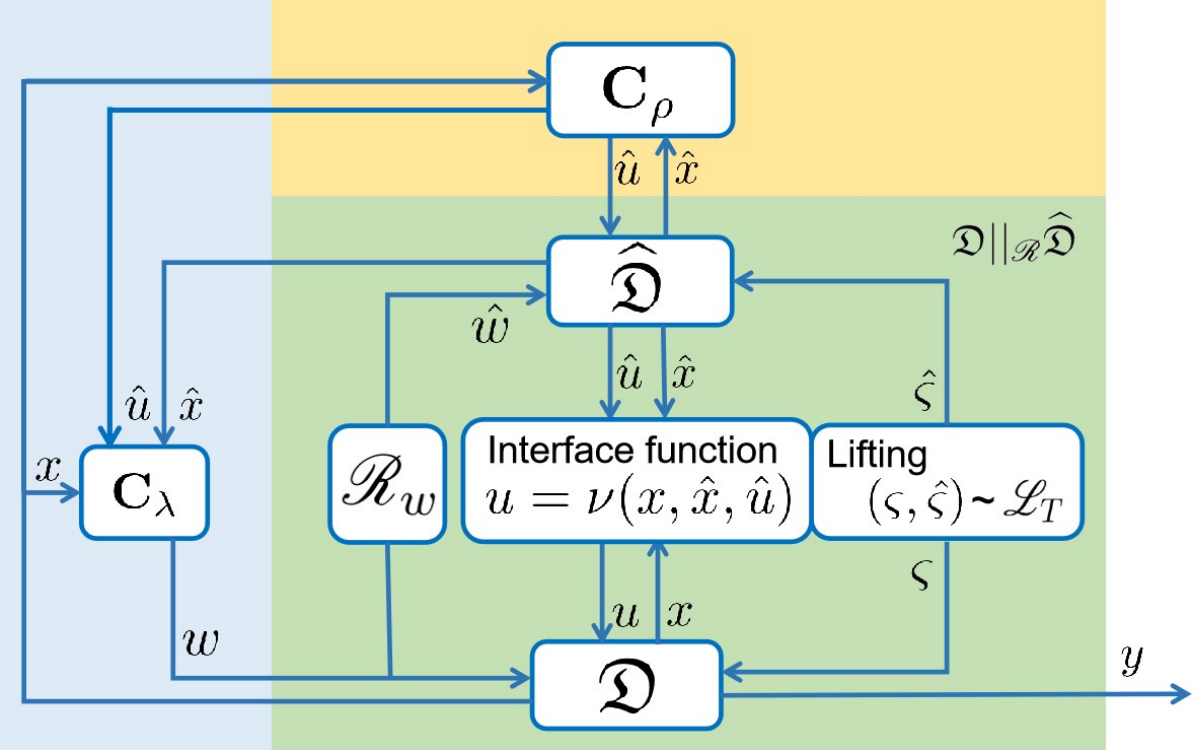}
	\caption{Coupling gDTSG $\mathfrak{D}||_{\mathscr{R}}\widehat{\mathfrak{D}}$ (green region) controlled by $\mathbf{C}_{\rho}$ (yellow region) and $\mathbf{C}_{\lambda}$ (blue region).}
	\label{fig:controller_equivalent}
\end{figure}
All ingredients of Definitions~\ref{Def: apr} and~\ref{Def:couplingmodel} are schematically depicted in Figure~\ref{fig:controller_equivalent}. 
Here, $\mathscr{L}_{T}$ characterizes the transition of states in $\mathfrak{D}||_{\mathscr{R}}\widehat{\mathfrak{D}}$ and specifies the relation of stochasticities between $\mathfrak{D}$ and $\widehat{\mathfrak{D}}$. 
Moreover, given an input $w$ from $\mathbf{C}_{\lambda}$, $\hat{w}$ is selected such that $(w,\hat{w})\in \mathscr{R}_w$ and fed to $\widehat{\mathfrak{D}}$.
In practice, we define a function 
\begin{equation}
\Pi_w:W\rightarrow \hat{W},\label{eq:pi_w}
\end{equation}
that matches each $w\in W$ to a $\hat{w}\in \hat{W}$. 
With $\Pi_w$, the stochastic kernel $\mathscr{L}_{T}$ as in Definition~\ref{Def: apr} can be written as $\mathscr{L}_{T}(\mathsf dx'\times d\hat{x}'~|~ x,\hat{x},\hat{u},w)$.
Moreover, according to~\cite[Corollary 3.1.2]{Borkar2012Probability}, one can decompose $\mathscr{L}_{T}$ as 
\begin{align}
\mathscr{L}_{T}(\mathsf dx'|x,\hat{x},\hat{x}',\nu(x,\hat{x},\hat{u}),w)\hat{T}(d\hat{x}'|\hat{x},\hat{u},\Pi_w(w))\label{eq:condition_kernel},
\end{align}
where $\mathscr{L}_{T}(\mathsf dx'|x,\hat{x},\hat{x}',\nu(x,\hat{x},\hat{u}),w)$ is a conditional stochastic kernel on $x'$ given $x$, $\hat{x}$, $\hat{x}'$, $\hat{u}$, and $w$.

\section{Abstraction Synthesis for a Class of Nonlinear gDTSG}\label{sec:nonlinear_case}
In this section, we focus on a particular class of nonlinear gDTSG, which is used to model many physical applications, such as fixed-joint robot~\cite{Fan2003Observer}, magnetic bearing~\cite{Arcak2001Observer}, etc.
This class of systems can be modeled as:
\begin{equation}\label{eq:nonlinsys}
\!\mathfrak{D}\!:\!
\left\{
\begin{aligned}
&x(k\!+\!1)\! =\! Ax(k)\!+\!Bu(k)\!+\!E\varphi(Fx(k))\!+\!Dw(k)\!+\!R\varsigma(k),\\
&y(k)=Cx(k),\quad \quad k\in\mathbb N,
\end{aligned}\\\right.
\end{equation}
with $A\in\R^{s\times s}$, $B\in\R^{s\times m}$, $E\in\R^{s\times 1}$,  $F\in\R^{1\times s}$,  $D\in\R^{s\times p}$,  $R\in\R^{s\times d}$, and $C\in\R^{q\times s}$.  
We assume that the stochasticity $\varsigma:\mathbb{N}\rightarrow\mathbb{R}^d$ in~\eqref{eq:nonlinsys} is a sequence of independent random vectors with multivariate standard normal distributions.
Moreover, the nonlinearity $\varphi:\R\rightarrow\R$ satisfies
\begin{align}
\underline{b}&\leq\frac{\varphi(c)-\varphi(d)}{c-d}\leq \bar{b},\label{ineq:varphi}
\end{align}
for all $c,d\!\in\!\R$, $c\!\neq\! d$, with some $\underline{b}, \bar{b}\!\in\! \R$ where $\underline{b}\!\leq\! \bar{b}$.
In the remainder of this paper, we use the tuple
\begin{align}\label{eq:tuple_for_system}
\mathfrak{D}=(A,B,C,D,E,F,R,\varphi),
\end{align}
to the systems as in~\eqref{eq:nonlinsys}.
Next, we first discuss the construction of finite abstractions for this class of systems.
Then, we propose how to establish an $(\epsilon,\delta)$-approximate probabilistic relation between original models and their corresponding abstractions.
\begin{remark}
		Note that although the main focus of this section is on establishing $(\epsilon,\delta)$-approximate probabilistic relations for a particular class of nonlinear gDTSGs as in~\eqref{eq:nonlinsys}, the proposed results in Sections~\ref{sec:apr_general} and~\ref{sec:safetyctr} on synthesizing controllers are independent of the form of dynamics and are applicable
		to the general setting of gDTSG.
\end{remark}

\subsection{Construction of Finite Abstractions}\label{subsec:ab_subsys}
Consider a gDTSG $\mathfrak{D}$ as in \eqref{eq:tuple_for_system}.
We first introduce the construction of a reduced-order version of $\mathfrak{D}$, denoted by $\widehat{\mathfrak{D}}_{\textsf r}\!=\!(\hat A_{\textsf r},\hat B_{\textsf r},\hat C_{\textsf r},\hat D_{\textsf r},\hat E_{\textsf r},\hat F_{\textsf r},\hat R_{\textsf r},\varphi)$, where the index ${\textsf r}$ signifies the reduced-order version of the original game throughout the paper. 
Then, we discuss how to build a finite abstraction, denoted by  $\widehat{\mathfrak{D}}$, for $\widehat{\mathfrak{D}}_{\textsf r}$.
Note that the reduced-order gDTSG $\widehat{\mathfrak{D}}_{\textsf r}$ is a simplified model of $\mathfrak{D}$, whose state and input sets are still continuous but with lower dimensions~\cite{Schilders2008Introduction}.
As a result, synthesizing controllers over reduced-order systems is more tractable than the original ones due to having less computational complexity (cf. Remark~\ref{moti_rn}).

To construct the reduced-order model $\widehat{\mathfrak{D}}_{\textsf r}$ for the gDTSG $\mathfrak{D}$ as in \eqref{eq:tuple_for_system}, we first need to select an abstraction matrix $P\in \R^{s\times \hat s}$ that maps the states of the abstraction to the $\mathfrak{D}$ as follows
\begin{equation}\label{eq:abs_map}
x = P \hat x.
\end{equation} 
Here, $\hat s$ denotes the dimension of the state space for $\widehat{\mathfrak{D}}_{\textsf r}$ ,
$x\in X\subseteq \R^s$ is the state of $\mathfrak{D}$, and $\hat x\in \hat X\subseteq\R^{\hat s}$ is the state of $\widehat{\mathfrak{D}}_{\textsf r}$.
With abstraction matrix $P$, we can construct the reduced-order game $\widehat{\mathfrak{D}}_{\textsf r}$ as long as the following equations hold for some matrices $G$, $Q$, and $S$ with appropriate dimensions:
\begin{align}
\hat{C}_{\textsf r} &= CP,\label{eq:infabs_cond1}\\
\hat{F}_{\textsf r}& = FP,\label{eq:infabs_cond2}\\
E &= P\hat{E}_{\textsf r}-BG,\label{eq:infabs_cond3}\\
AP &= P\hat{A}_{\textsf r}-BQ,\label{eq:infabs_cond4}\\
D &= P\hat{D}_{\textsf r}-BS.\label{eq:infabs_cond5}
\end{align}
Conditions~\eqref{eq:infabs_cond1} to~\eqref{eq:infabs_cond5} are similar to~\cite[conditions (5.5b) to (5.5f)]{Lavaei2021Compositional}.
We discuss later (cf.~\eqref{eq:good_Rr}) how to select $\hat R_{\textsf r}$ such that it is easier to establish an approximate probabilistic relation between the original game and its abstraction.
Additionally, we do not impose any restriction on the choice of $\hat B_{\textsf r}$.
For instant, one can choose $\hat B_{\textsf r} = I_{\hat s}$ so that $\widehat{\mathfrak{D}}_{\textsf r}$ is fully actuated, and, hence, solving the synthesis problem over it get easier. 
\begin{remark}\label{P_remark}
	Note that consider matrices $A$, $E$, $B$, and $P$.
	There exist matrices $\hat{A}_{\textsf r}$, $\hat{E}_{\textsf r}$, $G$, $Q$, and $S$ satisfying~\eqref{eq:infabs_cond3} to~\eqref{eq:infabs_cond5} if and only if~\cite[Lemma 5.10 and Lemma 5.12]{Zamani2018Compositional}
	$\textsl{im}\,AP \subseteq \textsl{im}\,P\,+\, \textsl{im}\, B$,
	$\textsl{im}\,D \subseteq \textsl{im}\,P\,+\, \textsl{im}\, B$, and
	$\textsl{im}\,E \subseteq \textsl{im}\, P\,+\,\textsl{im}\,B$.	
\end{remark}

We proceed with the construction of a finite abstraction $\widehat{\mathfrak{D}}$ of $\widehat{\mathfrak{D}}_{\textsf r}$. 
To this end, we introduce the~\emph{region of interest}, denoted by $\hat{X}_{rs}$, which is a compact subset of $\hat{X}_{\textsf r}$.
Note that this is usually the case for physical systems in practice, where variables evolve in a bounded domain.
Accordingly, we assume that $\widehat{\mathfrak{D}}_{\textsf r}$ will not come back to $\hat{X}_{rs}$ once it leaves $\hat{X}_{rs}$.
Instead, it will stay in a single absorbing state, denoted by $\phi$.
With these notions, we first partition $\hat{X}_{\textsf r}$ with $\hat{X}_{\textsf r}=\cup_{i\in \N} X_i$ and correspondingly select representative points $\hat{x}_i\in X_i$ for each cell, where $X_i$ are bounded cells.
Then, we use $\hat{X}=\{\Pi'(X_i)~|~X_i\cap (\hat{X}_{\textsf r}\backslash\hat{X}_{rs})=\emptyset\}\cup\{\phi\}$ as the state set of $\widehat{\mathfrak{D}}$, with $\Pi'$ being a function that maps $X_i$ to its representative points, and $\phi$ represents the aggregation of representative points in the set
\begin{equation}\label{eq:phi}
\big\{\Pi'(X_i)~|~X_i\cap (\hat{X}_{\textsf r}\backslash\hat{X}_{rs})\neq\emptyset\big\}.
\end{equation}
For the sake of succinctness, we use $\hat{X}=\{\hat x_i\}_{i=1}^{n_x}\cup \{\phi\}$ with $n_x$ being the number of representative points in the set $\{\Pi'(X_i)~|~X_i\cap (\hat{X}_{\textsf r}\backslash\hat{X}_{rs})=\emptyset\}$.
Additionally, we define $\tilde{\Pi}_x$ that maps any $\hat x_{\textsf r}\in \hat X_{\textsf r}$ to $\hat x_i=\Pi'(X_i)$ with $\hat x_{\textsf r}\in X_i$, based on which we define the set
\begin{equation}\label{eq:Delta}
\Delta:=\big\{\tilde{\Pi}_x(\hat x_{\textsf r})-\hat x_{\textsf r}\,|\, \hat x_{\textsf r}\,\in \hat X_{\textsf r}\big\}.
\end{equation} 
Since the partitions of $\hat{X}_{\textsf r}$ are bounded, $\Delta$ is also bounded, namely, there exists $\mathbf{\delta}'\in(-\infty,+\infty)$ such that $\forall \beta\in\Delta, \lVert\beta\rVert_{\infty}\leq \mathbf{\delta'}$.

Following the same idea for constructing the finite state set, we construct the finite input set of Player~\uppercase\expandafter{\romannumeral1} and Player~\uppercase\expandafter{\romannumeral2} by first selecting bounded partitions $\hat U_{\textsf r} = \cup_{n_u} U_i$ and $\hat W_{\textsf r} = \cup_{n_w} W_i$, and then choosing representative points $\hat u_i\in U_i$ and $\hat w_i\in W_i$.
Accordingly, we have $\hat U = \{\hat u_m\}_{m=1}^{n_u}$ being the input set for Player~\uppercase\expandafter{\romannumeral1} and $\hat W = \{\hat w_l\}_{l=1}^{n_w}$ being the input set for Player~\uppercase\expandafter{\romannumeral2}.
Similar to $\tilde{\Pi}_x$ as in~\eqref{eq:Delta}, we also define a function $\Pi_w:\hat{W}_{\textsf r}\rightarrow\hat{W}$ that maps any $\hat{w}_{\textsf r}\in\hat{W}_{\textsf r}$ to its representative point $\hat{w}\in\hat{W}$ of the partition that contains $\hat{w}_{\textsf r}$, and define a bounded set
\begin{equation}\label{eq:Delta_w}
\Delta_w:=\big\{\Pi_w(\hat w_{\textsf r})-\hat w_{\textsf r}\,|\, \hat w_{\textsf r}\,\in \hat W_{\textsf r}\big\}.
\end{equation} 
The dynamic of $\widehat{\mathfrak{D}}$ is constructed according to the dynamic of $\widehat{\mathfrak{D}}_{\textsf r}$ and the characteristic of $\phi$, \ie, $\hat x_{\textsf r} (k+1) = \hat f_{\textsf r} (\hat{x}_{\textsf r} (k),\hat{u}_{\textsf r}(k),$ $\hat{w}_{\textsf r} (k),\varsigma(k))=A_{\textsf r} \hat{x}_{\textsf r} (k)\!+\!E_{\textsf r} \varphi(F_{\textsf r} \hat{x}_{\textsf r} (k))\!+\!D_{\textsf r}\hat{w}_{\textsf r}(k)+B_{\textsf r}\hat{u}_{\textsf r}(k)+R_{\textsf r}\varsigma(k)$ when $\hat{x}_{\textsf r}(k)\!\in\! \hat{X}_{rs}$ and $\hat x_{\textsf r} (k+1)=\hat f_{\textsf r} (\hat{x}_{\textsf r} (k),\hat{u}_{\textsf r} (k),\hat{w}_{\textsf r} (k),\varsigma(k))=\phi$  when $\hat{x}_{\textsf r}(k) = \phi$.
Concretely, the dynamic of $\widehat{\mathfrak{D}}$ is given by
\begin{align}
\label{eq:abs_dyn}
\!\!\!\!\hat f(\hat{x}(k),\hat{u}(k),\hat{w}(k),\varsigma(k))\!:=\! \Pi_x(\hat f_{\textsf r} (\hat{x}_{\textsf r} (k),\hat{u}_{\textsf r} (k),\hat{w}_{\textsf r} (k),\varsigma(k)),	
\end{align}
where $\Pi_x:\hat X_{\textsf r}\rightarrow \hat X$ is the map that assigns to any $\hat x_{\textsf r}\in \hat X_{\textsf rs}$ the representative point $\hat x\in\hat X$ of the corresponding partition set containing $\hat x_{\textsf r}$, and assigns any $\hat x_{\textsf r}\in \hat X^c_{\textsf rs}$ to $\phi$. 
The output map is $\hat y = \hat C_{\textsf r} \hat x$ when $\hat{x}\neq \phi$, and $\hat y = \phi_y$ when $\hat{x} = \phi$, where $\phi_y$ represents the output when $\hat{x}= \phi$.
Then, we rewrite~\eqref{eq:abs_dyn} as 
\begin{align*}
\hat f(\hat{x}&(k),\hat{u}(k),\hat{w}(k),\varsigma(k))\!:=\! \hat f_{\textsf r} (\hat{x}_{\textsf r} (k),\hat{u}_{\textsf r} (k),\hat{w}_{\textsf r} (k),\varsigma(k))\!+\!\beta,~\beta\in\Delta.
\end{align*}
Finally, the initial state set of $\widehat{\mathfrak{D}}$ is defined as $\hat{X}_0 := \{\hat{x}_0 \in \hat{X}~|~\hat{x}_0 = \Pi_x(\hat{x}_{\textsf r0}),\hat{x}_{\textsf r0} \in \hat{X}_{\textsf r0}\}$, where $\hat{X}_{\textsf r0}$ is the initial state set of $\widehat{\mathfrak{D}}_{\textsf r}$, and the stochastic kernel $\hat{T}$ is computed as
\begin{align*}
\hat{T}(\hat{x}_h\,|\,&\hat{x}_{h'},\hat{u}_m,\hat{w}_l)\!=\left\{
\begin{aligned} 
&T(X_h|\hat{x}_{h'},\hat{u}_m,\hat{w}_l),\quad \text{if $\hat{x}_{h'},\hat{x}_h\in\{\hat{x}_i\}_{i=1}^{n_x}$},\\
&T(\hat X_{rs}^{c}|\hat{x}_{h'},\hat{u}_m,\hat{w}_l),\quad\! \text{if $\hat{x}_{h'}\!\in\!\{\hat{x}_i\}_{i=1}^{n_x}$,\! $\hat{x}_h\!=\phi$},\\
&\quad\quad\quad\quad 1, \quad\quad\quad \text{if $\hat{x}_{h'},\hat{x}_h=\phi$},\\
&\quad\quad\quad\quad 0,\quad\quad \quad\text{if $\hat{x}_{h'}\!=\phi$,\! $\hat{x}_h\!\in\!\{\hat{x}_i\}_{i=1}^{n_x}$}, 
\end{aligned}\right.
\end{align*}
with $h,h'\in[1,n_x]$, $\hat{x}_h=\Pi'(X_h)$, $\hat{u}_m\in\hat{U}$, and $\hat{w}_l\in\hat{W}$.
\begin{remark}\label{moti_rn}
		If the finite abstraction is directly constructed from original gDTSG, the size of $\hat{T}$ grows exponentially with the dimension of original state and input sets. As a promising alternative, 
		by constructing a reduced-order version of original gDTSG, the finite abstraction can be built based on gDTSG with a lower dimension, which alleviates the encountered computational complexity (cf. Section~\ref{cs_running}).
		One can also apply compositional techniques proposed in~\cite{Lavaei2021Compositional} for constructing finite abstractions of large-scale gDTSGs via abstractions of smaller subsystems, and utilize the techniques proposed in~\cite[Section 4.2]{Lavaei2020AMYTISS} to further reduce the memory usage required for storing the stochastic kernel of finite abstractions.
\end{remark}

\subsection{Conditions for Establishing Approximate Probabilistic Relations}\label{subs:construct_lifting}
In this subsection, we show under which conditions $\widehat{\mathfrak{D}}$ is $(\epsilon,\delta)$-stochastically simulated by $\mathfrak{D}$, denoted by $\widehat{\mathfrak{D}}\preceq^{\delta}_{\epsilon}\mathfrak{D}$, with respect to relations $\mathscr{R}$ and $\mathscr{R}_w$ defined as
\begin{align}
\mathscr{R} &= \big\{(x,\hat{x})\,|\,(x-P\hat{x})^TM(x-P\hat{x})\leq \epsilon^2\big\}\label{eq:Rx},\\
\mathscr{R}_w &= \big\{(w,\hat{w})\,|\,(w-\hat{w})^T\tilde{M}(w-\hat{w})\leq\tilde{\epsilon}^2\big\},\label{eq:Rw}
\end{align}
where $M$ and $\tilde{M}$ are positive-definite matrices with appropriate dimensions, 
and $\epsilon$, $\tilde{\epsilon}\in\R_{>0}$.
Prior to proposing the required conditions, we raise the following definition.
\begin{definition}\label{assmp_f}
		Consider a gDTSG $\mathfrak{D}=(A,B,C,D,E,F,R,$ $\varphi)$ as in~\eqref{eq:tuple_for_system}, its reduced-order version $\widehat{\mathfrak{D}}_{\textsf r}=(\hat A_{\textsf r},\hat B_{\textsf r},\hat C_{\textsf r},\hat D_{\textsf r},$ $\hat E_{\textsf r},\hat F_{\textsf r},\hat R_{\textsf r},\varphi)$ with the same additive noise, a finite abstraction $\widehat{\mathfrak{D}}$ constructed from $\widehat{\mathfrak{D}}_{\textsf r}$, and relations $\mathscr{R}$ and $\mathscr{R}_w$ as in~\eqref{eq:Rx} and~\eqref{eq:Rw}, respectively.
		For any $\kappa\in \R_{\geq0}$, and matrices $K,L\in\mathbb{R}^{m\times n}$, consider the following conditions:
		\begin{align}
		&M\succeq C^TC,\label{eq:liftcond1}\\
		&(A+BK)^TM(A+BK)\preceq\kappa M,\label{eq:liftcondM0}\\
		&\bar{A}^TM\bar{A}\preceq\kappa M,\label{eq:liftcondMu}\\
		&\underline{A}^TM\underline{A}\preceq\kappa M,\label{eq:liftcondMl}\\
		& \sqrt{\kappa}\leq 1-\tilde{\gamma}/\epsilon\leq 1,\label{eq:liftcond2}
		\end{align}
		in which $\bar{A}:=A+BK+\bar{b}(BL+EF)$ and $\underline{A}:=A+BK+\underline{b}(BL+EF)$ with $\underline{b}$ and $\overline{b}$ as appeared in~\eqref{ineq:varphi}, respectively, and $\tilde{\gamma} := \gamma_0 + \gamma_1 + \gamma_2 + \gamma_3 + \gamma_4$ with
		\begin{align}
		\gamma_0 &:= \mathop{\arg\max}_{\bar{w},\,\lVert\bar{w}\rVert_{\tilde{M}}\leq\tilde{\epsilon}}\lVert D\bar{w}\rVert_M,\label{eq:gamma0}\\
		\gamma_1 &:= \mathop{\arg\max}_{\hat{u}\in\hat{U}'}\lVert (B\tilde{R}-P\hat{B}_{\textsf r})\hat{u}\rVert_M,\label{eq:gamma1}\\
		\gamma_2 &:= \mathop{\arg\max}_{\varsigma,\,\lVert\varsigma\rVert\leq \chi^{-1}_d(1-\delta)}\lVert (R-P\hat{R}_{\textsf r})\varsigma\rVert_M,\label{eq:gamma2}\\
		\gamma_3 &:= \mathop{\arg\max}_{\beta\in\Delta}\lVert P\beta\rVert_M,\label{eq:gamma3}\\
		\gamma_4 &:= \mathop{\arg\max}_{\hat{w}\in\hat{W}}\lVert BS\hat{w}\rVert_M.\label{eq:gamma4}
		\end{align}
		In~\eqref{eq:gamma0}-\eqref{eq:gamma4}, $\chi^{-1}_d\!\!:\!\![0,1]\!\!\rightarrow\!\!\R$ is the chi-square inverse cumulative distribution function with $d$ degrees of freedom~\cite{Bernardo2009Bayesian}, $\lVert \bar{x}\rVert_M := \sqrt{\bar{x}^TM\bar{x}}$, $\Delta$ is as in~\eqref{eq:Delta}, and $\hat{U}'\!\!\subseteq\!\hat{U}$ is the input set for $\widehat{\mathfrak{D}}$.
\end{definition}

With Definition~\ref{assmp_f}, we are ready to introduce the required conditions under which one has $\widehat{\mathfrak{D}}\preceq^{\delta}_{\epsilon}\mathfrak{D}$ with respect to the relations as in~\eqref{eq:Rx} and~\eqref{eq:Rw}.
\begin{theorem}\label{thm:PRS}
	Consider a gDTSG $\mathfrak{D}$ and its finite abstraction $\widehat{\mathfrak{D}}$ constructed from $\widehat{\mathfrak{D}}_{\textsf r}$.
	For any $x_0\in X_0$ and $\hat{x}_0 \in \hat{X}_0$ with $(x_0,\hat{x}_0)\in\mathscr{R}$, $\widehat{\mathfrak{D}}$ is $(\epsilon,\delta)$-stochastically simulated by $\mathfrak{D}$ (i.e., $\widehat{\mathfrak{D}}\preceq^{\delta}_{\epsilon}\mathfrak{D}$) with respect to the relations as in~\eqref{eq:Rx} and~\eqref{eq:Rw}, if 
	\begin{itemize}
		\setlength{\itemsep}{0pt}
		\setlength{\parsep}{0pt}
		\setlength{\parskip}{0pt}
		\item \emph{(\textbf{Cd.1})} there exist $\kappa\in \R_{\geq0}$ and $K,L\in\mathbb{R}^{m\times n}$, such that conditions in~\eqref{eq:liftcond1}-\eqref{eq:liftcond2} holds;
		\item \emph{(\textbf{Cd.2})} the associated interface function is 
		\begin{align}
		\nu(x,\hat{x},\hat{u})\!:=\!(K+b(x,\hat{x})L)(x\!-\!P\hat{x})\!+\! Q\hat{x}\!+\!\tilde{R}\hat{u}\!+\!G\varphi(FP\hat{x}),\label{eq:interface}
		\end{align}
		with $P$, $Q$, and $G$ being as in~\eqref{eq:abs_map},~\eqref{eq:infabs_cond4}, and~\eqref{eq:infabs_cond3}, respectively, $K$ and $L$ being as in~\emph{(\textbf{Cd.1})} above, $\tilde{R}$ being a matrix with an appropriate dimension, 
		$$b(x,\hat{x}) = \frac{\varphi(Fx)-\varphi(FP\hat{x})}{F(x-P\hat{x})}\in[\underline{b},\overline{b}],$$ 
		if $x\neq P\hat{x}$, with $\underline{b}$ and $\overline{b}$ appeared in~\eqref{ineq:varphi}, and $b(x,\hat{x})=0$ otherwise;
		\item \emph{(\textbf{Cd.3})} and $\hat{U}'$ in~\eqref{eq:gamma1} is constructed such that $\forall \hat{u}\in\hat{U}'$ and $\forall (x,\hat{x})\in\mathscr{R}$, one has $\nu(x,\hat{x},\hat{u})\in U$.
	\end{itemize}
\end{theorem} 
The proof of Theorem~\ref{thm:PRS} is provided in Appendix~\ref{proof1}.
\begin{remark}\label{rem:init_and_Rtilde}
	Given an initial state $x_0$ of $\mathfrak{D}$, if there exists $\hat{x}_0\in\hat{X}$ such that $(x_0,\hat{x}_0)\in\mathscr{R}$, one can choose $\hat{x}_0 = \Pi_x((P^TMP)^{-1}P^TMx_0)$, which minimizes $\lVert x_0-P\hat{x}_0\rVert_M$.
	Moreover, we do not have any restriction on $\tilde{R}$ in~\eqref{eq:interface} in general. 
	However, we recommend using $\tilde{R}=(B^TB)^{-1}B^TP\hat{B}_{\textsf r}$ to obtain a smaller $\gamma_1$ as in~\eqref{eq:gamma1}.
	Then, it gets easier to find $\tilde{\gamma}$ and $\epsilon$ such that an approximate probabilistic relation exists (cf.~\eqref{eq:gamma_max}).
\end{remark}

Next, we propose another result to show that under which conditions, conditions~\eqref{eq:liftcond1}-\eqref{eq:liftcond2} hold.

\begin{corollary}\label{col:stable}
	Consider a gDTSG $\mathfrak{D}=(A,B,C,D,E,$ $F,R,\varphi)$.
	There exist $M$, $K$, $L$, $\epsilon$, and $\tilde{\gamma}$ such that~\eqref{eq:liftcond1}-\eqref{eq:liftcond2} hold if and only if for all $b'\in\{\underline{b},\bar{b},0\}$, the pair $(A+b'EF,B)$ is stabilizable, where $\underline{b}$ and $\overline{b}$ appeared in~\eqref{ineq:varphi}.
\end{corollary}
The proof of Corollary~\ref{col:stable} is given in Appendix~\ref{proof1}.
So far, we have introduced conditions under which there exists an approximate probabilistic relation between a gDTSG and its finite abstraction.
Next, we discuss how to establish such a relation based on those conditions.

\subsection{Algorithmic Procedure for Establishing Approximate Probabilistic Relation}
In this subsection, we propose an algorithmic procedure to search for $M$, $K$, $L$, and $\epsilon$ in Definition~\ref{assmp_f} given the following items s.t. (\textbf{Cd.1})-(\textbf{Cd.3}) in Theorem~\ref{thm:PRS} hold:
\begin{enumerate}[(i)]
	\setlength{\itemsep}{0pt}
	\setlength{\parsep}{0pt}
	\setlength{\parskip}{0pt}
	\item $\delta$ in the approximate probabilistic relation;
	\item a tolerable range for $\epsilon$, denoted by $[\epsilon_{min},\epsilon_{max}]$;
	\item the finite abstraction constructed as in Section~\ref{subsec:ab_subsys};
	\item  the set $\hat{U}'$ as in Definition~\ref{assmp_f} for synthesizing the controller over the finite abstraction.
\end{enumerate}
Here, we first discuss how to accommodate $\hat{U}'$ when searching for $M$, $K$, $L$, and $\epsilon$ so that (\textbf{Cd.3}) holds.
Then, we investigate how to jointly compute $M$, $K$, and $L$ given candidates $\epsilon\in [\epsilon_{min},\epsilon_{max}]$ and $\kappa\in[0,1]$, with $\kappa$ appeared in Definition~\ref{assmp_f}.
Finally, we formally propose the algorithmic procedure for establishing the approximate probabilistic relations.  

{\bf Accommodating $\mathbf{\hat{U}'}$.}
Here, we assume that all $\hat{u} \in\hat{U}$ are within a polytope defined by a matrix inequality
\begin{equation}
A_u\hat{u}\leq b_u\label{eq:input_con},
\end{equation}
where $A_u\in\mathbb{R}^{r\times m}$ and $b_u\in\mathbb{R}^{r\times 1}$.
Note that the input set of the form of~\eqref{eq:input_con} is appropriate for many physical systems.
Next, we substitute the interface function as in~\eqref{eq:interface} into~\eqref{eq:input_con}, and rewrite~\eqref{eq:input_con} as
\begin{equation}
A_u\bar{u}\leq \tilde{b}_u(\hat{x},\hat{u}),\label{eq:input_con_xu}
\end{equation}
with $\tilde{b}_u(\hat{x},\hat{u}) = b_u-A_u(Q\hat{x}+\tilde{R}\hat{u}+G\varphi(FP\hat{x}))$ and $\bar{u}=(K+bL)\bar{x}$, where $\bar{x}=x-P\hat{x}$.
One can readily see that every pair $(\hat{x},\hat{u})$ corresponds to a polytope for $\bar{u}$ specified by $A_u$ and $\tilde{b}_u(\hat{x},\hat{u})$, with $\hat{x}\in\hat{X}$ and $\hat{u}\in\hat{U}'$.
Here, we denote by $\mathsf{A}$ the set of all possible polytopes of the form of~\eqref{eq:input_con_xu} given $\hat{X}$ and $\hat{U}'$, and by
\begin{equation}
\tilde{A}\bar{u}\leq\tilde{b},\label{ineq:total_con_ubar}
\end{equation}
in which $\tilde{A}\in\R^{r\times m}$ and $\tilde{b}\in\R^{r\times1}$, a polytope $\tilde{\mathsf{A}}:= \cap_{r=1}^{a}\tilde{\mathsf{A}}_r$, with $a$ the number of polytopes within $\mathsf{A}$.
This polytope can be computed by multi-parametric toolbox \texttt{MPT}~\cite{Herceg2013Multia}.
We now rewrite the polytope in~\eqref{ineq:total_con_ubar} as:
\begin{equation}
\alpha_i\bar{u}\leq 1,~~i \in\{1,\ldots,r\},\label{ineq:constraints}
\end{equation}
with $\alpha_i = \frac{1}{\tilde{b}_i}\tilde{A}_i$, where $\tilde{A}_i$ and $\tilde{b}_i$ are the $i$-th row of $\tilde{A}$ and $\tilde{b}$, respectively.
Now, we are ready to introduce Theorem~\ref{thm:constraints_ineq}, which accommodates~\eqref{ineq:constraints} in the search for $M$, $K$, $L$, and $\epsilon$.

\begin{theorem}\label{thm:constraints_ineq}
	Consider a series of constraints as in~\eqref{ineq:constraints} for $\bar{u}:=(K+bL)\bar{x}$, with $\bar{x}\in\R^s$. 
	For all $\bar{x}\in E_x$ with $E_x :=\{\bar{x}~|~\bar{x}^TM\bar{x}\leq \epsilon^2\}$, $M\in\R^{s\times s}$, and $\epsilon\in\R_{>0}$, constraints as in~\eqref{ineq:constraints} are satisfied for all $b\in[\underline{b},\bar{b}]\cup \{0\}$ if and only if 
	\begin{align}
	\alpha_i K \bar{M} K^T\alpha^T_i\leq 1/\epsilon^2,\label{concond1}\\
	\alpha_i (K+\underline{b}L) \bar{M} (K+\underline{b}L)^T\alpha^T_i\leq 1/\epsilon^2,\label{concond2}\\
	\alpha_i (K+\bar{b}L) \bar{M} (K+\bar{b}L)^T\alpha^T_i\leq 1/\epsilon^2,\label{concond3}
	\end{align}
	with $\underline{b}$ and $\overline{b}$ appeared in~\eqref{ineq:varphi}, $\bar{M} = M^{-1}$, and $i \in\{1,\ldots,r\}$.
\end{theorem}
The proof of Theorem~\ref{thm:constraints_ineq} is provided in Appendix~\ref{proof1}.
Next, we proceed with studying how to apply Theorem~\ref{thm:constraints_ineq} when searching for $M$, $K$, and $L$.

{\bf Jointly Computing $\mathbf{M}$, $\mathbf{K}$, and $\mathbf{L}$.}
Consider~\eqref{eq:liftcond1}-\eqref{eq:liftcondMl} and~\eqref{concond1}-\eqref{concond3}.
When $\epsilon$ and $\kappa$ are fixed, $M$, $K$, and $L$ can be computed (if existing) by solving a semidefinite (SDP) programming problem~\cite{Toh1999SDPT3}.
Accordingly, one can first uniformly select samples from $[\epsilon_{min},\epsilon_{max}]$ and $[0,1]$ as candidates for $\epsilon$ and $\kappa$, respectively, and then try to compute $M$, $K$, and $L$ for each ($\epsilon$, $\kappa$) sample pairs. 
The next corollary shows how to compute $M$, $K$, and $L$ jointly, given $\delta$ and  a sample pair ($\epsilon$, $\kappa$).
\begin{corollary}~\label{cor:MKL}
	Consider a gDTSG $\mathfrak{D}=(A,B,C,D,E,$ $F,R,\varphi)$, input constraints as in~\eqref{ineq:constraints}, $\delta$ as in the approximate probabilistic relation, candidates $\epsilon\in[\epsilon_{min},\epsilon_{max}]$, and $\kappa \in [0,1]$.
	Matrix $M$ as in~\eqref{eq:Rx} as well as $K$ and $L$ as in~\eqref{eq:interface} can be computed jointly by solving the convex optimization problem:
	\begin{align*}
	\min_{\bar{M}} &~~~-\log(\det(\bar{M}))  \\
	\mbox{s.t.}&~~~\bar{M}\succ0;\\
	&~~~\begin{bmatrix}
	\bar{M}\ &\bar{M}C^T\\C\bar{M}\ &I_q
	\end{bmatrix}\succeq0;\\
	&~~~\begin{bmatrix}\bar{M}\ &\bar{A}_b\\\bar{A}^T_b\ &\kappa\bar{M}\end{bmatrix}\succeq 0,\,b\!\in\!\{\underline{b},\bar{b},0\};\\
	&~~~\begin{bmatrix}\!\!1/\epsilon^2&\alpha_i(\bar{K}\!+\!b\bar{L})\\(\bar{K}\!+\!b\bar{L})^T\alpha^T_i&\bar{M}\end{bmatrix}\succeq0,\\
	&~~~\text{with } i \in\{1,\ldots,r\}\text{ and }b\in\{\underline{b},\bar{b},0\};
	\end{align*}
	where $\bar{A}_b = (A+bEF)\bar{M}+B(\bar{K}+b\bar{L})$ and $\det(\bar{M})$ is the determinate of $\bar{M}$, with $\bar{M}$, $\bar{K}$, and $\bar{L}$ being matrices with appropriate dimensions. 
	If there is a solution for this optimization problem, one can compute $M$, $K$, and $L$ as $M=\bar{M}^{-1}$, $K=\bar{K}M$, and $L=\bar{L}M$, respectively, and we have $\widehat{\mathfrak{D}}\preceq^{\delta}_{\epsilon}\mathfrak{D}$, if
	\begin{equation}
	\tilde{\gamma}\leq\epsilon(1-\sqrt{\kappa}),\label{eq:gamma_max}
	\end{equation}
	with $\tilde{\gamma}$ being computed as in Definition~\ref{assmp_f}.
\end{corollary}
Corollary~\ref{cor:MKL} is a direct result of Theorems~\ref{thm:PRS} and~\ref{thm:constraints_ineq} with Schur complement~\cite{Boyd2004Convex}.
Additionally, one can design $\hat{R}_r$ as
\begin{equation}
\hat{R}_r = (P^TMP)^{-1}P^TMR,\label{eq:good_Rr}
\end{equation}
to minimize $\gamma_2$ for the selected $\delta$.
Finally, we summarize in Algorithm~\ref{alg:lifting} our solution to systematically establish an approximate probabilistic relation.
\begin{remark}
		The number of constraints in the optimization problem in Corollary~\ref{cor:MKL} grows linearly with the dimension of the system and $r$ as in~\eqref{ineq:constraints}.
		In practice, this problem can be solved efficiently with existing SDP solvers such as \texttt{SDPT3}~\cite{Toh1999SDPT3}.
\end{remark}
{\bf Running example (continued)}.
For constructing the finite abstraction, we select $P = [0.6199;0.4443;0.6219]$, and construct the reduced-order game with $\hat{A}_{\textsf r}=0.55$, $\hat{B}_{\textsf r}=1$, $\hat{D}_{\textsf r}=1$, $\hat{E}_{r}=0.32$, $\hat{F}_{r}=0.7957$, and $\hat{C}_{\textsf r}=0.1686$.
We therefore have $G = [-0.0334;-0.0311;-0.0342]$, $Q =[-0.1617;-0.1269;0.1877]$, and $S =[0.0021;0.0038;-0.0014]$ as in~\eqref{eq:infabs_cond3} to~\eqref{eq:infabs_cond5}. 
The finite abstraction for the reduced-order game is constructed as in Table~\ref{tbl:synthesis}, with $[-1.5,1.5]$ and $[-0.5,0.5]$ being selected as the the input set of Player~\uppercase\expandafter{\romannumeral1} and~\uppercase\expandafter{\romannumeral2} respectively.
Based on the discretization of the Player~\uppercase\expandafter{\romannumeral2}'s input set, we select $\tilde{M}=1$ and $\tilde{\epsilon}=0.05$.
As for establishing the ($\epsilon$,$\delta$)-approximate probabilistic relation, we set $\hat{U}'=\hat{U}$, 
$\delta = 0.001$, and the set for $\epsilon$ as $[0.05,1]$.
Then, by applying Algorithm~\ref{alg:lifting}, the finite abstraction is $(\epsilon,\delta)$-stochastically simulated by the original model with $\epsilon = 0.1509$, 
\begin{align*}
M=\begin{bmatrix}\begin{smallmatrix}0.0132\ &0.0082\ &0.0146\\0.0082\ &0.0110\ &0.0074\\0.0146\ &0.0074\ &0.0188\end{smallmatrix} \end{bmatrix}\!,
\end{align*}
$\hat{R}_{\textsf r}\!=\!0.8256$, and the interface function as in~\eqref{eq:interface} with 
\begin{align*}
&K\!=\!\begin{bmatrix}\begin{smallmatrix}-0.1163\ &-0.0355\ &-0.0999\\-0.0367\ &-0.0499\ &-0.0514\\0.0222\ &-0.0215\ &0.0125\end{smallmatrix} \end{bmatrix}\!,
~L\!=\!\begin{bmatrix}\begin{smallmatrix}-0.0450\ &-0.0824\ &-0.0200\\-0.0682\ &-0.0761\ &-0.0573\\0.0524\ &0.0666\ &0.0378\end{smallmatrix} \end{bmatrix}\!,
\end{align*}
and $\tilde{R} =[0.0422;0.0213;0.0562]$.	

\begin{algorithm}[h]
	\caption{Establishing an approximate probabilistic relation between a stochastic game and its abstraction}
	\label{alg:lifting} 
	\begin{center}
		\begin{algorithmic}[1]
			\State
			Select matrix $P$ as in Remark~\ref{P_remark}, compute $\hat C_{\textsf r}$, $\hat F_{\textsf r}$, $\hat E_{\textsf r}$, $\hat A_{\textsf r}$, and $\hat D_{\textsf r}$ following~\eqref{eq:infabs_cond1}-\eqref{eq:infabs_cond5}, and choose $\hat B_{\textsf r}$ freely;  
			\State
			Discretize the state set as well as the input sets of Player~\uppercase\expandafter{\romannumeral1} and Player~\uppercase\expandafter{\romannumeral2}, and then select $\tilde{M}$ and $\tilde{\epsilon}$ in~\eqref{eq:Rw} according to the discretization of $\hat{W}_{\textsf r}$;
			\State
			Select $\hat{U}'\subseteq \hat{U}$ for synthesizing controllers over the finite abstraction, compute $\tilde{R}$ in~\eqref{eq:interface} according to Remark~\ref{rem:init_and_Rtilde}, and compute constraints in~\eqref{ineq:constraints};
			\State 
			Select $\delta$ and appropriate interval $[\epsilon_{min},\epsilon_{max}]$. Then, uniformly select samples of $\epsilon$ within $[\epsilon_{min},\epsilon_{max}]$ and $\kappa$ within [0,1]. 
			For each $(\epsilon,\kappa)$,
			\begin{enumerate}[(i)]
				\setlength{\itemsep}{0pt}
				\setlength{\parsep}{0pt}
				\setlength{\parskip}{0pt}
				\item Compute $M$, $K$, and $L$ as in Corollary~\ref{cor:MKL};
				\item If there are solutions in Step $4$(i), compute $\hat{R}_r$ as in~\eqref{eq:good_Rr}, then compute $\tilde{\gamma}$ as in Definition~\ref{assmp_f} and check~\eqref{eq:gamma_max} accordingly;
				\item If~\eqref{eq:gamma_max} in Step $4$(ii) holds, solutions for $M$,$K$,$L$, and $\epsilon$ are founded for establishing the relation.
			\end{enumerate}	
		\end{algorithmic}
	\end{center}
\end{algorithm}

\section{Controller Synthesis Problem}\label{sec:safetyctr}
In this section, we discuss the synthesis of controller $\tilde{\mathbf{C}}_{\rho}$ for a gDTSG $\mathfrak{D}\!\!=\!\! (X,U,W,X_0,T\!,Y\!,h)$ for Problems~\ref{problem1} and~\ref{problem2}, given a finite abstraction $\widehat{\mathfrak{D}}\!\!=\!\! (\hat X,\hat U, \hat W,\hat{X}_0, \hat T, Y,\hat h)$ of $\mathfrak{D}$ with $\widehat{\mathfrak{D}}\!\preceq_{\epsilon}^{\delta}\!\mathfrak{D}$, and a property $(\mathcal{A},H)$, with $\mathcal{A}\!=\!(Q, q_0, \Pi,\tau, F)$.
\begin{figure}[htbp]
	\centering
	\subfigure{
		\includegraphics[width=0.14\textwidth]{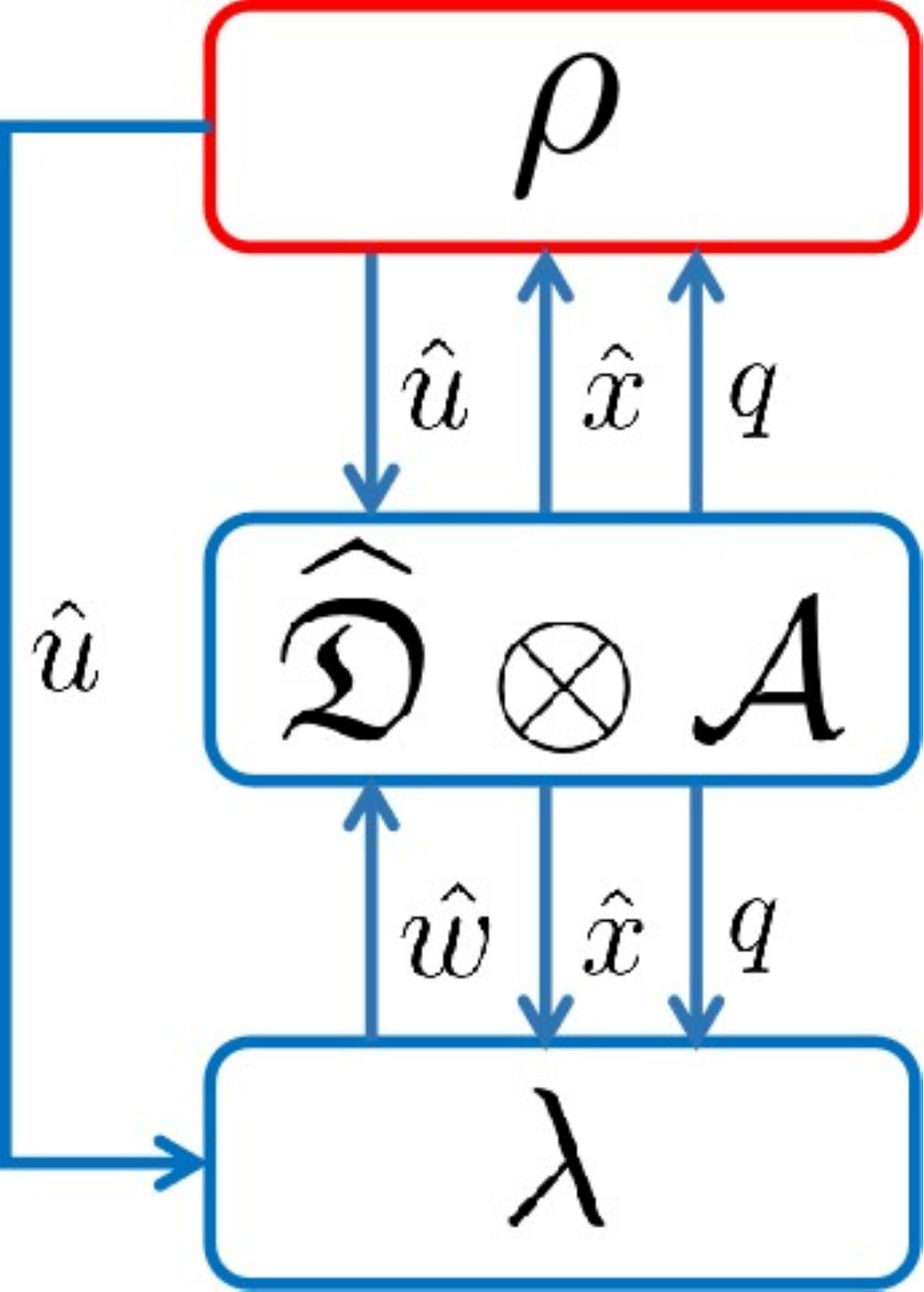}
	}
	\hspace{0.8cm}
	\subfigure{
		\includegraphics[width=0.35\textwidth]{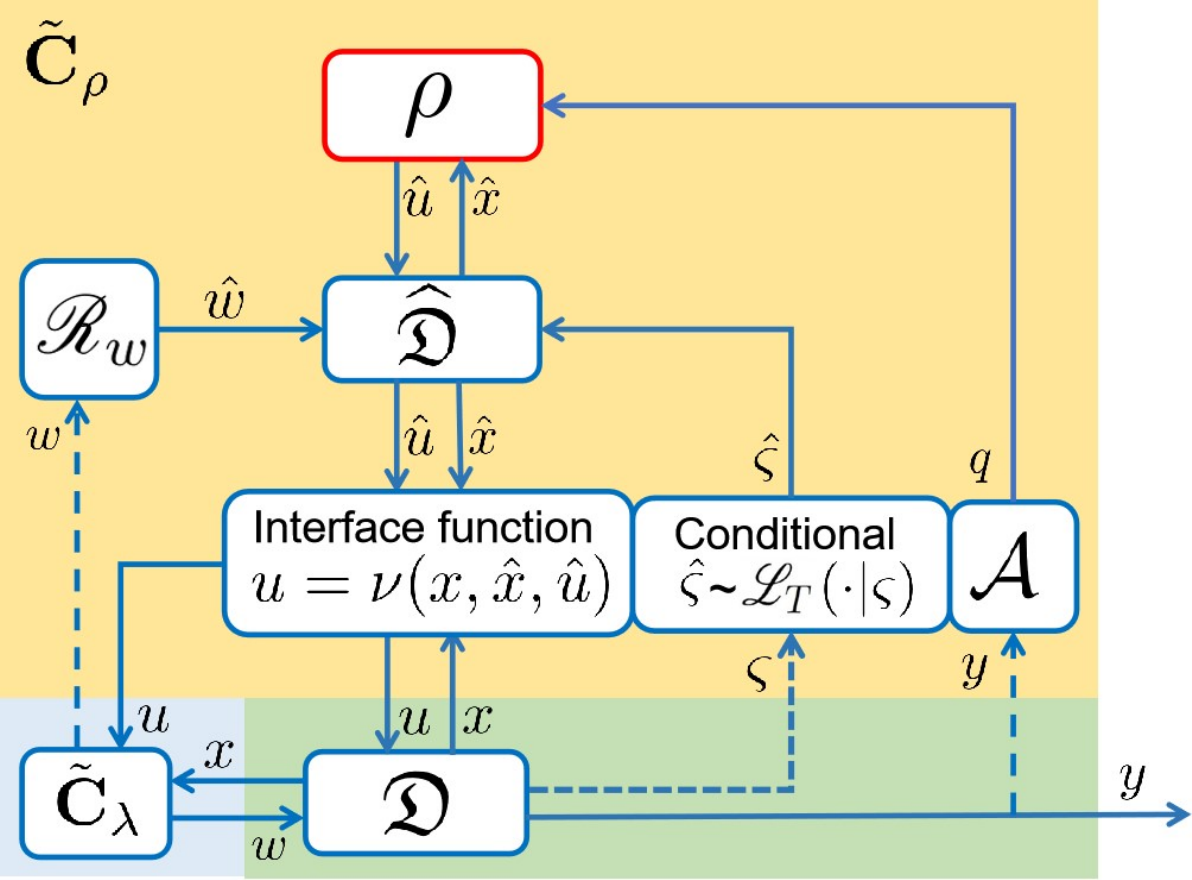}
	}
	\caption{\textbf{Left:} Synthesizing Markov policy for $\widehat{\mathfrak{D}}\otimes\mathcal{A}$.~\textbf{Right:} Construction of $\tilde{\mathbf{C}}_{\rho}$ (yellow region).}
	\label{fig:idea}
\end{figure}
The general idea of our methods is depicted in Figure~\ref{fig:idea} and summarized as follows:
\begin{itemize}
	\setlength{\itemsep}{0pt}
	\setlength{\parsep}{0pt}
	\setlength{\parskip}{0pt}
	\item As shown in Figure~\ref{fig:idea} (left), we first synthesize a Markov policy $\rho$ for Player~\uppercase\expandafter{\romannumeral1} of the gDTSG $\widehat{\mathfrak{D}}\otimes\mathcal{A}$, assuming that Player~\uppercase\expandafter{\romannumeral2} of the gDTSG selects its actions in a rational fashion against the choice of Player~\uppercase\expandafter{\romannumeral1}.
	The outcomes are the Markov policy $\rho$ and the robust satisfaction probability $\mathbf{s}$ for Problem~\ref{problem1} (resp. worst-case violation probability $\mathbf{v}$ for Problem~\ref{problem2});
	\item We then construct $\tilde{\mathbf{C}}_{\rho}$ based on $\rho$ (cf. Definition~\ref{def:C_rho}) as depicted in Figure~\ref{fig:idea} (right).
	At runtime, when a state $x$ of $\mathfrak{D}$ is fed to $\tilde{\mathbf{C}}_{\rho}$:
	\begin{enumerate}[(i)]
		\setlength{\itemsep}{0pt}
		\setlength{\parsep}{0pt}
		\setlength{\parskip}{0pt}
		\item  State $\hat{x}$ of $\widehat{\mathfrak{D}}$ is first updated according to $x$, the conditional stochastic kernel $\mathscr{L}_T$, and the action $w$ of $\tilde{\mathbf{C}}_{\lambda}$ in the previous time instant.
		Then, the state $q$ of $\mathcal{A}$ are updated according to the output function $h(x)$ of $\mathfrak{D}$ and the transition function $\tau$ of $\mathcal{A}$;
		\item  Afterwards, a $\hat{u}$ is provided by $\rho$ based on $\hat{x}$ and $q$, and refined to $\mathfrak{D}$ by virtue of the interface function $\nu$; 
		\item  $\tilde{\mathbf{C}}_{\lambda}$ selects $w$ according to $x$ and $u$, and feeds $w$ to $\mathfrak{D}$. 
	\end{enumerate}
\end{itemize}
Here, we formally present the construction of $\tilde{\mathbf{C}}_{\rho}$.
\begin{definition}\label{def:C_rho}
	\emph{(Construction of $\tilde{\mathbf{C}}_{\rho}$)} 
	Consider gDTSGs $\mathfrak{D}=(X,U,W,X_0,T,Y,h)$ and $\widehat{\mathfrak{D}}\!=\!(\hat X,\hat U, \hat W,\hat{X}_0, \hat T,$ $Y,\hat h)$ with $\widehat{\mathfrak{D}}\preceq_{\epsilon}^{\delta}\mathfrak{D}$.
	Given a Markov policy $\rho\!=\!(\rho_{0},\rho_{1},\ldots,$ $\rho_{H-1})$ for Player~\uppercase\expandafter{\romannumeral1} of $\widehat{\mathfrak{D}}\otimes\mathcal{A}$, we construct $\tilde{\mathbf{C}}_{\rho}= (\tilde{\mathsf{M}},\tilde{\mathsf{U}},\tilde{\mathsf{Y}},\tilde{\mathsf{H}},\tilde{\mathsf{M}}_{0},\tilde{\pi}_{\mathsf{M}},$ $\tilde{\pi}_{\mathsf{Y}})$ for Player~\uppercase\expandafter{\romannumeral1} of $\mathfrak{D}$ with $\tilde{\mathsf{M}}=X\times\hat{X}\times Q\times W\times\hat{W}$, $\tilde{\mathsf{U}}=X\times W$, $\tilde{\mathsf{Y}}=U$, $\tilde{\mathsf{H}}=[0,H-1]$,
	\begin{itemize}
		\setlength{\itemsep}{0pt}
		\setlength{\parsep}{0pt}
		\setlength{\parskip}{0pt}
		\item $\tilde{\mathsf{m}}_0\!\!=\!\!\big(\tilde{\mathsf{m}}_{X}(0),\tilde{\mathsf{m}}_{\hat{X}}(0),\tilde{\mathsf{m}}_{Q}(0),\tilde{\mathsf{m}}_{W}(0),\tilde{\mathsf{m}}_{\hat{W}}(0)\big)\!\!\in\!\tilde{\mathsf{M}}_0$, with
		$\tilde{\mathsf{m}}_{X}(0)=x_0$, where $x_0\in X_0$; $\tilde{\mathsf{m}}_{\hat{X}}(0)=\hat{x}_0$ such that $(x_0,\hat{x}_0)\in \mathscr{R}$, where $\mathscr{R}$ is as in~\eqref{eq:Rx}; $\tilde{\mathsf{m}}_{Q}(0)=\tau\big(q_0,L\circ h(\tilde{\mathsf{m}}_{X}(0))\big)$;
		$\tilde{\mathsf{m}}_{W}(0)$ is initialized as $\tilde{\mathsf{m}}_{W}(0)=w(0)$ after Player~\uppercase\expandafter{\romannumeral2} of $\mathfrak{D}$ has chosen $w(0)$, and $\tilde{\mathsf{m}}_{\hat{W}}(0)$ is accordingly initialized as $\tilde{\mathsf{m}}_{\hat{W}}(0)=\Pi_w(w(0))$ with $\Pi_w$ as in~\eqref{eq:pi_w};
		\item $\tilde{\pi}_{\mathsf{M}}$ updates $\big(\tilde{\mathsf{m}}_{X}(k),\tilde{\mathsf{m}}_{\hat{X}}(k),\tilde{\mathsf{m}}_{Q}(k),\tilde{\mathsf{m}}_{W}(k),\tilde{\mathsf{m}}_{\hat{W}}(k)\big)$ $ \in\tilde{\mathsf{M}}$ at all time instants $k\in\mathsf{H}\backslash\{0\}$, with the following steps:
		\begin{enumerate}[(i)]
			\setlength{\itemsep}{0pt}
			\setlength{\parsep}{0pt}
			\setlength{\parskip}{0pt}
			\item update $\tilde{\mathsf{m}}_{\hat{X}}(k)$ according to the conditional kernel:
			\begin{align*}
			\mathscr{L}_{T}\big(d\hat{x}|\tilde{\mathsf{m}}_{\hat X}(k\!-\!1),&\tilde{\mathsf{m}}_{X}(k\!-\!1),x(k),\hat{u}(k\!-\!1),\tilde{\mathsf{m}}_{W}(k\!-\!1)\big),
			\end{align*}
			where $x(k)$ is the state of $\mathfrak{D}$, $\hat{u}(k-1)=\tilde{\rho}_k(\tilde{\mathsf{m}}_X(k-1),\tilde{\mathsf{m}}_{\hat X}(k-1),\tilde{\mathsf{m}}_{Q}(k-1))$, and $\mathscr{L}_{T}(\cdot)$ as in~\eqref{eq:condition_kernel}; 
			\item update $\tilde{\mathsf{m}}_{X}(k)$ with $\tilde{\mathsf{m}}_{X}(k)=x(k)$;
			\item update $\tilde{\mathsf{m}}_{Q}(k)$ with $\tilde{\mathsf{m}}_{Q}(k)\!=\!\tau\big(\tilde{\mathsf{m}}_{Q}(k\!-\!1),L\!\circ\! h(\tilde{\mathsf{m}}_{X}(k))\big)$;
			\item update $\tilde{\mathsf{m}}_W(k)$ with $\tilde{\mathsf{m}}_W(k)=w(k)$ after Player~\uppercase\expandafter{\romannumeral2} of $\mathfrak{D}$ has selected $w(k)$, and accordingly update $\tilde{\mathsf{m}}_{\hat{W}}(k)$ as $\tilde{\mathsf{m}}_{\hat{W}}(k) = \Pi_w(w(k))$ with $\Pi_w$ as in~\eqref{eq:pi_w};
		\end{enumerate}
		\item $\tilde{\pi}_{\mathsf{Y}}$ updates $\mathsf{y}(k)\in\mathsf{Y}$ at the time instant $k\in\mathsf{H}$ with $\mathsf{y}(k)=$ $\nu\big(\tilde{\mathsf{m}}_X(k),$ $\tilde{\mathsf{m}}_{\hat X}(k),\rho_k(\tilde{\mathsf{m}}_{\hat X}(k),\tilde{\mathsf{m}}_{Q}(k))\big)$, where $\nu$ is the interface function associated with the ($\epsilon$,$\delta$)-approximate probabilistic relation.	
	\end{itemize}
\end{definition}
The remaining problem is how to synthesize the Markov policy $\rho$ for $\widehat{\mathfrak{D}}\!\otimes\!\mathcal{A}$. 
In Sections~\ref{sec:robust_max} and~\ref{sec:robust_min}, we propose new Bellman operators to synthesize $\rho$ for Problems~\ref{problem1} and~\ref{problem2}, respectively.
Prior to introducing these operators, we point out that these operators require the following assumption.
\begin{assumption}\label{asp1}
	Consider gDTSGs $\mathfrak{D}= (X,U,W,X_0,T,Y,h)$ and $\widehat{\mathfrak{D}}= (\hat X,\hat U, \hat W,\hat{X}_0, \hat T, Y,\hat h)$ with $\widehat{\mathfrak{D}}\preceq^{\delta}_{\epsilon}\mathfrak{D}$ regarding relations $\mathscr{R}$ and $\mathscr{R}_w$ as in Definition~\ref{Def: apr}.
	For all $\hat{w}\in \hat{W}$ and $w\in W$ with $(w,\hat w) \in \mathscr{R}_w$, we assume that 
	\begin{equation*}
	\int_{\bar{\mathscr{R}}_{\hat{x}'}}\mathscr{L}_{T}(\mathsf dx'|x,\hat{x},\hat{x}',\nu(x,\hat{x},\hat{u}),w,\hat{w})\geq 1-\delta,
	\end{equation*}
	holds $\forall \hat{x},\hat{x}'\!\in\!\hat{X}$, with $\bar{\mathscr{R}}_{\hat{x}'}\!=\!\{x' \in X | (x',\hat{x}')\in \mathscr{R}\}$ and $\mathscr{L}_{T}(\mathsf dx'|x,$ $\hat{x},\hat{x}',\nu(x,\hat{x},\hat{u}),w,\hat{w})$ as the conditional probability of $x'\in X$ given $x\in X$, $\hat{x}$, $\hat{x}'$, $w$, $\hat{w}$, and the interface function $\nu(x,\hat{x},\hat{u})$.
\end{assumption}
\begin{remark}\label{cmp_curre}
		Assumption~\ref{asp1} presumes that all states of $\widehat{\mathfrak{D}}$ are coupled into the $\delta$-lifted relation, and at every time instant $k$, $\mathbb{P}\{(x',\hat{x}')\!\in\!\mathscr{R}|(x,\hat{x})\!\in\! \mathscr{R},\forall(w,\hat{w})\!\in\!\mathscr{R}_w\}\!\geq\! 1\!-\!\delta$ holds for all $\hat{x}\in\hat{X}$ via the interface function used in controller refinement, with $(x,\hat{x})$ and $(x',\hat{x}')$ being the state pairs at time instants $k$ and $k+1$, respectively. 
		Given the existing results on ($\epsilon$,$\delta$)-approximate probabilistic relations~\cite{Haesaert2020Robust,Lavaei2021Compositional,Huijgevoort2020Similarity}, 
		Assumption~\ref{asp1} does not introduce extra subtlety in practice.
		In fact, although the results in~\cite{Haesaert2020Robust,Lavaei2021Compositional,Huijgevoort2020Similarity} do not explicitly require such an assumption, the existence of an ($\epsilon$,$\delta$) approximate probabilistic relation is guaranteed by enforcing Assumption~\ref{asp1} (cf.~\cite[Condition A3]{Haesaert2020Robust},~\cite[Theorem 5.5]{Lavaei2021Compositional} and~\cite[Theorem 3]{Huijgevoort2020Similarity}).
\end{remark}

\subsection{Robust Satisfaction Problem}\label{sec:robust_max}
We start with discussing how to synthesize the Markov policy $\rho$ for the problem of robust satisfaction as in Problem~\ref{problem1}.
Consider a gDTSG $\mathfrak{D}= (X,U,W,X_0, T,Y,h)$ and its finite abstraction $\widehat{\mathfrak{D}}= (\hat X,\hat U,\hat W,\hat{X}_0, \hat T,Y,\hat h)$, a property $(\mathcal{A},H)$ with $\mathcal{A}=(Q, q_0, \Pi,\tau, F)$, and the product gDTSG $\widehat{\mathfrak{D}}\otimes\mathcal{A} = \{\bar{X},\bar{U},\bar{W},\bar{X}_0,\bar{T},\bar{Y},\bar{h}\}$ as in Definition~\ref{def:product_gmdp_dfa}.
Given a Markov policy $\rho=(\rho_{0},\rho_{1},$ $\ldots,\rho_{H-1})$ for Player~\uppercase\expandafter{\romannumeral1} and $\lambda=(\lambda_{0},\lambda_{1},\ldots,\lambda_{H-1})$ for Player~\uppercase\expandafter{\romannumeral2} of $\widehat{\mathfrak{D}}\otimes\mathcal{A}$, we define a cost-to-go function $\bar{V}^{\rho,\lambda}_n:\hat{X}\times Q \rightarrow [0,1]$ that assigns a real number to states of $\widehat{\mathfrak{D}}\otimes\mathcal{A}$ at the time instant $H-n$.
We initialize $\bar{V}^{\rho,\lambda}_{n+1}(\hat{x},q)$ with $\bar{V}_0^{\rho,\lambda}(\hat{x},q)=1$ when $q\in F$ and $\bar{V}_0^{\rho,\lambda}(\hat{x},q)=0$, otherwise, and recursively compute it  as
\begin{equation}\label{eq:P_general}
\bar{V}_{n+1}^{\rho,\lambda}(\hat{x},q) = \mathfrak{P}(\bar{V}_{n}^{\rho,\lambda})(\hat{x},q).
\end{equation}
\vspace{-0.2cm}
Here, $\mathfrak{P}$ is a Bellman operator defined as 
\begin{align}
\mathfrak{P}(\bar{V}_{n}^{\rho,\lambda})(\hat{x},q):=\left\{
\begin{aligned} 
&(1-\delta)\sum_{\hat{x}'\in \hat{X}}\bar{V}^{\rho,\lambda}_{n}(\hat{x}',\underline{q}(\hat{x}',q))\hat{T}(\hat{x}'|\hat{x},\hat{u},\hat{w}), ~\text{ if } q\notin F;\\
& \quad \quad \quad \quad 1, \quad \quad \quad\quad \quad \quad\quad \quad \quad\quad \quad \quad\quad\quad \text{if } q\in F,
\end{aligned}\right.\label{eq:value_gen_sac_ab}
\end{align}
with  $\hat{u}=\rho_{H-n-1}(\hat{x},q)$, $\hat{w}=\lambda_{H-n-1}(\hat{x},q,\hat{u})$, and
\begin{equation}\label{eq:underline_p}
\underline{q}(\hat{x}',q) = \mathop{\arg\min}_{q'\in Q'_{\epsilon}(\hat{x}')}\bar{V}^{\rho,\lambda}_{n}(\hat{x}',q'),
\end{equation}
where 
\begin{align}\label{eq:Q_epsilon}
Q'_{\epsilon}(\hat{x}') \!:=\! \Big\{q'\!\in\! Q\mid\!\exists x\!\in\! X,q'\!=\! \tau(q,L\!\circ\! h(x)),\text{with}\ h(x)\!\in\! {N}_{\epsilon}(\hat{h}(\hat{x}'))\Big\},
\end{align}
and $\mathcal{N}_{\epsilon}(\hat{y}):=\{y\in Y\,|\, \lVert y-\hat{y}\rVert\leq \epsilon \}$.
Moreover, given a Markov policy $\rho$ for Player~\uppercase\expandafter{\romannumeral1}, the corresponding worst-case adversarial policy $\lambda_*(\rho)$ for Player~\uppercase\expandafter{\romannumeral2} can be computed as
\begin{align}
\lambda_{*_{H-n-1}}(\rho)\!\in\!\!\!\!\min_{\lambda'_{H-n-1}\in \Lambda}\!\!(1\!-\!\delta)\!\!\sum_{\hat{x}'\in \hat{X}}\!\bar{V}^{\rho,\lambda_*(\rho)}_{n}\!(\hat{x}'\!,\underline{q}(\hat{x}'\!,q))\hat{T}(\hat{x}'|\hat{x},\hat{u},\hat{w})\label{eq:policy_max_worst_case},
\end{align}
for all $n\in[0,H-1]$, with $\hat{u}=\rho_{H-n-1}(\hat{x},q)$ and $\hat{w}=\lambda'_{H-n-1}(\hat{x},$ $q,\hat{u})$.
Now, we are ready to propose one of the main results for the problem of robust satisfaction.

\begin{theorem}\label{thm:gua_prmax}
	Consider gDTSGs $\mathfrak{D}\!\!=\!\! (X,U,W,X_0,T,Y,h)$ and $\widehat{\mathfrak{D}}=(\hat X,\hat U, \hat W,\hat{X}_0, \hat T, Y,\hat h)$ with $\widehat{\mathfrak{D}}\preceq_{\epsilon}^{\delta}\mathfrak{D}$, and a property $(\mathcal{A},H)$ with $\mathcal{A}=(Q, q_0, \Pi,\tau, F)$.
	Given a Markov policy $\rho$ designed for Player~\uppercase\expandafter{\romannumeral1} of $\widehat{\mathfrak{D}}\otimes\mathcal{A}$ and a control strategy $\tilde{\mathbf{C}}_{\rho}$ for Player~\uppercase\expandafter{\romannumeral1} of $\mathfrak{D}$ that is constructed based on $\rho$ as in Definition~\ref{def:C_rho}, for any control strategy $\tilde{\mathbf{C}}_{\lambda}$ for Player~\uppercase\expandafter{\romannumeral2} of $\mathfrak{D}$, we have
	\begin{equation}\label{eq:thm_sad_pmax}
	\mathbb{P}_{(\tilde{\mathbf{C}}_{\rho},\tilde{\mathbf{C}}_{\lambda})\times \mathfrak{D}}\big\{\exists k\leq H, y_{\omega k}\models \mathcal{A}\big\}\geq \bar{V}_H^{\rho,\lambda_*(\rho)}(\hat{x}_0,\bar{q}_0),
	\end{equation}
	where $\hat{x}_0\in \hat{X}_0$ and $x_0\in X_0$, with $(x_0,\hat{x}_0)\in\mathscr{R}$ and $\mathscr{R}$ as in~\eqref{eq:Rx}, $\bar{V}_H^{\rho,\lambda_*(\rho)}(\hat{x}_0,\bar{q}_0)$ is computed as in~\eqref{eq:P_general}, with $\lambda_*(\rho)$ as in~\eqref{eq:policy_max_worst_case} and $\bar{q}_0 = \tau(q_0,L\circ h(x_0))$.
\end{theorem}
The proof of Theorem~\ref{thm:gua_prmax} is provided in Appendix~\ref{proof2.1}.
In practice, we are interested in constructing a $\rho$ that maximizes the robust satisfaction probability, \ie, $\bar{V}_H^{\rho,\lambda_*(\rho)}(\hat{x}_0,\bar{q}_0)$ as in~\eqref{eq:thm_sad_pmax}. 
We discuss in the following proposition how to synthesize such a policy.

\begin{proposition}\label{prop:maxmin}
	Consider gDTSGs $\mathfrak{D}=(X,U,W,X_0,T,Y,h)$ and $\widehat{\mathfrak{D}}\!\!=\!\!(\hat X,\hat U, \hat W,\hat{X}_0, \hat T, Y,\hat h)$ with $\widehat{\mathfrak{D}}\!\!\preceq_{\epsilon}^{\delta}\!\!\mathfrak{D}$, and a property $(\mathcal{A},H)$ with $\mathcal{A}=(Q, q_0, \Pi,\tau, F)$.
	Considering that Player~\uppercase\expandafter{\romannumeral2} minimizes $\bar{V}_{n+1}^{\rho,\lambda}\!(\hat{x},q)$ according to $\rho$, the Markov policy $\rho^*\!=\!(\rho_{0}^*,\rho_{1}^*,\ldots,\rho_{H-1}^*)$ for Player~\uppercase\expandafter{\romannumeral1} maximizes $\bar{V}_{n+1}^{\rho,\lambda}(\hat{x},q)$, with
	\begin{align}
	\rho^*_{H-n-1}\!\in\!\mathop{\arg}\mathop{\max}_{\rho_{H-n-1}\in \mathcal P}\min_{\lambda_{H-n-1}\in \Lambda}(1\!-\!\delta)\!\!\sum_{\hat{x}'\in \hat{X}}\!\!\bar{V}^*_{n}(\hat{x}',\underline{q}^*(\hat{x}',q))\hat{T}(\hat{x}'|\hat{x},\hat{u},\hat{w})\label{eq:policy_opt_sac_ab},
	\end{align}
	for all $n\in[0,H-1]$, where $\hat{u}=\rho_{H-n-1}(\hat{x},q)$ and $\hat{w}=\lambda_{H-n-1}(\hat{x},q,\hat{u})$.
	Here, we denote by 
	\begin{equation}\label{eq:value_function_opt_ab}
	\bar{V}_H^*(\hat{x},q):=\max_{\rho \in \mathcal P^H}\min_{\lambda \in \Lambda^H}\bar{V}_H^{\rho,\lambda}(\hat{x},q),
	\end{equation}
	the cost-to-go function associated with $\rho^*$.
\end{proposition}
Similar to~\eqref{eq:P_general}, by initializing $\bar{V}_0^*(\hat{x},q)=1$ when $q\in F$, and $\bar{V}_0^*(\hat{x},q)=0$ otherwise, $\bar{V}^*_n(x,q)$ in~\eqref{eq:value_function_opt_ab} can be recursively computed as
\begin{equation}\label{eq:P_opt_sac}
\bar{V}_{n+1}^*(\hat{x},q) = \mathfrak{P}^*(\bar{V}_{n}^*)(\hat{x},q),
\end{equation}
with $\mathfrak{P}^*$ being a Bellman operator defined as
\begin{align}
&\mathfrak{P}^*\!(\bar{V}_{n}^*)(\hat{x},q)\!\!:=\!\left\{
\begin{aligned} 
&\!\max_{\rho_{H-n-1}\in \mathcal P}\!\min_{\lambda_{H-n-1}\in \Lambda}\!\!(1-\delta)\!\!\sum_{\hat{x}'\in \hat{X}}\!\!\bar{V}^*_{n}(\hat{x}',\underline{q}^*(\hat{x}',q))\hat{T}(\hat{x}'|\hat{x},\hat{u},\hat{w}),\quad\text{if } q\notin F;\\
& \quad \quad \quad \quad 1,\quad \quad \quad \quad \quad\,\, \quad \quad \quad\quad \quad \quad\quad \quad \quad\quad \quad \quad\quad \quad \quad\quad  \text{if } q\in F,
\end{aligned}\right.\label{eq:value_opt_sac_ab}
\end{align}
where $\hat{u}=\rho_{H-n-1}(\hat{x},q)$,  $\hat{w}=\lambda_{H-n-1}(\hat{x},q,\hat{u})$, and
\begin{equation}\label{eq:underline_p_star}
\underline{q}^*(\hat{x}',q) = \mathop{\arg\min}_{q'\in Q'_{\epsilon}(\hat{x}')}\bar{V}^{*}_{n}(\hat{x}',q'),
\end{equation}
with $Q'_{\epsilon}(\hat{x}')$ being the set as in~\eqref{eq:Q_epsilon}.
With these notions, we are ready to show the following corollary that associates $\rho^*$ as in~\eqref{eq:policy_opt_sac_ab} with its corresponding robust satisfaction probability.
\begin{corollary}~\label{cor:max}
	Consider gDTSGs $\mathfrak{D}= (X,U,W,X_0,T,Y,h)$ and $\widehat{\mathfrak{D}}= (\hat X,\hat U, \hat W,\hat{X}_0, \hat T, Y,\hat h)$ with $\widehat{\mathfrak{D}}\preceq_{\epsilon}^{\delta}\mathfrak{D}$, and the desired property $(\mathcal{A},H)$ with $\mathcal{A}=(Q, q_0, \Pi,\tau, F)$.
	Given a Markov policy $\rho^*$ synthesized for Player~\uppercase\expandafter{\romannumeral1} of $\widehat{\mathfrak{D}}\otimes\mathcal{A}$ as in~\eqref{eq:policy_opt_sac_ab}, and a control strategy $\tilde{\mathbf{C}}_{\rho^*}$ for Player~\uppercase\expandafter{\romannumeral1} of $\mathfrak{D}$ that is constructed based on $\rho^*$ as in Definition~\ref{def:C_rho},
	for any control strategy $\tilde{\mathbf{C}}_{\lambda}$ for Player~\uppercase\expandafter{\romannumeral2} of $\mathfrak{D}$, we have
	\begin{equation}\label{eq:thm_sad_pmax_opt}
	\mathbb{P}_{(\tilde{\mathbf{C}}_{\rho^*},\tilde{\mathbf{C}}_{\lambda})\times \mathfrak{D}}\Big\{\exists k\leq H, y_{\omega k}\models \mathcal{A}\Big\}\geq \bar{V}_H^*(\hat{x}_0,\bar{q}_0),
	\end{equation}
	where $\hat{x}_0\in \hat{X}_0$ and $x_0\in X_0$, with $(x_0,\hat{x}_0)\in\mathscr{R}$ and $\mathscr{R}$ as in~\eqref{eq:Rx}, $\bar{V}_H^*(\hat{x}_0,\bar{q}_0)$ is as in~\eqref{eq:P_opt_sac} with $\bar{q}_0 = \tau(q_0,L\circ h(x_0))$.
\end{corollary}

Note that Corollary~\ref{cor:max} holds since Theorem~\ref{thm:gua_prmax} is valid for any arbitrary Markov policy $\rho$ for Player~\uppercase\expandafter{\romannumeral1} of $\widehat{\mathfrak{D}}\otimes\mathcal{A}$.
Therefore, the probabilistic guarantee associated with $\rho^*$ as in~\eqref{eq:policy_opt_sac_ab} can also be preserved for $\mathfrak{D}$.
\begin{remark}\label{det_Mar}
		Given the zero-sum Stackelberg game setting with Player~\uppercase\expandafter{\romannumeral1} as leader (cf. Remark~\ref{rem:asy_info}), Markovian stochastic kernel of $\mathfrak{D}\otimes\mathcal{A}$ as in Definition~\ref{def:product_gmdp_dfa}, and sum-multiplicative utility function as in~\eqref{eq:value_opt_sac_ab}, there always exists a deterministic~\cite[Section 5.1]{Breton1988Sequential} and Markovian~\cite[Section 4]{Rieder1991Non} policy as in~\eqref{eq:policy_opt_sac_ab}.
		In particular, considering Markov policy is sufficient in our case thanks to the sum-multiplicative utility function as constructed in~\eqref{eq:value_opt_sac_ab} and the Markovian stochastic kernel $T$ of $\mathfrak{D}$, which results in a Markovian stochastic kernel for the product $\widehat{\mathfrak{D}}\otimes\mathcal{A}$.
		Note that a similar deduction can also be applied to the corresponding policy for the worst-case violation problem, which is introduced later (cf.~\eqref{eq:value_opt_vio_ab} and~\eqref{eq:policy_opt_vio_ab}).
\end{remark}
Finally, it is also worth noting that operators in~\eqref{eq:value_gen_sac_ab} and~\eqref{eq:value_opt_sac_ab} can readily be applied to synthesis problems for stochastic systems \emph{without rational adversarial inputs}.
In this case, thanks to Assumption~\ref{asp1}, we are able to consider all states of finite abstraction $\widehat{\mathfrak{D}}$ in the proposed Bellman operators (instead of only a part of these states as the setting in~\cite{Haesaert2020Robust}).
Accordingly, the operator in~\eqref{eq:value_gen_sac_ab} provides less conservative probabilistic guarantees than the one proposed in~\cite[equation (41)]{Haesaert2020Robust}, which is formally shown with the following lemma.
\begin{lemma}\label{lem:less_con}
	Consider a property $(\mathcal{A},H)$ in which $\mathcal{A}=(Q, q_0, \Pi,\tau, F)$, gDTSGs $\mathfrak{D}= (X,U,W,X_0,T,Y,h)$ and $\widehat{\mathfrak{D}}= (\hat X,\hat U, \hat W,\hat{X}_0, \hat T, Y,\hat h)$ with $\widehat{\mathfrak{D}}\preceq_{\epsilon}^{\delta}\mathfrak{D}$ and $W=\hat{W}=\{\mathbf{0}_p\}$.
	Given a Markov policy $\rho$ designed for Player~\uppercase\expandafter{\romannumeral1} of $\widehat{\mathfrak{D}}\otimes\mathcal{A}$ and a control strategy $\tilde{\mathbf{C}}_{\rho}$ for Player~\uppercase\expandafter{\romannumeral1} of $\mathfrak{D}$ that is constructed based on $\rho$ as in Definition~\ref{def:C_rho}, we have
	\begin{align}\label{eq:less_con}
	\mathbb{P}_{(\tilde{\mathbf{C}}_{\rho},\tilde{\mathbf{C}}_{\lambda})\times \mathfrak{D}}\big\{\exists& k\leq H, y_{\omega k}\models \mathcal{A}\big\}\!\geq\! \bar{V}_H^{\rho,\lambda}(\hat{x}_0,\bar{q}_0)\geq \mathcal{S}(\hat{x}_0),
	\end{align}
	where $\hat{x}_0\in \hat{X}_0$ and $x_0\in X_0$, with $(x_0,\hat{x}_0)\in\mathscr{R}$ and $\mathscr{R}$ as in~\eqref{eq:Rx}, $\tilde{\mathbf{C}}_{\lambda}(\mathsf{m})\equiv \mathbf{0}_p$ with $\mathsf{m}$ as the memory state of $\tilde{\mathbf{C}}$, $\bar{V}_H^{\rho,\lambda}(\hat{x}_0,\bar{q}_0)$ is as in~\eqref{eq:P_general} with $\lambda(\hat{x},\hat{u})\equiv 0$, $\bar{q}_0 = \tau(q_0,L\circ h(x_0))$, and $\mathcal{S}(\hat{x}_0)$ is the probabilistic guarantee provided by the operator in~\cite[equation (41)]{Haesaert2020Robust}.
\end{lemma}

The proof of Lemma~\ref{lem:less_con} is provided in in Appendix~\ref{proof2.2}.
Similarly, the following corollary shows that the operator in~\eqref{eq:value_opt_sac_ab} also provides less conservative probabilistic guarantees than the one proposed in~\cite[equation (42)]{Haesaert2020Robust}.

\begin{corollary}\label{co:less_con}
	Given a Markov policy $\rho^*$ synthesized for Player~\uppercase\expandafter{\romannumeral1} of $\widehat{\mathfrak{D}}\otimes\mathcal{A}$ as in~\eqref{eq:policy_opt_sac_ab}, and a control strategy $\tilde{\mathbf{C}}_{\rho^*}$ for Player~\uppercase\expandafter{\romannumeral1} of $\mathfrak{D}$ that is constructed based on $\rho^*$ as in Definition~\ref{def:C_rho}, we have
	\begin{align}\label{eq:less_con_opt}
	\mathbb{P}_{(\tilde{\mathbf{C}}_{\rho^*},\tilde{\mathbf{C}}_{\lambda})\times \mathfrak{D}}\Big\{\exists k\leq H, y_{\omega k}\models \mathcal{A}\Big\}\geq \bar{V}_H^*(\hat{x}_0,\bar{q}_0)\geq \mathcal{S}^*(\hat{x}_0),
	\end{align}
	where $\hat{x}_0\in \hat{X}_0$ and $x_0\in X_0$, with $(x_0,\hat{x}_0)\in\mathscr{R}$ and $\mathscr{R}$ as in~\eqref{eq:Rx}, $\tilde{\mathbf{C}}_{\lambda}(\mathsf{m})\equiv \mathbf{0}_p$ with $\mathsf{m}$ as the memory state of $\tilde{\mathbf{C}}$, $\bar{V}_H^*(\hat{x}_0,\bar{q}_0)$ is as in~\eqref{eq:P_opt_sac} with $\lambda(\hat{x},\hat{u})\equiv 0$, $\bar{q}_0 = \tau(q_0,L\circ h(x_0))$, and $\mathcal{S}^*(\hat{x}_0)$ is the probabilistic guarantee provided by the operator in~\cite[equation (42)]{Haesaert2020Robust}.
\end{corollary}
The proof of Corollary~\ref{co:less_con} is similar to that of Lemma~\ref{lem:less_con}. 
We illustrate the results in Corollary~\ref{co:less_con} with an example in Section~\ref{sec:comp2}.
Next, we proceed with proposing the results for the problem of worst-case violation.  

\subsection{Worst-case Violation Problem}\label{sec:robust_min}
Here, we discuss the controller synthesis for Problem~\ref{problem2} in details.
Consider a gDTSG $\mathfrak{D}\!\!=\!\! (X,U,W,X_0,T,Y,h)$ and its finite abstraction $\widehat{\mathfrak{D}}\!\!=\!\! (\hat X,\hat U\!,\hat W\!,\hat{X}_0,\hat T\!,Y\!,\hat h)$ with $\widehat{\mathfrak{D}}\!\preceq_{\epsilon}^{\delta}\!\mathfrak{D}$, a property $(\mathcal{A},H)$, and a product gDTSG $\widehat{\mathfrak{D}}\otimes\mathcal{A}$ $ =\{\bar{X},\bar{U},\bar{W},\bar{X}_0,\bar{T},\bar{Y},\bar{h}\}$.
Given a Markov policy $\rho=(\rho_{0},\rho_{1},\ldots,\rho_{H-1})$ for Player~\uppercase\expandafter{\romannumeral1} and $\lambda=(\lambda_{0},\lambda_{1},\ldots,\lambda_{H-1})$ for Player~\uppercase\expandafter{\romannumeral2} of $\widehat{\mathfrak{D}}\otimes\mathcal{A}$, we define a cost-to go function $\ul{V}^{\rho,\lambda}_n:\hat{X}\times Q\rightarrow [0,1]$ which maps each state of $\widehat{\mathfrak{D}}\otimes\mathcal{A}$ at the time instant $H-n$ to a real number.
Then, $\ul{V}^{\rho,\lambda}_{n+1}(\hat{x},q)$ is recursively computed as
\begin{equation}\label{eq:T_general}
\ul{V}_{n+1}^{\rho,\lambda}(\hat{x},q) = \mathfrak{T}(\ul{V}_{n}^{\rho,\lambda})(\hat{x},q),
\end{equation}
initialized by $\ul{V}_0^{\rho,\lambda}(\hat{x},q)\!=\!1$ when $q\!\in\! F$, and $\ul{V}_0^{\rho,\lambda}(\hat{x},q)\!=\!0$, otherwise.
Here, $\mathfrak{T}$ is a Bellman operator defined as
\begin{align}
\mathfrak{T}(\ul{V}_{n}^{\rho,\lambda})(\hat{x},q):=
&\left\{
\begin{aligned} 
&(1-\delta)\sum_{\hat{x}'\in \hat{X}}\ul{V}^{\rho,\lambda}_{n}(\hat{x}',\bar{q}(\hat{x}',q))\hat{T}(\hat{x}'|\hat{x},\hat{u},\hat{w})+\delta, \text{ if } q\notin F;\\
& \quad \quad \quad \quad 1, \quad \quad \quad \quad \quad\, \quad \quad \quad\quad \quad \quad\quad \quad \quad\quad\text{ if } q\in F,
\end{aligned}\right.\label{eq:value_gen_vio_ab}
\end{align}
where $\hat{u}=\rho_{H-n-1}(\hat{x},q)$, $\hat{w}=\lambda_{H-n-1}(\hat{x},q,\hat{u})$, and
\begin{equation}\label{eq:overline_p}
\overline{q}(\hat{x}',q) = \mathop{\arg\max}_{q'\in Q'_{\epsilon}(\hat{x}')}\ul{V}^{\rho,\lambda}_{n}(\hat{x}',q'),
\end{equation}
with $Q'_{\epsilon}(\hat{x}')$ as in~\eqref{eq:Q_epsilon}.
Additionally, one can compute the worst-case adversarial policy $\lambda^*(\rho)$  for Player~\uppercase\expandafter{\romannumeral2} with respect to the Markov policy $\rho$ for Player~\uppercase\expandafter{\romannumeral1} as
\begin{align}
\lambda^*_{H-n-1}(\rho)\!\in\!\max_{\lambda_{H-n-1}\in \Lambda}\!\!\big((1\!-\!\delta)\!\!\sum_{\hat{x}'\in \hat{X}}\!\ul{V}^{\rho,\lambda^*(\rho)}_{n}(\hat{x}',\overline{q}(\hat{x}',q))\hat{T}(\hat{x}'|\hat{x},\hat{u},\hat{w})\!+\!\delta\big)\label{eq:policy_min_worst_case},
\end{align}
for all $n\in[0,H-1]$, with $\hat{u}=\rho_{H-n-1}(\hat{x},q)$, and $\hat{w}=\lambda_{H-n-1}(\hat{x},q,\hat{u})$.
Now, we propose in the next theorem the main result corresponding to the problem of worst-case violation.
\begin{theorem}\label{thm:gua_prmin}
	Consider gDTSGs $\mathfrak{D}= (X,U,W,X_0,T,Y,h)$ and $\widehat{\mathfrak{D}}\!=\! (\hat X,\hat U, \hat W,\hat{X}_0, \hat T, Y,\hat h)$ with $\widehat{\mathfrak{D}}\!\preceq_{\epsilon}^{\delta}\!\mathfrak{D}$, and a property $(\mathcal{A},H)$ in which $\mathcal{A}=(Q, q_0, \Pi,\tau, F)$.
	Given a Markov policy $\rho$ for Player~\uppercase\expandafter{\romannumeral1} of $\widehat{\mathfrak{D}}\otimes\mathcal{A}$, and a control strategy $\tilde{\mathbf{C}}_{\rho}$ for Player~\uppercase\expandafter{\romannumeral1} of $\mathfrak{D}$ that is constructed based on $\rho$ as in Definition~\ref{def:C_rho}, for any control strategy $\tilde{\mathbf{C}}_{\lambda}$ for Player~\uppercase\expandafter{\romannumeral2} of $\mathfrak{D}$, we have
	\begin{equation}\label{eq:thm_sad_pmin}
	\mathbb{P}_{(\tilde{\mathbf{C}}_{\rho},\tilde{\mathbf{C}}_{\lambda})\times\mathfrak{D}}\Big\{\exists k\leq H, y_{\omega k}\models\mathcal{A}\Big\}\leq \ul{V}_H^{\rho,\lambda^*(\rho)}(\hat{x}_0,\bar{q}_0),
	\end{equation}
	where $\hat{x}_0\in \hat{X}_0$ and $x_0\in X_0$, with $(x_0,\hat{x}_0)\in\mathscr{R}$ and $\mathscr{R}$ as in~\eqref{eq:Rx}, $\ul{V}_H^{\rho,\lambda_*(\rho)}(\hat{x}_0,\bar{q}_0)$ is computed as in~\eqref{eq:T_general}, with $\lambda^*(\rho)$ as in~\eqref{eq:policy_min_worst_case} and $\bar{q}_0 = \tau(q_0,L\circ h(x_0))$.
\end{theorem}
The proof of Theorem~\ref{thm:gua_prmin} is provided in Appendix~\ref{proof2.3}.
In practice, synthesizing a $\rho$ that minimizes the worst-case violation probability, \ie, $\ul{V}_H^{\rho,\lambda^*(\rho)}(\hat{x}_0,\bar{q}_0)$ as in~\eqref{eq:thm_sad_pmin}, is of particular interest. 
The following proposition shows how such a Markov policy can be synthesized.

\begin{proposition}
	Consider gDTSGs $\mathfrak{D}= (X,U,W,X_0,T,Y,h)$ and $\widehat{\mathfrak{D}}= (\hat X,\hat U, \hat W,\hat{X}_0, \hat T, Y,\hat h)$ with $\widehat{\mathfrak{D}}\preceq_{\epsilon}^{\delta}\mathfrak{D}$, and a property $(\mathcal{A},H)$ in which $\mathcal{A}=(Q, q_0, \Pi,\tau, F)$.
	Consider that Player~\uppercase\expandafter{\romannumeral2} is assumed to be able to maximize $\ul{V}_{n+1}^{\rho,\lambda}(\hat{x},q)$ according to $\rho$.
	The Markov policy  $\rho_*=(\rho_{*_0},\rho_{*_1},\ldots,\rho_{*_{H-1}})$ for Player~\uppercase\expandafter{\romannumeral1} minimizes $\ul{V}_{n+1}^{\rho,\lambda}(\hat{x},q)$, with 
	\begin{align}\label{eq:policy_opt_vio_ab}
	&\rho_{*_{H-n-1}}\in\mathop{\arg}\mathop{\min}_{\rho_{H-n-1}\in \mathcal P}\max_{\lambda_{H-n-1}\in \Lambda}\big((1-\delta)\sum_{\hat{x}'\in \hat{X}}\ul{V}_{*,n}(\hat{x}',\overline{q}_*(\hat{x}',q))\hat{T}(\hat{x}'|\hat{x},\hat{u},\hat{w})+\delta\big),
	\end{align}
	for all $n\in[0,H-1]$, where $\hat{u}=\rho_{H-n-1}(\hat{x},q)$ and $\hat{w}=\lambda_{H-n-1}(\hat{x},q,\hat{u})$.
	Here, we denote by 
	\begin{equation}\label{eq:value_function_wor_ab}
	\ul{V}_{*,H}(\hat{x},q):=\min_{\rho \in \mathcal P^H}\max_{\lambda \in \Lambda^H}\ul{V}_H^{\rho,\lambda}(\hat{x},q),
	\end{equation}
	the cost-to-go function associated with $\rho_*$.
\end{proposition}
Analogous to~\eqref{eq:T_general}, we initialize $\ul{V}_{*,n}(\hat{x},q)$ with $\ul{V}_{*,0}(\hat{x},q)=1$ when $q\in F$, and $\ul{V}_{*,0}(\hat{x},q)=0$ when $q\notin F$.
Then, $\ul{V}_{*,n}(\hat{x},q)$ can be recursively computed as
\begin{equation}\label{eq:P_opt_vio}
\ul{V}_{*,n+1}(\hat{x},q) = \mathfrak{T}_*(\ul{V}_{*,n})(\hat{x},q), 
\end{equation}
where $\mathfrak{T}_*$ is a Bellman operator defined as
\begin{align}
&\mathfrak{T}_*(\ul{V}_{*,n})(\hat{x},q)\!:=\!\left\{
\begin{aligned} 
&\min_{\rho_{H-n-1}\in \mathcal P}\max_{\lambda_{H-n-1}\in \Lambda}\big((1-\delta)\sum_{\hat{x}'\in \hat{X}}\ul{V}_{*,n}(\hat{x}',\overline{q}_*(\hat{x}',q)) \hat{T}(\hat{x}'|\hat{x},\hat{u},\hat{w})+\delta\big),\text{if } q\notin F;\\
& \quad \quad \quad \quad 1, \quad \quad \quad \quad \quad  \quad \quad \quad \quad\quad \quad \quad\quad \quad \quad\quad \quad \quad\quad \quad \quad\quad \quad \quad\quad \text{if } q\in F,
\end{aligned}\right.\label{eq:value_opt_vio_ab}
\end{align}
with $\hat{u}=\rho_{H-n-1}(\hat{x},q)$, $\hat{w}=\lambda_{H-n-1}(\hat{x},q,\hat{u})$, 
\begin{equation}\label{eq:overline_p_star}
\overline{q}_*(\hat{x}',q) = \mathop{\arg\max}_{q'\in Q'_{\epsilon}(\hat{x}')}\ul{V}_{*,n}(\hat{x}',q'),
\end{equation}
and $Q'_{\epsilon}(\hat{x}')$ as in~\eqref{eq:Q_epsilon}.
Note that Theorem~\ref{thm:gua_prmin} holds for any arbitrary Markov policy $\rho$ for Player~\uppercase\expandafter{\romannumeral1} of $\widehat{\mathfrak{D}}\otimes\mathcal{A}$.
Thus, the probabilistic guarantee associated with $\rho_*$ as in~\eqref{eq:policy_opt_vio_ab} can also be preserved for $\mathfrak{D}$.
This preservation is formally proposed in the following corollary.
\begin{corollary}~\label{cor:min}
	Consider gDTSGs $\mathfrak{D}= (X,U,W,X_0,T,Y,h)$ and $\widehat{\mathfrak{D}}= (\hat X,\hat U, \hat W,\hat{X}_0,\hat T, Y,\hat h)$ with $\widehat{\mathfrak{D}}\preceq_{\epsilon}^{\delta}\mathfrak{D}$, and a property $(\mathcal{A},H)$ with $\mathcal{A}=(Q, q_0, \Pi,\tau, F)$.
	Given a Markov policy $\rho_*$ synthesized for Player~\uppercase\expandafter{\romannumeral1} of $\widehat{\mathfrak{D}}\otimes\mathcal{A}$ as in~\eqref{eq:policy_opt_vio_ab}, and a control strategy $\tilde{\mathbf{C}}_{\rho_*}$ for Player~\uppercase\expandafter{\romannumeral1} of $\mathfrak{D}$ that is constructed based on $\rho_*$ as in Definition~\ref{def:C_rho},
	for any control strategy $\tilde{\mathbf{C}}_{\lambda}$ for Player~\uppercase\expandafter{\romannumeral2} of $\mathfrak{D}$, one has 
	\begin{equation}\label{eq:thm_sad_pmin_opt}
	\mathbb{P}_{(\tilde{\mathbf{C}}_{\rho_*},\tilde{\mathbf{C}}_{\lambda})\times \mathfrak{D}}\big\{\exists k\leq H, y_{\omega k}\models \mathcal{A}\big\}\leq \underline{V}_{*,H}(\hat{x}_0,\bar{q}_0),
	\end{equation}
	where $\hat{x}_0\in \hat{X}_0$ and $x_0\in X_0$, with $(x_0,\hat{x}_0)\in\mathscr{R}$ and $\mathscr{R}$ as in~\eqref{eq:Rx},  $\underline{V}_{*,H}(\hat{x}_0,\bar{q}_0)$ is as in~\eqref{eq:P_opt_vio} and $\bar{q}_0 = \tau(q_0,L\circ h(x_0))$.
\end{corollary}
Finally, we summarize the controller synthesis procedure as follows:
\begin{itemize}
		\setlength{\itemsep}{0pt}
		\setlength{\parsep}{0pt}
		\setlength{\parskip}{0pt}
		\item For the problem of robust satisfaction, we first synthesize a Markov policy $\rho^*$ as in~\eqref{eq:policy_opt_sac_ab}.
		Then, we construct the controller $\tilde{\mathbf{C}}_{\rho^*}$ as in Definition~\ref{def:C_rho} based on $\rho^*$.
		\item As for the problem of worst-case violation, we construct a control strategy $\tilde{\mathbf{C}}_{\rho_*}$ as in Definition~\ref{def:C_rho} based on a Markov policy $\rho_*$ synthesized as in~\eqref{eq:policy_opt_vio_ab}.
\end{itemize}
\begin{remark}\label{complexity_rem}
		Note that given the product gDTSG $\widehat{\mathfrak{D}}\otimes\mathcal{A} = \{\bar{X},\bar{U},\bar{W},\bar{X}_0,\bar{T},\bar{Y},\bar{h}\}$, both $\rho^*$ as in~\eqref{eq:policy_opt_sac_ab} and $\rho_*$ as in~\eqref{eq:policy_opt_vio_ab} are (offline) look-up tables, 
		whose sizes grow linearly with the time horizon $H$ and the cardinality of $\bar{X}$.
		Moreover, the number of operations required for computing~\eqref{eq:policy_opt_sac_ab} and~\eqref{eq:policy_opt_vio_ab} is proportional to $H$ and the cardinality of $\bar{X}$, $\bar{U}$, and $\bar{W}$.
		It is also worth noting that, for all $n\in[0,H-1]$, the computations of $\rho_{*_{n}}(\hat{x},q)$ and $\rho^*_{n}(\hat{x},q)$ for all $(\hat{x},q)\in \bar{X}$ are independent from each other and can be done in a parallel fashion.
\end{remark}
\section{Case Studies}\label{Case_study}
In this section, we apply our proposed approaches to two case studies, including the running example and a control problem for a Quadrotor helicopter. 
We simulate each case study with $1.0 \times 10^5$ different realizations of noise, in which inputs of Player~\uppercase\expandafter{\romannumeral2} are randomly selected from their input sets following a uniform distribution.
Here, we do not consider that Player~\uppercase\expandafter{\romannumeral2} selects adversarial inputs rationally since it is challenging to obtain closed-form solutions for such case.
Meanwhile, the probabilistic guarantees provided by our results are still valid regardless of how Player~\uppercase\expandafter{\romannumeral2} chooses inputs (cf. Remark~\ref{rem:asy_info}).
To show the applicability of our results, in all case studies, we summarize the required memory\footnote{In this section, we allocate $4$ bytes for each entry of matrices to be stored as a single-precision floating-point. } for storing stochastic kernels and synthesized controllers, and report the average execution time of these controllers.
All experiments are performed via MATLAB 2019b, on a machine with Ubuntu 20.04 (Intel(R) Xeon(R) Gold 6254 CPU (3.1 GHz) and 378 GB of RAM).

\subsection{Running Example (continued)}\label{cs_running}
\begin{figure}[htbp]
	\centering
	\includegraphics[width=0.5\textwidth]{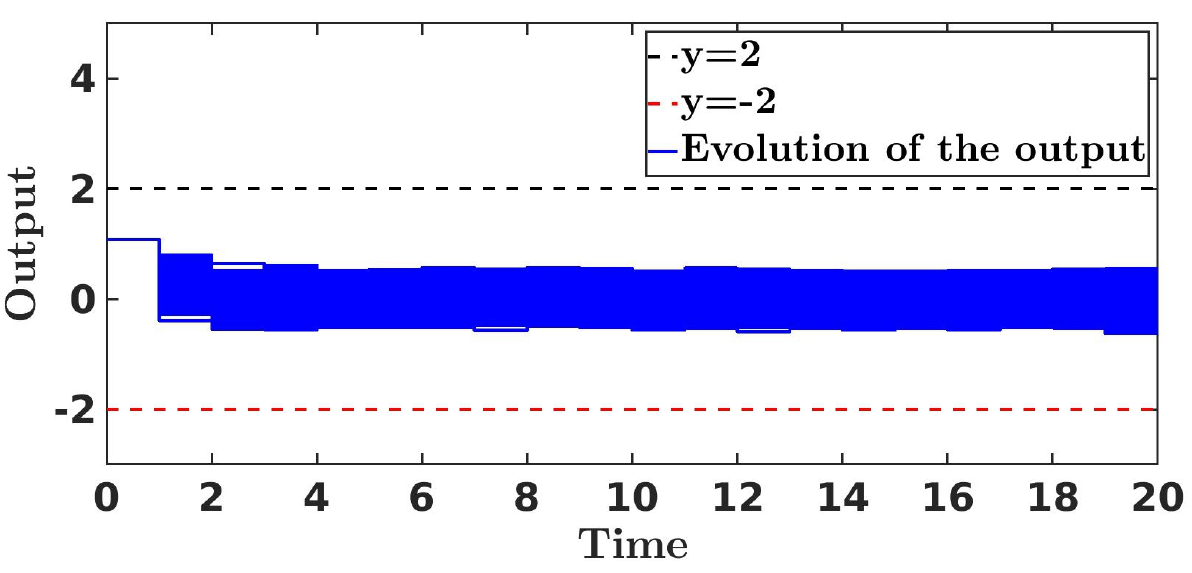}
	\caption{Simulation of the running example with respect to $\psi$.}\label{fig:temp_simulation}
\end{figure}

Here, we synthesize the controller following~\eqref{eq:policy_opt_vio_ab}-\eqref{eq:overline_p_star}. 
The simulation setting and results are summarized in Table~\ref{tbl:sim} and depicted in Figure~\ref{fig:temp_simulation}.
One can readily observe that the probabilistic guarantee of satisfaction is respected.
Additionally, we also show how the reduced-order game improves the scalability issue (cf. Remark~\ref{moti_rn}) via the running example. To do so, we first build the finite abstraction of the original game without performing any model order reduction by considering $[-12,12]^3$ as the region of interest.
We uniformly partition this region with girds whose sizes are $(0.24,0.24,0.24)$ for a fair comparison with the reduced-order model setting (cf. Table~\ref{tbl:synthesis}).
For the same reason, we uniformly divide $U$ and $W$ with grids whose sizes are $(0.06,0.06)$ and $0.1$, respectively.
Under this setting, when a reduced-order game is built, we only need around $19$ MB to store the stochastic kernel.	
On the other hand, without constructing a reduced-order game, the finite abstraction contains $10^6$ states, $1.25\times10^5$ inputs for Player~\uppercase\expandafter{\romannumeral1}, and $10$ inputs for Player~\uppercase\expandafter{\romannumeral2}.
As a result, one needs $4.65\times10^9$ GB to store the stochastic kernel, which is not practical.

\subsection{Controller Synthesis for a Quadrotor}\label{sec:quadrotor}
\begin{figure}[htbp]
	\centering
	\subfigure{
		\includegraphics[width=0.2\textwidth]{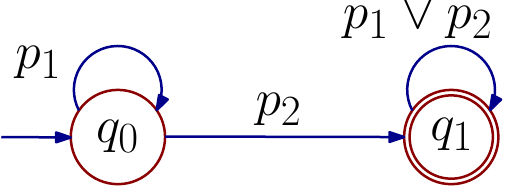}
	}\hspace{-0.4cm}
	\quad
	\subfigure{
		\includegraphics[width=0.2\textwidth]{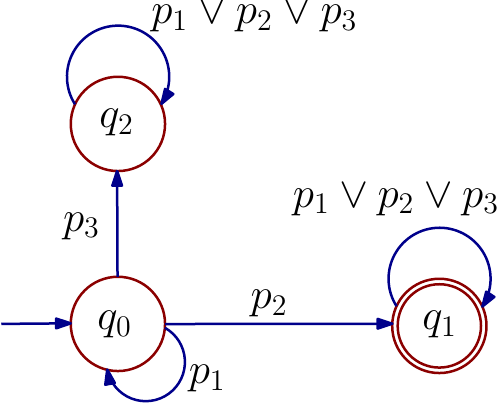}
	}
	\caption{\textbf{Left:} DFA for modeling $\psi_1$, with accepting state $q_1$, alphabet $\Pi=\{p_1,p_2\}$, and labeling function $L:Y \rightarrow \Pi$ with $L(y)=p_1$ when $y \in [-0.7, 0.7]$, and $L(y)=p_2$ when $y \in (-\infty, -0.7)\cup (0.7,+\infty)$. ~\textbf{Right:} DFA for modeling $\psi_2$, with accepting state $q_1$, alphabet $\Pi=\{p_1,p_2,p_3\}$, and labeling function $L:Y \rightarrow \Pi$ with $L(y)=p_1$ when $y \in [-1,-0.5)\cup (0.5, 1]$, $L(y)=p_2$ when $y \in [-0.5, 0.5]$, and $L(y)=p_3$ when $y \in (-\infty,-1)\cup (1, +\infty)$.}
	\label{fig:quad_dfa}
\end{figure}

\begin{figure}[htbp]
	\centering
	\includegraphics[width=0.4\textwidth]{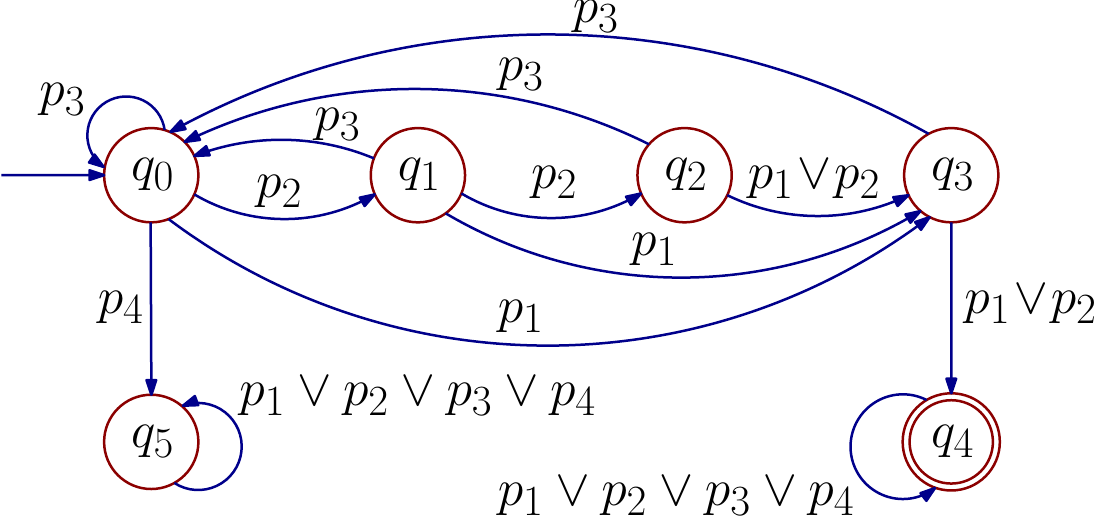}
	\caption{DFA for modeling $\psi_3$ with accepting state $q_4$, alphabet $\Pi=\{p_1,p_2,p_3,$ $p_4\}$, and labeling function $L\!:\!Y\!\rightarrow\!\Pi$ with $L(y)\!=\!p_1$ when $y\!\in\![-0.1, 0.1]$; $L(y)=p_2$ when $y\!\in\![-0.45, -0.1)\cup (0.1,0.45]$; $L(y)=p_3$ when $y\!\in\![-0.8, -0.45)\cup (0.45,0.8]$, and $L(y)=p_4$ when $y\!\in\!(-\infty, -0.8)\cup(0.8,+\infty)$.
		Transitions $q_5=\tau(q_j,p_4)$, with $j \in\{1,2,3\}$, are omitted to keep the figure less crowded.}
	\label{fig:psi3}
\end{figure}

\begin{figure*}[htbp]
	\centering
	\subfigure{
		\includegraphics[width=0.30\textwidth]{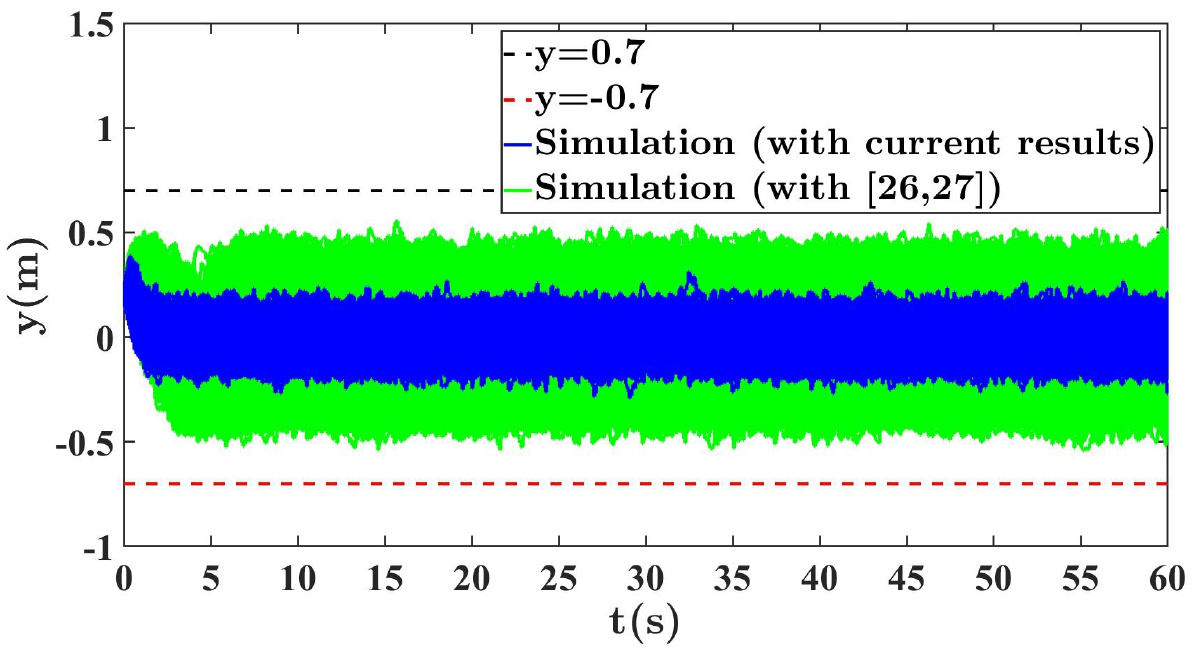}
	}\hspace{-0.4cm}
	\quad
	\subfigure{
		\includegraphics[width=0.30\textwidth]{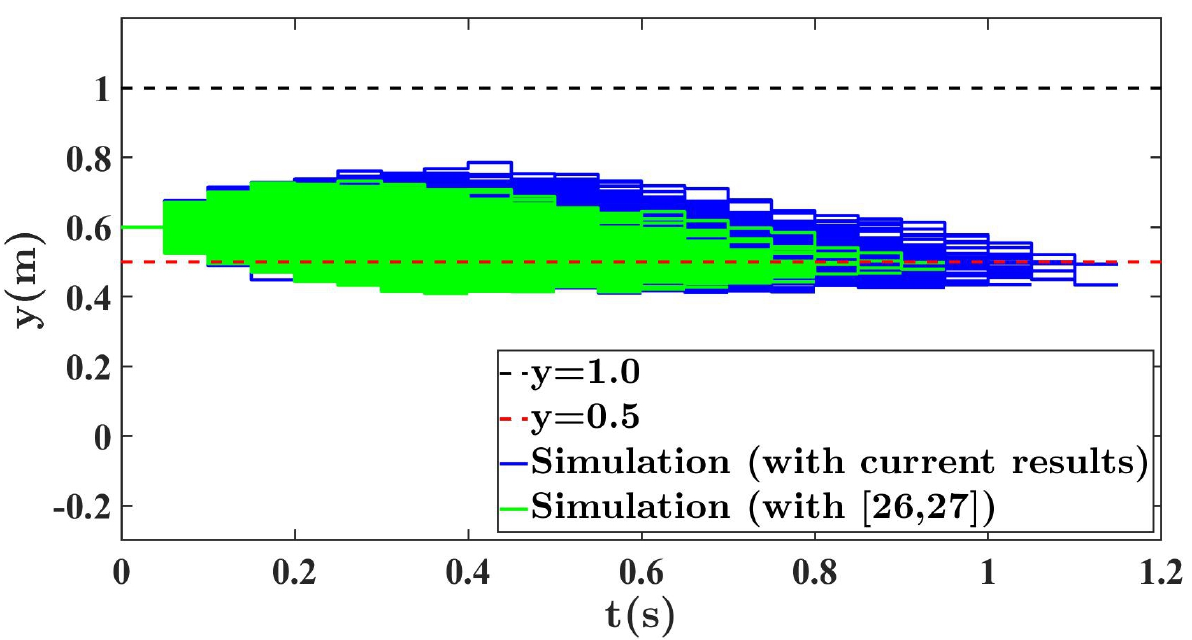}
	}\hspace{-0.4cm}
	\quad
	\subfigure{ 
		\includegraphics[width=0.30\textwidth]{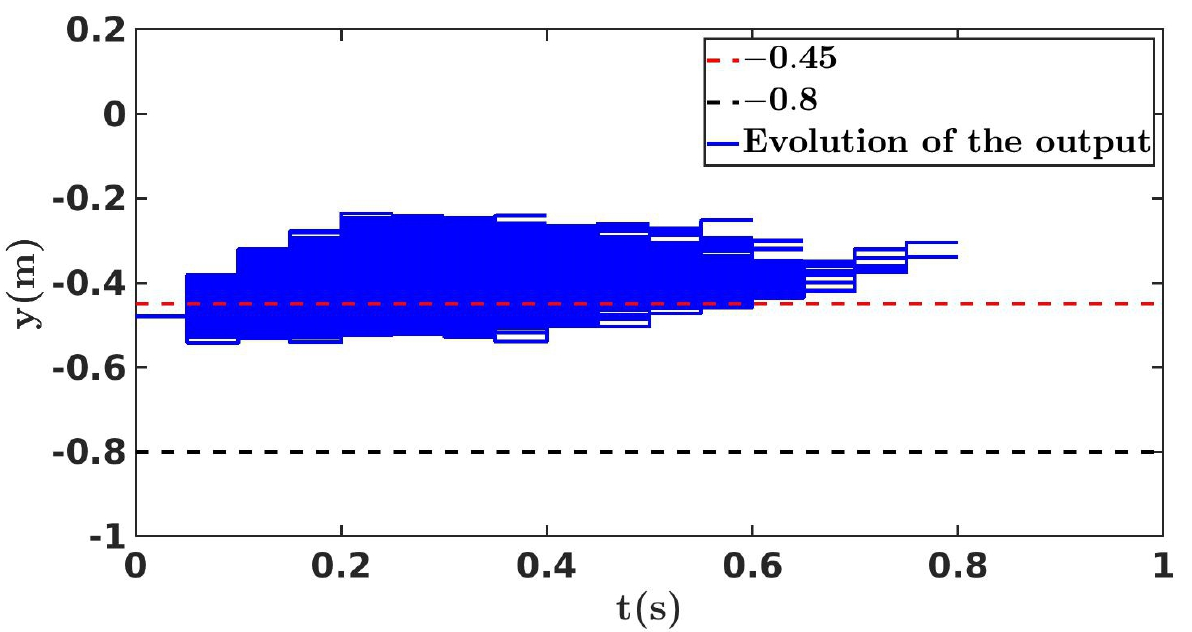}
	}
	\caption{Simulation results for $\psi_1$ (left), $\psi_2$ (middle), and $\psi_3$ (right).}
	\label{fig:sim}
\end{figure*}

Here, we apply our proposed results to a quadrotor tracking a moving vehicle on the ground.
Consider a quadrotor moving on a 2-dimensional planar.
As discussed in~\cite{Kamgarpour2011Discrete}, the control of a quadrotor can be decoupled into the control on different dimensions. 
Hence, we borrow the model from~\cite{Kamgarpour2011Discrete} which models the relative motion between the quadrotor and the ground vehicle:
\begin{align*}
\mathfrak{D}\!:
\left\{\hspace{-0.15cm}\begin{array}{l}
x(k+1) = Ax(k)+Bu(k)+Dw(k)+R\varsigma(k),\\
y(k)=Cx(k),\quad \quad \quad \quad \quad \quad \quad k\in\mathbb N,\end{array}\right.
\end{align*}
where 
$A \!=\! \begin{bmatrix}\begin{smallmatrix}1\,&\Delta t\\0\,&1\end{smallmatrix}\end{bmatrix}$,
$R \!=\! \begin{bmatrix}\begin{smallmatrix}0.4\Delta t\,&0\\0\,&0.4\Delta t\end{smallmatrix}\end{bmatrix}$,
$B \!=\! [\frac{\Delta t^2g}{2};\Delta tg]$, $D \!=\! [\frac{\Delta t^2}{2};\Delta t]$, and $C = [1;0]^T$,
with $\Delta t = 0.05s$ being the sampling time and $g\!=\!9.8m/s$ being the gravitational constant.
Here, $x(k) = [x_1(k);$ $x_2(k)]$ with $x_1(k)$ and $x_2(k)$ being the relative position and velocity between the quadrotor and the vehicle, respectively;
$u(k)\in[-0.25,0.25] (m/s^2)$ denotes the acceleration of the quadrotor as the control input;
$w(k)\in[-0.6,0.6] (m/s^2)$ denotes the acceleration that can be (rationally) chosen by the vehicle;
$\varsigma(k)$ is a standard Gaussian random variable;
and $y$ is the output of the system.
Here, we are interested in the following properties:
\begin{enumerate}[(i)]
	\setlength{\itemsep}{0pt}
	\setlength{\parsep}{0pt}
	\setlength{\parskip}{0pt}
	\item $\psi_1$: $y$ should stay in $[-0.7,0.7]$ for $1$ minute (\emph{i.e.,} time horizon $H=1200$). The DFA for modeling $\psi_1$ is shown in~Figure~\ref{fig:quad_dfa} (left), and we focus on the problem of worst-case violation concerning this DFA.
	\item $\psi_2$: starting from $[-1.0,1.0]$, $y$ should reach $[-0.5,0.5]$ within $5$ seconds (\emph{i.e.,} time horizon $H=100$). 
	Here, we construct a DFA for characterizing $\psi_2$ as in Figure~\ref{fig:quad_dfa} (right). 
	Accordingly, we are interested in the problem of robust satisfaction regarding this DFA.
	\item $\psi_3$: within $2$ seconds (\emph{i.e.,} time horizon $H\!=\!40$), (1) $y$ should reach $[-0.45,0.45]$ and then stay within $[-0.45,0.45]$ for 3 time instants after it reaches $[-0.45,$ $0.45]$; (2) if it reaches $[-0.1,0.1]$, it only needs to stay within $[-0.45,0.45]$ for $1$ time instant after it reaches $[-0.1,0.1]$; (3) $y$ is not allowed to leave $[-0.8,0.8]$.
	The DFA for modeling $\psi_3$ is depicted in Figure~\ref{fig:psi3} and we focus on the problem of robust satisfaction accordingly.
\end{enumerate}
\vspace{-0.25cm}
First, we construct the finite abstraction of the model.
Since we do not apply any model order reduction to this model, we select $P = \mathit{I}_2$. 
Therefore, we have $\hat{A}_{\textsf r} = A$, $\hat{D}_{\textsf r} = D$, $\hat{R}_{\textsf r} = R$, $\hat{C}_{\textsf r} = C$ and $Q=S = 0_{2\times 1}$.
The finite abstraction is constructed as in Table~\ref{tbl:synthesis}.
Accordingly, we select $\tilde{M}=1$ and $\tilde{\epsilon}=0.05$.
As for establishing the relation between the constructed abstraction and the original game, we set $\hat{U}'=\{\hat{u}\in\hat{U}~|~-0.12\leq\hat{u}\leq 0.12\}$, $\delta = 0$, and the tolerable range of $\epsilon$ as $[0.05,0.4]$.
By applying Algorithm~\ref{alg:lifting}, the finite abstraction is $(\epsilon,\delta)$-stochastically simulated by the original model with $\delta = 0$, $M = \begin{bmatrix}\begin{smallmatrix}1.7699\ &0.5494\\0.5494\ &0.3920\end{smallmatrix}\end{bmatrix}$, and $\epsilon = 0.2911$, when the interface function in~\eqref{eq:interface} is applied with $\tilde{R} =1$ and $K=[-0.4294;-0.2773]$. 
Now we are ready to synthesize a controller enforcing $\psi_1$ following~~\eqref{eq:policy_opt_vio_ab}-\eqref{eq:overline_p_star}, and controllers enforcing $\psi_2$ and $\psi_3$ following~~\eqref{eq:policy_opt_sac_ab}-\eqref{eq:underline_p_star}.
The setting and results of the simulation for $\psi_1$, $\psi_2$, and $\psi_3$ are summarized in Table~\ref{tbl:sim} and depicted in Figure~\ref{fig:sim}.
In all case studies, the probabilistic guarantees of satisfaction are well respected.

\begin{table}[]
	\centering
	\caption{Simulation results with respect to properties $\psi$, $\psi_1$, $\psi_2$, $\psi_3$, with $P_f$ denoting the formal probabilistic guarantees, and $P_e$ being the empirical satisfaction probability.}	\label{tbl:sim}
	{\small\begin{tabular}{|l|l|l|l|l|}
			\hline
			& $x_0$  & $P_f$ & $P_e$ & Execution time (ms) \\ \hline
			$\psi$  & $[3.8;4.1;2.9]$                                                & $\geq 99.90$\%                                                    & $100$\%                                                                          & 0.0755                                                           \\ \hline
			$\psi_1$ & $[0.2;0.2]$                                                    & $\geq 99.26$\%                                                    & $100$\%                                                                          & 0.0684                                                            \\ \hline
			$\psi_2$ & $[0.6;0.1]$                                                    & $\geq 94.77$\%                                                    & $100$\%                                                                          & 0.0683                                                             \\ \hline
			$\psi_3$ & $[-0.48;0.45]$                                                    & $\geq 98.75$\%                                                    & $100$\%                                                                          & 0.0766                                                            \\ \hline
	\end{tabular}}
\end{table}
\begin{table*}[]
	\centering
	\caption{Construction of finite abstractions against properties $\psi$, $\psi_1$, $\psi_2$, $\psi_3$, where $X_{rs}$ and $\hat{X}$ denote the region of interest (see Section~\ref{subsec:ab_subsys} for its definition) and its corresponding finite state set, respectively; $\hat{T}$ denotes the stochastic kernel of the finite abstraction; $\rho^*$ (or $\rho_*$) denotes the look-up tables for constructing the desired controllers (cf. Remark~\ref{complexity_rem}); $U$ (resp. $W$) denotes the input set of Player~\uppercase\expandafter{\romannumeral1} (resp. Player~\uppercase\expandafter{\romannumeral2}) to be partitioned, and $\hat{U}$ (resp. $\hat{W}$) denotes its corresponding finite sets.
		\vspace{0.2cm}}	\label{tbl:synthesis}
	{\small\begin{tabular}{|c|c|ccc|cccc|c|cc|}
			\hline
			\multirow{2}{*}{} & \multirow{2}{*}{$X_{rs}$} & \multicolumn{3}{c|}{Grids' size}                                                                               & \multicolumn{4}{c|}{Number of Elements}                                                                                    & \multirow{2}{*}{\begin{tabular}[c]{@{}l@{}}Time\\ horizon\end{tabular}}  & \multicolumn{2}{c|}{Required Memory (GB)}                        \\ \cline{3-9} \cline{11-12} 
			&                                 & \multicolumn{1}{c|}{$X_{rs}$}        & \multicolumn{1}{c|}{$U$}                   & $W$                  & \multicolumn{1}{c|}{$\hat{X}$} & \multicolumn{1}{c|}{$\hat{U}$}           & \multicolumn{1}{c|}{$\hat{W}$}           & $Q$ &                               & \multicolumn{1}{c|}{$\hat{T}$}           &  $\rho^*$ (or $\rho_*$)  \\ \hline
			$\psi$            & $[-12,12]$                      & \multicolumn{1}{c|}{0.24}                  & \multicolumn{1}{c|}{0.06}                  & 0.1                  & \multicolumn{1}{c|}{101}       & \multicolumn{1}{c|}{50}                  & \multicolumn{1}{c|}{10}                  & 4   & 20                            & \multicolumn{1}{c|}{$1.9\times 10^{-2}$} & $3.01\times 10^{-5}$  \\ \hline
			$\psi_1$          & $[-0.7,0.7]\times[-0.5,0.5]$    & \multicolumn{1}{c|}{\multirow{3}{*}{(0.02, 0.02)}} & \multicolumn{1}{c|}{\multirow{3}{*}{0.02}} & \multirow{3}{*}{0.1} & \multicolumn{1}{c|}{3501}      & \multicolumn{1}{c|}{\multirow{3}{*}{25}} & \multicolumn{1}{c|}{\multirow{3}{*}{12}} & 2   & 1200                          & \multicolumn{1}{c|}{$7.13 $}            & $3.13\times 10^{-2}$  \\ \cline{1-2} \cline{6-6} \cline{9-12} 
			$\psi_2$          & $[-1,1]\times[-0.75,0.75]$      & \multicolumn{1}{c|}{}                      & \multicolumn{1}{c|}{}                      &                      & \multicolumn{1}{c|}{7501}      & \multicolumn{1}{c|}{}                    & \multicolumn{1}{c|}{}                    & 3   & 100                           & \multicolumn{1}{c|}{$32.70$}             & $8.38 \times 10^{-3}$ \\ \cline{1-2} \cline{6-6} \cline{9-12} 
			$\psi_3$          & $[-0.8,0.8]\times[-0.55,0.55]$  & \multicolumn{1}{c|}{}                      & \multicolumn{1}{c|}{}                      &                      & \multicolumn{1}{c|}{4401}      & \multicolumn{1}{c|}{}                    & \multicolumn{1}{c|}{}                    & 6   & 40                            & \multicolumn{1}{c|}{$11.26$}             & $3.93\times 10^{-3}$  \\ \hline
	\end{tabular}}
\end{table*}

\subsection{Comparison with Results in~\cite{Kamgarpour2011Discrete,Ding2013stochastic}}\label{sec:comp1}
By virtue of the grid-based approximation framework introduced in~\cite{Abate2010Approximate}, results in~\cite{Kamgarpour2011Discrete,Ding2013stochastic} can be applied to the synthesis problem for (nonlinear) stochastic games with continuous state and input sets.
In this subsection, we compare our approaches with these results in the sense of the conservativeness of probabilistic guarantees associated with the synthesized controllers. 
Note that providing less conservative probabilistic guarantees are crucial in correct-by-construction synthesis techniques.
The ultimate goal for employing these techniques is to obtain formal (probabilistic) guarantees for satisfying the desired properties, instead of performing exhaustive testing, which is heuristic, costly, and time-consuming.

Under the grid-based approximation framework in~\cite{Abate2010Approximate}, the probabilistic guarantee for a desired property is provided in terms of a \emph{probabilistic closeness}, denoted by $\mathbf{e}$, between the finite abstraction and the original system, with:
\begin{equation}
|p-\hat{p}|\leq \mathbf{e},
\end{equation}
where $\hat{p}$ and $p$ denote the probabilities of satisfaction over the finite abstraction and the original system, respectively. 
Moreover,~\cite{Soudjani2015Dynamic} shows that $\mathbf{e}$ is proportional to the size of discretization parameters, denoted by $\eta_i$, $i = \{1,\ldots,s\}$, with $s$ being the dimension of the state set.
Roughly speaking, the quantity $\eta_i$ is the maximum diameter of partition cells along with the $i^{th}$ dimension of the state set.
We refer the interested reader to~\cite[Theorem 9]{Soudjani2015Dynamic} for the formal definition.
Here, by employing  the results in~\cite[Section 5]{Soudjani2015Dynamic}, we have $\mathbf{e}=3.586\times 10^4$,  $\mathbf{e}=4.356\times 10^3$, and $\mathbf{e}=1.345\times 10^3$ for $\psi_1$, $\psi_2$, and $\psi_3$\footnote{Although results in ~\cite{Kamgarpour2011Discrete,Ding2013stochastic} only solve the reachability problem over continuous sets, enforcing DFA properties can be cast as a reachability problem over state set of the product system between the DFA and the original system. 
		Therefore, results in~\cite[Section 5]{Soudjani2015Dynamic} can readily be used to compute $\mathbf{e}$ for $\psi_3$. }, respectively, when grid-size parameters are $(\eta_1,\eta_2)=(0.02,0.02)$ (as the discretization setting in Table~\ref{tbl:synthesis}).

In all cases, $\mathbf{e}$ is significantly larger than 1. 
Notably, the results in~\cite[Section 5]{Soudjani2015Dynamic} only consider the effect of state set's discretization on $\mathbf{e}$.
According to results in~\cite{Soudjani2014Formal,Tkachev2013Quantitative}, the discretization of input sets would make $\mathbf{e}$ even larger.
Since probability should be a real number between $0$ and $1$, the probabilistic guarantees for the original system are very conservative in all cases.
To show this, we first synthesize controllers with the results in~\cite{Kamgarpour2011Discrete,Ding2013stochastic} enforcing $\psi_1$ and $\psi_2$\footnote{We are not able to synthesize controllers enforcing $\psi_3$ using the results in~\cite{Kamgarpour2011Discrete,Ding2013stochastic}, since they do not provide any operator that handle general DFA properties like $\psi_3$.}.
By deploying these controllers, one gets formally that the probabilities of satisfying $\psi_1$ and $\psi_2$ will be within $[-3.586\times 10^4,3.586\times 10^4]$ and $[-4.356\times 10^3,4.356\times 10^3]$, respectively, which are very conservative.
Then, starting from the same initial states as in Table~\ref{tbl:sim}, we simulate both cases with $10^5$ different noise realizations.
In both cases, as depicted in Figure~\ref{fig:sim}, trajectories under different noise realizations satisfy the desired properties with probability 1 in the experiments.
Hence, the formal probabilistic guarantees associated with both controllers are very conservative considering the empirical results.
In comparison, as shown in Table~\ref{tbl:sim}, our controllers empirically perform as good as those controllers synthesized with the results in~\cite{Kamgarpour2011Discrete,Ding2013stochastic}.
On the other hand, our results provide formal probabilistic guarantees which are much less conservative.
Note that one may select smaller $\eta_i$ such that $\mathbf{e}$ becomes smaller. 
Here, we summarize in Table~\ref{tab_scalbility} the required $(\eta_1,\eta_2)$ and the corresponding memory for storing the stochastic kernels of finite abstractions such that we have reasonable $\mathbf{e}$. 
In terms of required memory, it is computationally expensive to provide a reasonable guarantee under the grid-based approximation framework proposed in~\cite{Abate2010Approximate}.

\begin{table}[h!]
	\centering
	\caption{Required $(\eta_1,\eta_2)$ and corresponding required memory for different properties when applying the results of~\cite{Kamgarpour2011Discrete,Ding2013stochastic}. \vspace{0.2cm}}\label{tab_scalbility}
	{\small 	\begin{tabular}{|p{1.2cm}|p{3.2cm}|p{3cm}|}
			\hline
			Properties& Required $(\eta_1,\eta_2)$ ($\times 10^{-6}$) &  Required memory (GB)\\
			\hline
			$\psi_1$ & $(0.492,0.644)$ & $2.184\times 10^{19}$ \\
			\hline
			$\psi_2$ & $(4.132,5.166)$ & $2.207\times 10^{16}$ \\
			\hline
			$\psi_3$ & $(12.915,17.512)$ & $6.768\times 10^{13}$ \\
			\hline
	\end{tabular}}
\end{table}

\subsection{Comparison with Operators in~\cite{Haesaert2020Robust}}\label{sec:comp2}
\begin{figure}[htbp]
	\centering
	\includegraphics[width=0.4\textwidth]{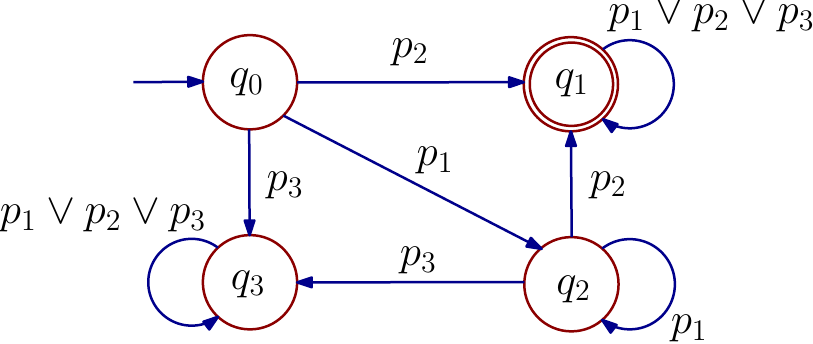}
	\caption{DFA for modelling $\psi'$, with accepting state $q_1$, alphabet $\Pi=\{p_1,p_2,p_3\}$, and labelling function $L\!:\!Y\!\rightarrow\!\Pi$ with $L(y)\!=\!p_1$ when $y\!\in\![-2, -0.3)\cup (0.3,2]$, $L(y)=p_2$ when $y\!\in\![-0.3, 0.3]$, and $L(y)=p_3$ when $y\!\in\!(-\infty, -2)\cup(2,+\infty)$. }
	\label{fig:compare_operator1}
\end{figure}

Here, we show Corollary~\ref{co:less_con} with an example.
To this end, we focus on the following system: 
\begin{align*}
\mathfrak{D}\!:
\left\{\hspace{-0.15cm}\begin{array}{l}
x(k+1) = Ax(k)+Bu(k)+R\varsigma(k),\\
y(k)=Cx(k),\quad \quad \quad k\in\mathbb N,\end{array}\right.
\end{align*}
where 
\begin{align*}
&A \!=\!  \begin{bmatrix}
\begin{smallmatrix}0.91\ &0.47\ &0.76\\0.65\ &0.71\ &0.93\\0.69\ &0.28\ &0.53\end{smallmatrix}
\end{bmatrix}\!,
~~B \!=\! \begin{bmatrix}
\begin{smallmatrix}9.5\ &0.6\ &4.1\\2.4\ &12.4\ &2.9\\5.7\ &5.4\ &5.8\end{smallmatrix}
\end{bmatrix}\!,
\end{align*}
$R=[0.63;0.28;0.48]$, and $C =[0.1;0.1;0.1]^T$.
Here, we have $x(t) = [x_1(k);x_2(k);x_3(k)]$ and $u(k)\in[-3,3]^3$.
We focus on a co-safe linear temporal logic property~\cite{Faruq2018Simultaneous} $\psi'$ that can be handled by the operators proposed in~\cite{Haesaert2020Robust}: starting from $[-2,2]$, the output of the system should reach $[-0.3,0.3]$ while avoiding $(-\infty,-2)\cup(2,+\infty)$ within 90 time steps (i.e., $H=90$).
Accordingly, we synthesize the controller by solving the problem of robust satisfaction corresponding to the DFA in Figure~\ref{fig:compare_operator1}.

\begin{figure}[htbp]
	\centering
	\includegraphics[width=0.45\textwidth]{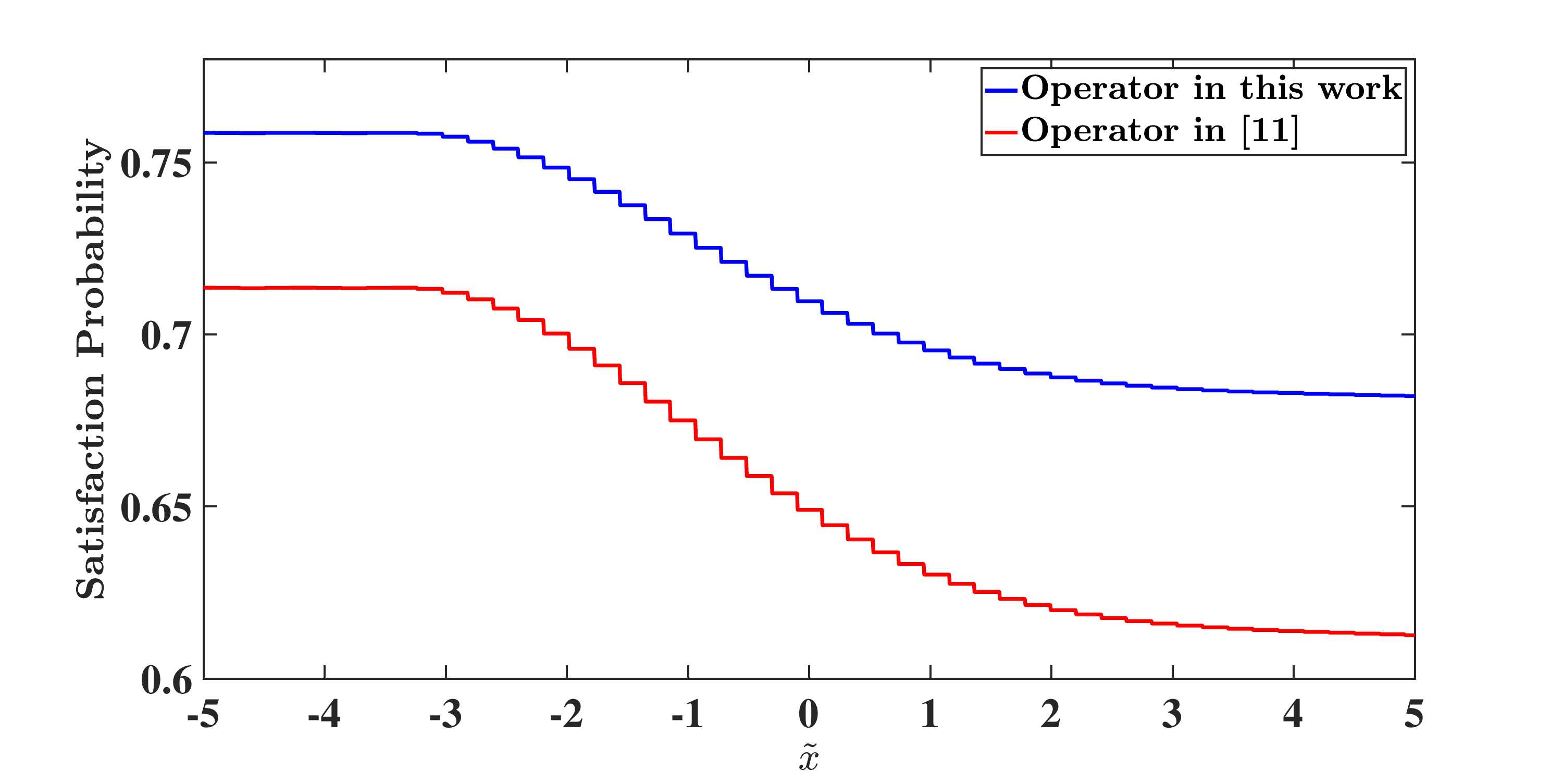}
	\caption{Comparison of probabilistic guarantees between the operator in~\eqref{eq:value_opt_vio_ab} and the one proposed in~\cite{Haesaert2020Robust}.}
	\label{fig:compare_operator2}
\end{figure}

For constructing the finite abstraction, we select $P = [0.5;0.4;0.5]$ and accordingly construct a reduced-order model with $\hat{A}_{\textsf r}=0.62$, $\hat{B}_{\textsf r}=1$, $\hat{C}_{\textsf r}=0.14$, $\hat{R}_{\textsf r}=0.9939$, and $Q = [-0.1179;-0.0694;0.1094]$ as proposed in~\eqref{eq:infabs_cond1}-\eqref{eq:infabs_cond4}. 
The finite abstraction is then constructed by uniformly dividing the region of interest, i.e. $[-15,15]$, of the reduced-order model's state set into partitions whose lengths are $0.15$, and partitioning the input set, i.e. $[-3,3]$, for the reduced-order model uniformly with $48$ cells.
Here, we set $\hat{U}'=\hat{U}$, and the finite abstraction is $(\epsilon,\delta)$-stochastically simulated by original model with $\delta = 0.1$, $\epsilon = 0.2466$, with the associated $\nu(x,\hat{x},\hat{u}):= K(x-P\hat{x}) + Q\hat{x} +\tilde{R}\hat{u}$ with $\tilde{R}=[0.0369;0.0172;0.0340]$,
\begin{align*}
M=\begin{bmatrix}\begin{smallmatrix}0.0107\ &0.0106\ &0.0108\\0.0106\ &0.0105\ &0.0106\\0.0108\ &0.0106\ &0.0108\end{smallmatrix} \end{bmatrix},
K=\begin{bmatrix}\begin{smallmatrix}-0.3225\ &-0.1899\ &-0.3033\\0.2199\ &0.2355\ &0.2094\\-0.0441\ &-0.1894\ &-0.0532\end{smallmatrix} \end{bmatrix}.
\end{align*}
Then, we synthesize controllers with the operator in~\eqref{eq:value_opt_sac_ab} and the one proposed in~\cite[equation (42)]{Haesaert2020Robust}.
As an example, we depict in Figure~\ref{fig:compare_operator2} the lower bounds for the probability of satisfaction associated with both controllers when the original system's initial state is $x = [5;5;\tilde{x}]$ where $\tilde{x}\in[-5,5]$ (correspondingly, original system's output $y\in[0.5,1.5]$).	 
One can readily observe that our proposed operator provides a less conservative lower bound than the one proposed in~\cite{Haesaert2020Robust}.

\section{Conclusions}\label{sec:discussion}
In this paper, we consider a notion of ($\epsilon$,$\delta$)-approximate probabilistic relations to quantify the similarity between two stochastic games.
Based on this notion, we proposed new Bellman operators to synthesize controllers for stochastic games enforcing complex logical properties modeled by deterministic finite automata.
To do so, we first synthesized a controller based on a finite abstraction that is $(\epsilon,\delta)$-stochastically simulated by the original game.
Then, this controller is refined to the original game based on the approximate probabilistic relation between the original game and its finite abstraction, which is the key to providing probabilistic guarantees.
Moreover, we proposed a systematic algorithm to establish such a relation for a particular class of nonlinear stochastic games with slope restrictions on the nonlinearity. 
The empirical results show that our method is less conservative than the existing methods in the literature.
Providing a systematic approach to establish approximate probabilistic relations for the general class of nonlinear stochastic games is under investigation as future work.
 
\bibliographystyle{elsarticle-num}   
\bibliography{mybibfile}

\appendix

\section{Proof of Statements: Section~\ref{sec:nonlinear_case}}\label{proof1}
The following proposition is required to show the results of Section~\ref{sec:nonlinear_case}.
\begin{proposition}\label{prop:positivedef}
	Consider a positive (semi)definite matrix $M_0\in\R^{s\times s}$. 
	Given $a,b\in\R$ with $a\leq b$, and a matrix $M\in\R^{s\times s}$, if $M_0+aM$ and $M_0+bM$ are positive (semi)definite, then for all $t\in[a,b]$, $M_0+tM$ is positive (semi)definite. 
\end{proposition}
{\bf Proof:}
For any $t\in[a,b]$,
\begin{itemize}
	\item If $a\leq t<0$, one has
	\begin{align*}
	M_0+tM 	&=\ (1-\frac{t}{a})M_0+\frac{t}{a}M_0+tM=\ (1-\frac{t}{a})M_0 + \frac{t}{a}(M_0+aM).
	\end{align*}
	Since $1-\frac{t}{a}\geq 0$ and $\frac{t}{a}\geq 0$, both $(1-\frac{t}{a})M_0$ and $ \frac{t}{a}(M_0+aM)$ are positive (semi)definite, so that $M_0+tM$ is also positive (semi)definite.
	\item If $0<t\leq b$, one has
	\begin{align*}
	M_0+tM &=\ (1-\frac{t}{b})M_0+\frac{t}{b}M_0+tM	=\ (1-\frac{t}{b})M_0 + \frac{t}{b}(M_0+bM).
	\end{align*}
	Since $1-\frac{t}{b}\geq 0$ and $\frac{t}{b}\geq 0$, both $(1-\frac{t}{b})M_0$ and $ \frac{t}{b}(M_0+bM)$ are positive (semi)definite, so that $M_0+tM$ is also positive (semi)definite.
\end{itemize}
Additionally, $M_0+tM$ is positive (semi)definite when $t=0$, which completes the proof. $\hfill\blacksquare$

Now we are ready to show the results of Section~\ref{sec:nonlinear_case}.

{\bf Proof of Theorem~\ref{thm:PRS}:}
Since $\mathfrak{D}$ and $\widehat{\mathfrak{D}}$ are affected by the same additive noise $\varsigma\sim\mathcal{N}(\mathbf{0}_d,I_d)$, one can readily define an lifting $\mathscr{L}_T$ based on  $\varsigma\sim\mathcal{N}(\mathbf{0}_d,I_d)$ for the approximation probabilistic relation. 
Now, we need to check the conditions in Definition~\ref{Def: apr}.
Note that the third condition in Definition~\ref{Def: apr} holds trivially since we focus on initial states $x_0 \in X_0$ and $\hat{x}_0\in \hat{X}_0$ such that $(x_0,\hat{x}_0)\in\mathscr{R}$.
We show that the first condition holds for all $(x,\hat{x})\in\mathscr{R}$. 
With~\eqref{eq:infabs_cond1} and~\eqref{eq:liftcond1}, we have
\begin{align*}
\lVert y-\hat{y}\rVert^2=\lVert Cx-\hat{C}_{\textsf r}\hat{x}\rVert^2= (x-P\hat{x})^TC^TC(x-P\hat{x})\leq (x-P\hat{x})^TM(x-P\hat{x})\leq \epsilon^2,
\end{align*}
for any $(x,\hat{x})\in\mathscr{R}$. 
Then, the first condition holds.
Next, we proceed with showing the second condition. 
This condition requires that $\forall (x,\hat{x})\in\mathscr{R}$, $\forall \hat{u}\in\hat{U}$, $\exists u \in U$ s.t. $\forall w\in W$, $\exists\hat{w}\in\hat{W}$ with $(w,\hat{w})\in\mathscr{R}_w$ s.t. the next state $(x',\hat{x}')$ is also in the relation $\mathscr{R}$ with a probability of at least $1-\delta$.
According to Assumption~\ref{asp1}, the following should hold:
\begin{equation}\label{eq:cond2}
\mathbb{P}\{(x'-P\hat{x}')^TM(x'-P\hat{x}')\leq\epsilon^2 \}\geq 1-\delta.
\end{equation}
From the slope restriction of $\varphi$ as in~\eqref{ineq:varphi}, we have
\begin{equation}\label{eq:sloperet}
\varphi(Fx)-\varphi(FP\hat{x}) =b(Fx-FP\hat{x}) = bF(x-P\hat{x}),
\end{equation}
with $b\in[\underline{b},\bar{b}]$ if $x\neq P\hat{x}$, and $b=0$ otherwise. 
Then, by applying the dynamics of $\mathfrak{D}$ as in~\eqref{eq:nonlinsys} and $\widehat{\mathfrak{D}}$ as in~\eqref{eq:abs_dyn}, we have
\begin{align}
&x'-P\hat{x}'=\ Ax\!+\!\!E\varphi(Fx)\!+\!Dw\!+\!B\nu(x,\hat{x},\hat{u})\!+\!R\varsigma- P(\hat{A}_{\textsf r}\hat{x}+\hat{E}_{\textsf r}\varphi(\hat{F}_{\textsf r}\hat{x})+\hat{D}_{\textsf r}\hat{w}+\hat{B}_{\textsf r}\hat{u}+\hat{R}_{\textsf r}\varsigma)+P\beta.\label{eq:simplify1}
\end{align}
Additionally, one can simplify~\eqref{eq:simplify1} to
\begin{align*}
\big(A+BK&+b(BL+EF)\big)(x-P\hat{x})+(B\tilde{R}-P\hat{B}_{\textsf r})\hat{u}+D(w-\hat{w})+(R-P\hat{R}_{\textsf r})\varsigma+P\beta-BS\hat{w},
\end{align*}
by employing~\eqref{eq:infabs_cond2}-\eqref{eq:infabs_cond5},~\eqref{eq:sloperet} and~\eqref{eq:interface}.
Note that here we use $b$ to denote $b(x,\hat{x})$ as in~\eqref{eq:interface} for succinctness, and it is clear from the context.
Therefore,~\eqref{eq:cond2} is fulfilled when 
\begin{align}
\lVert&\big(A\!+\!BK+b(BL\!+\!EF)\big)(x\!-\!P\hat{x})\!+\!(B\tilde{R}-P\hat{B}_{\textsf r})\hat{u}+D(w\!-\!\hat{w})\!+\!(R\!-\!P\hat{R}_{\textsf r})\varsigma\!+\!P\beta\!-\!BS\hat{w}\rVert_M\leq\epsilon\label{eq:cond2_2}
\end{align}
holds for all $b\in[\underline{b},\bar{b}]\cup \{0\}$, for all $\beta \in \Delta$ as in~\eqref{eq:Delta}, and for all $\varsigma$ s.t. $\mathbb{P}\{\varsigma^T\varsigma\leq c_{\varsigma}^2\}\geq 1-\delta$ with $c_{\varsigma}=\chi^{-1}_d(1-\delta)$, since $\varsigma \sim (\mathbf{0}_d,I_d)$ so that $\varsigma^T\varsigma$ has chi-square distribution with $d$ degrees of freedom.
Considering the left-hand side of~\eqref{eq:cond2_2}, we have
\begin{align}
&\lVert\big(A\!+\!BK+\!\!b(BL\!+\!EF)\big)(x\!-\!P\hat{x})\!+\!(B\tilde{R}-P\hat{B}_{\textsf r})\hat{u}+D(w\!-\!\hat{w})\!+\!(R\!-\!P\hat{R}_{\textsf r})\varsigma+\!P\beta\!-\!BS\hat{w}\rVert_M\nonumber\\
&\leq \lVert\big(A\!+\!BK+b(BL\!+\!EF)\big)(x-P\hat{x})\rVert_M+\!\lVert D(w\!-\!\hat{w})\rVert_M\!+\!\lVert(B\tilde{R}\!-\!P\hat{B}_{\textsf r})\hat{u}\rVert_M+\lVert(R-P\hat{R}_{\textsf r})\varsigma\rVert_M\nonumber\\
&~~~~~+\lVert P\beta\rVert_M+\lVert BS\hat{w}\rVert_M\nonumber\\	
&\leq \lVert\big(A\!+\!BK+b(BL\!+\!EF)\big)(x-P\hat{x})\rVert_M+\gamma_0+\gamma_1+\gamma_2+\gamma_3+\gamma_4\nonumber\\
&= \lVert\big(A\!+\!BK+b(BL\!+\!EF)\big)(x-P\hat{x})\rVert_M\!+\tilde{\gamma},\label{eq:proof3.501}
\end{align}
with $\tilde{\gamma}$ as in~\eqref{eq:liftcond2}, $\gamma_0$ as in~\eqref{eq:gamma0}, $\gamma_1$ as in~\eqref{eq:gamma1}, $\gamma_2$ as in~\eqref{eq:gamma2}, $\gamma_3$ as in~\eqref{eq:gamma3}, and $\gamma_4$ as in~\eqref{eq:gamma4}. 
According to S-procedure~\cite{Boyd2004Convex}, for all $\lVert x-P\hat{x}\rVert_M\leq\epsilon$, $\lVert\big(A\!+\!BK+b(BL\!+\!EF)\big)(x-P\hat{x})\rVert_M\!+\tilde{\gamma}\leq\epsilon$ holds for all $b\in[\underline{b},\bar{b}]\cup \{0\}$ if and only if there exists a $\kappa\geq0$ such that
\begin{align}
\begin{bmatrix}A_b^TMA_b&\mathbf{0}_s\\\mathbf{0}^T_s&-(\epsilon-\tilde{\gamma})^2\end{bmatrix}\preceq \kappa \begin{bmatrix}M&\mathbf{0}_s\\\mathbf{0}^T_n&-\epsilon^2\end{bmatrix}\label{eq:sp1}
\end{align}
holds for all $b\in[\underline{b},\bar{b}]\cup \{0\}$, with $A_b=A\!+\!BK+b(BL\!+\!EF)$.
Note that~\eqref{eq:sp1} holds if and only if $\kappa M-A_b^TMA_b$ is positive semidefinite and $-\kappa\epsilon^2+(\epsilon-\tilde{\gamma})^2\geq0$.
Therefore, we have~\eqref{eq:sp1} holds for all $b\in[\underline{b},\bar{b}]\cup \{0\}$ if and only if $\forall b\in[\underline{b},\bar{b}]\cup \{0\}$, there exists a $\kappa\in[0,(\epsilon-\tilde{\gamma})^2/\epsilon^2]$ such that 
\begin{equation}
A_b^TMA_b\preceq\kappa M.\label{eq:sp2}
\end{equation}
Using Schur complement~\cite{Boyd2004Convex}, we rewrite~\eqref{eq:sp2} as 
\begin{align*}
&\begin{matrix}
\underbrace{\begin{bmatrix}
	\bar{M}\ &A\bar{M}+B\bar{K}\\\bar{M}^TA^T+\bar{K}^TB^T\ &\kappa\bar{M}
	\end{bmatrix}}\\M_0
\end{matrix}\begin{matrix}
+b\underbrace{\begin{bmatrix}
	0_{s\times s}\ &B\bar{L}+EF\bar{M}\\\bar{L}^TB^T+\bar{M}^TF^TE^T\ &0_{s\times s}
	\end{bmatrix}}\succeq 0,\\M'
\end{matrix}
\end{align*}
with $\bar{M}=M^{-1}$, $\bar{K} = K\bar{M}$, and $\bar{L} = L\bar{M}$.
According to~\eqref{eq:liftcondM0}, $M_0$ is positive semidefinite.
Furthermore,~\eqref{eq:liftcondMu},~\eqref{eq:liftcondMl}, and~\eqref{eq:liftcond2} ensure that there exists a $\kappa$ with $0\leq \kappa \leq (\epsilon-\tilde{\gamma})^2/\epsilon^2$ such that~\eqref{eq:sp2} holds for $b = \{\bar{b},\underline{b}\}$.
As a result, according to Proposition~\ref{prop:positivedef}, there exists a $\kappa\in [0, (\epsilon-\tilde{\gamma})^2/\epsilon^2]$ such that $M_0+bM'$ is positive semidefinite for all $b\in[\underline{b},\bar{b}]\cup \{0\}$. 
Therefore, the second condition also holds, which completes the proof. $\hfill\blacksquare$

Next, we show the results of Corollary~\ref{col:stable}.

{\bf Proof of Corollary~\ref{col:stable}}
According to~\cite[Theorems 5,6]{Ogata1995Discrete}, for all $b'\in\{\underline{b},\bar{b},0\}$, the pair $(A+b'EF,B)$ is stabilizable if and only if there exist positive-definite matrix $M$ and $\kappa\in[0,1]$ such that~\eqref{eq:liftcondM0}-\eqref{eq:liftcondMl} hold. 
Next, we show that if there exist $M'$ and $\kappa'$ such that~\eqref{eq:liftcondM0}-\eqref{eq:liftcondMl} hold, then:
\begin{itemize}
	\item (\textbf{C1}) There exist $M$ and $\kappa$ such that~\eqref{eq:liftcond1}-\eqref{eq:liftcondMl};
	\item (\textbf{C2}) There exist $\tilde{\gamma}$ and $\epsilon$ so that~\eqref{eq:liftcond2} holds. 
\end{itemize}
Firstly, we start by showing (\textbf{C1}).
Suppose we have $M'$ and $\kappa'$ such that $(A+BK)^TM'(A+BK)\preceq\kappa' M'$, $\bar{A}^TM'\bar{A}\preceq\kappa' M'$, and $\underline{A}^TM'\underline{A}\preceq\kappa' M'$ (\ie,~\eqref{eq:liftcondM0}-\eqref{eq:liftcondMl} hold).
Then, for any $C$ as in~\eqref{eq:liftcond1}, there exists $m'\in\R_{>0}$ such that $m'M'\succeq C^TC$, since $M'$ is positive definite.
Meanwhile, one can readily verify that~\eqref{eq:liftcondM0}-\eqref{eq:liftcondMl} still hold with $M=m'M'$ and $\kappa = \kappa'$.
Therefore, (\textbf{C1}) holds. 
Next, we proceed with showing (\textbf{C2}).
Suppose we have $M$ and $\kappa$ such that~\eqref{eq:liftcondM0}-\eqref{eq:liftcondMl} hold, one can verify that~\eqref{eq:liftcondM0}-\eqref{eq:liftcondMl} also hold with $M$ and any $\kappa''$ such that 
\begin{equation}
\sqrt{\kappa''}\geq \max_{b\in[\underline{b},\bar{b}]\cup\{0\}}\lVert N A_bN^{-1}\rVert,\label{eq:minkappa}
\end{equation}
where $A_b = A+BK+b(BL+EF)$, and $N\in \mathbb{R}^{s\times s}$ is a positive-definite matrix such that $N^TN=M$.
Thus, if one has $1 - \tilde{\gamma}/\epsilon\in[\max_{b\in[\underline{b},\bar{b}]\cup\{0\}}\lVert N A_bN^{-1}\rVert, 1]$, then~\eqref{eq:liftcond2} holds.
In fact, for any $\epsilon\in\R_{>0}$, we have $\tilde{\gamma}$ such that $1- \tilde{\gamma}/\epsilon\in[\max_{b\in[\underline{b},\bar{b}]\cup\{0\}}\lVert N A_bN^{-1}\rVert, 1]$, when the finite abstraction is properly constructed.
On one hand, one always has $\gamma_1=\gamma_2=\gamma_4=0$ when there is no model order reduction involving in the abstraction since, in this case, one has $P = I_s$.
Then, one can select $\hat{B}_{\textsf r}=B$ and $\tilde{R}=I_m$ in~\eqref{eq:gamma1}, $R=\hat{R}_{\textsf r}$ in~\eqref{eq:gamma2}, and $S=0_{m\times p}$  in~\eqref{eq:gamma4}, so that one has $B\tilde{R}\!-\!P\hat{B}_{\textsf r}=0_{s\times m}$, $R-P\hat{R}_{\textsf r}=0_{s\times d}$ and $BS = 0_{s\times p}$.
On the other hand, $\gamma_0$ and $\gamma_3$ are proportional to the cardinality of $\Delta_w$ in~\eqref{eq:Delta_w} and $\Delta$ in~\eqref{eq:Delta}, respectively.
Therefore, we have (\textbf{C2}) also holds, which completes the proof. $\hfill\blacksquare$

To show the results of Theorem~\ref{thm:constraints_ineq}, the following proposition is required.
\begin{proposition}\label{prop:x_con}
	Consider a constraint $c\bar{x}\leq 1$ with $\bar{x}\in\R^s$ and a set $E_{\bar{x}} = \{\bar{x}~|~\bar{x}^TM\bar{x}\leq\epsilon^2\}$ with $M\in\R^{s \times s}$ and $\epsilon\in\R_{>0}$.
	Then, $c\bar{x}\leq 1$ holds for all $\bar{x}\in E_{\bar{x}}$ if and only if $cM^{-1}c^T\leq1/\epsilon^2$.
\end{proposition}
{\bf Proof:}
For all $\bar{x}\in E_{\bar{x}}$, $c\bar{x}\leq1$ if and only if $\max_{\bar{x}\in E_{\bar{x}}}c\bar{x}\leq1$.
Let $\bar{x}^*=\argmax_{\bar{x}\in E_{\bar{x}}}c\bar{x}$.
Then, $\bar{x}^*$ satisfies the Karush-Kuhn-Tucker conditions~\cite{Boyd2004Convex}:
\begin{align}
\lambda(\epsilon^2-\bar{x}^{*T}M\bar{x}^*)=0,\label{kkt1}\\
c-2\lambda M\bar{x}^*=0,\label{kkt2}
\end{align} 
with $\lambda\geq0$.
Solving~\eqref{kkt1} and~\eqref{kkt2}, we have
\begin{equation*}
\bar{x}^* = \epsilon\frac{M^{-1}c^T}{\sqrt{cM^{-1}c^T}}.
\end{equation*}
Therefore, 
$\max_{\bar{x}\in E_{\bar{x}}}c\bar{x}\!=\!c\bar{x}^*\!=\!\epsilon\sqrt{cM^{-1}c^T}\!\leq\! 1$
if and only if $cM^{-1}c^T\!\leq\!1/\epsilon^2$, which concludes the proof.$\hfill\blacksquare$

Now we are ready to show the results of Theorem~\ref{thm:constraints_ineq}.

{\bf Proof of Theorem~\ref{thm:constraints_ineq}:}
For any $i \in\{1,\ldots,r\}$, $\alpha_i\bar{u}\leq1$ implies that $\alpha_i(K+bL)\bar{x}\leq1$ for all $b\in[\underline{b},\bar{b}]\cup\{0\}$.
According to Proposition~\ref{prop:x_con}, $\forall \bar{x}\in E_{\bar{x}}$, $\alpha_i(K+bL)\bar{x}\leq1$ is fulfilled for all $b\in[\underline{b},\bar{b}]\cup\{0\}$ if and only if
\begin{equation}
\alpha_i(K+bL)M^{-1}(K+bL)^T\alpha_i^T\leq 1/\epsilon^2\label{eq:con1}
\end{equation}
holds for all $b\in[\underline{b},\bar{b}]\cup\{0\}$.
Using Schur complement~\cite{Boyd2004Convex},~\eqref{eq:con1} can be rewritten as
\begin{equation}
\begin{matrix}
\underbrace{\begin{bmatrix}	1/\epsilon^2\ &\alpha_i\bar{K}\\\bar{K}^T\alpha^T_i\ &\bar{M}\end{bmatrix}}\\M_0
\end{matrix}\begin{matrix}
+b\underbrace{\begin{bmatrix}0\ &\alpha_i\bar{L}\\\bar{L}^T\alpha^T_i\ &0\end{bmatrix}}\succeq0,\\M'
\end{matrix}\label{eq:con1_schurc}
\end{equation}
with $\bar{M}=M^{-1}$, $\bar{K}=K\bar{M}$, and $\bar{L}=L\bar{M}$.
Note that $M_0$ is positive semidefinite according to~\eqref{concond1}.
Moreover,~\eqref{concond2} and~\eqref{concond3} ensure that $M_0+\underline{b}M'$ and $M_0+\bar{b}M'$ are both positive semidefinite.
Then, according to Proposition~\ref{prop:positivedef},~\eqref{concond1} to~\eqref{concond3} guarantee that~\eqref{eq:con1_schurc} holds for all $b\in[\underline{b},\bar{b}]\cup\{0\}$, which completes the proof.$\hfill\blacksquare$

\section{Proof of Statements: Section~\ref{sec:safetyctr}}\label{proof2}
To show the results of Section~\ref{sec:safetyctr}, we need some additional definitions and lemmas for the product gDTSG $\mathfrak{D}||_{\mathscr{R}}\widehat{\mathfrak{D}}$ between the original gDTSG $\mathfrak{D}=(X,U,W,X_0,T,Y,h)$ and its finite abstraction $\widehat{\mathfrak{D}}= (\hat X,\hat U,\hat W,\hat{X}_0,\hat T, Y,\hat h)$ as in Definition~\ref{Def:couplingmodel}.
Given a DFA $\mathcal{A}= (Q, q_0, \Pi, \tau, F)$ that models the desired property, the reachability over the set $F$ of the gDTSG $(\mathfrak{D}||_{\mathscr{R}}\widehat{\mathfrak{D}})\otimes\mathcal{A}$ within the time horizon $[0,H]$ can be characterized by a value function defined as
\begin{align}
\tilde{V}_n^{\rho,\lambda}(x,\hat{x},q)= &\mathbb{E}[ \max_{H-n\leq t\leq H} \textbf{1}_F(q(t))|x(H-n)=x,\hat{x}(H-n)=\hat{x}, q(H-n) = q]\nonumber \label{eq:def_vfunction_couple}\\
= & \mathbb{P}_{(\rho,\lambda)\times (\mathfrak{D}||_{\mathscr{R}}\widehat{\mathfrak{D}})\otimes\mathcal{A}}\{\exists k\in[H-n, H], q(k)\in F\},
\end{align}
for all $n\in [0,H]$, with $\rho\in\mathcal P^H$ and $\lambda\in \Lambda^H$ being Markov policies for Players~\uppercase\expandafter{\romannumeral1} and~\uppercase\expandafter{\romannumeral2} of  $(\mathfrak{D}||_{\mathscr{R}}\widehat{\mathfrak{D}})\otimes\mathcal{A}$, respectively.
Given any Markov policy $\rho=(\rho_{0},\ldots,\rho_{H-1})$ and $\lambda=(\lambda_{0},\ldots,\lambda_{H-1})$, we initialize~\eqref{eq:def_vfunction_couple} with $\tilde{V}_0^{\rho,\lambda}(x,\hat{x},q)=1$ when $q\in F$, and $\tilde{V}_0^{\rho,\lambda}(x,\hat{x},q)=0$ when $q\notin F$, and recursively calculate it as  
\begin{align}\label{eq:coupling_inter}
\tilde{V}_{n+1}^{\rho,\lambda}(x,\hat{x},q)=&\!\!\!\sum_{q^+\in Q}\nonumber\int_{X\times\hat{X}}\!\!\!\!\!\tilde{V}^{\rho,\lambda}_{n}(x',\hat{x}',q^+)\bar{T}(\mathsf dx'\!\times\! d\hat{x}'\!\times\! q^+|x,\hat{x},q,\hat{u},w)\\
=& \int_{X\times\hat{X}}\tilde{V}^{\rho,\lambda}_{n}(x',\hat{x}',q')\mathscr{L}_T(\mathsf dx'\times d\hat{x}'|x,\hat{x},\hat{u},w),
\end{align}
where $\hat{u}=\rho_{H-n-1}(x,\hat{x},q)$, $w=\lambda_{H-n-1}(x,\hat{x},q,\hat{u})$, and $q'=\tau(q,L\circ h(x'))$.
In the case that $\lambda=\lambda_r$ is a randomized Markov policy over $[0,H-1]$,~\eqref{eq:coupling_inter} should be rewritten as
\begin{align}\label{eq:coupling_inter_r}
\tilde{V}_{n+1}^{\rho,\lambda_r}&(x,\hat{x},q)= \int_{W}\int_{X\times\hat{X}}\tilde{V}^{\rho,\lambda_r}_{n}(x',\hat{x}',q')\mathscr{L}_T(\mathsf dx'\times d\hat{x}'|x,\hat{x},\hat{u},w)\lambda_{r,H-n-1}(dw|x,\hat{x},q,\hat{u}).
\end{align}
In both~\eqref{eq:coupling_inter} and~\eqref{eq:coupling_inter_r}, we have
\begin{align}
\mathbb{P}_{(\rho,\lambda)\times (\mathfrak{D}||_{\mathscr{R}}\widehat{\mathfrak{D}})\otimes\mathcal{A}}\{\exists k\!\leq\! H, q(k)\!\in\! F\}\!=\!\tilde{V}_{n+1}^{\rho,\lambda}(x_0,\hat{x}_0,\bar{q}_0)\label{eq:proba_coupleV},
\end{align}
with $\bar{q}_0=\tau(q_0,L\circ h(x_0))$, $x_0\in X_0$, and $\hat{x}_0\in \hat{X}_0$ with $(x_0,\hat{x}_0)\in\mathscr{R}$. 

\begin{lemma}\label{lem:chosen_determine1}
	Consider a Markov policy $\rho$ over the time horizon $[0,H-1]$ for Player~\uppercase\expandafter{\romannumeral1} of the gDTSG $(\mathfrak{D}||_{\mathscr{R}}\widehat{\mathfrak{D}})\otimes \mathcal{A}$.
	For any randomized Markov policy $\lambda_r\in\Lambda^H$ for Player~\uppercase\expandafter{\romannumeral2} of $(\mathfrak{D}||_{\mathscr{R}}\widehat{\mathfrak{D}})\otimes \mathcal{A}$, one has
	\begin{equation}
	\tilde{V}_n^{\rho,\lambda'}(x_0,\hat{x}_0,\bar{q}_0)\leq \tilde{V}_n^{\rho,\lambda_r}(x_0,\hat{x}_0,\bar{q}_0)\label{eq:chosen_determine1}
	\end{equation}
	and 
	\begin{equation}
	\tilde{V}_n^{\rho,\lambda_r}(x_0,\hat{x}_0,\bar{q}_0)\leq \tilde{V}_n^{\rho,\lambda''}(x_0,\hat{x}_0,\bar{q}_0)\label{eq:chosen_determine2}
	\end{equation}
	for all $n\in[0,H]$, with $\bar{q}_0=\tau(q_0,L\circ h(x_0))$, $x_0\in X_0$, and $\hat{x}_0\in \hat{X}_0$ with $(x_0,\hat{x}_0)\in\mathscr{R}$.
	Here, $\lambda'$ and $\lambda''$ are nonrandomized Markov policies that are computed based on $\rho$, as
	\begin{align}
	\lambda'_{H-n-1}\in\inf_{\lambda_{H-n-1}\in \Lambda}&\int_{X\times\hat{X}}\tilde{V}^{\rho,\lambda}_{n}(x',\hat{x}',q')\mathscr{L}_T(\mathsf dx'\times d\hat{x}'|x,\hat{x},\hat{u},w),\label{eq:worst_case_couple}
	\end{align}
	and
	\begin{align}
	\lambda''_{H-n-1}\in\sup_{\lambda_{H-n-1}\in \Lambda}&\int_{X\times\hat{X}}\tilde{V}^{\rho,\lambda}_{n}(x',\hat{x}',q')\mathscr{L}_T(\mathsf dx'\times d\hat{x}'|x,\hat{x},\hat{u},w),\label{eq:worst_case_couple2}
	\end{align}
	for all $n\in[0,H]$, with $\hat{u}=\rho_{H-n-1}(x,\hat{x},q)$, and $w=\lambda_{H-n-1}(x,\hat{x},q,\hat{u})$.
\end{lemma}

{\bf Proof:}
First, we show~\eqref{eq:chosen_determine1} in Lemma~\ref{lem:chosen_determine1} by induction.
When $n=0$, according to the initialization of $\tilde{V}_0^{\rho,\lambda'}(x,\hat{x},q)$, we have $\tilde{V}_0^{\rho,\lambda'}(x,\hat{x},q)$ $= \tilde{V}_0^{\rho,\lambda_r}(x,\hat{x},q)$ so that~\eqref{eq:chosen_determine1} holds.
Suppose that~\eqref{eq:chosen_determine1} is met when $n=k$. Then, when $n=k+1$, we have 
\begin{align*}
\tilde{V}_{k+1}^{\rho,\lambda_r}(x,\hat{x},q)=&\int_{W}\int_{X\times\hat{X}}\tilde{V}^{\rho,\lambda_r}_{k}(x',\hat{x}',q')\mathscr{L}_T(\mathsf dx'\times d\hat{x}'|x,\hat{x},\hat{u},w)\lambda_{r,H-k-1}(dw|x,\hat{x},q,\hat{u})\\
\geq&\int_{W}\int_{X\times\hat{X}}\!\!\!\tilde{V}^{\rho,\lambda'}_{k}(x',\hat{x}',q')\mathscr{L}_T(\mathsf dx'\times d\hat{x}'|x,\hat{x},\hat{u},w)\lambda_{r,H-k-1}(dw|x,\hat{x},q,\hat{u})\tag{c1}\\
\geq&\int_{X\times\hat{X}}\!\!\!\tilde{V}^{\rho,\lambda'}_{k}(x',\hat{x}',q')\mathscr{L}_T(\mathsf dx'\times d\hat{x}'|x,\hat{x},\hat{u},w')\int_{W}\!\!\!\lambda_{r,H-k-1}(dw|x,\hat{x},q,\hat{u})\tag{c2}\\
=&\int_{X\times\hat{X}}\tilde{V}^{\rho,\lambda'}_{k}(x',\hat{x}',q')\mathscr{L}_T(\mathsf dx'\times d\hat{x}'|x,\hat{x},\hat{u},w')=\tilde{V}_{k+1}^{\rho,\lambda'}(x,\hat{x},q).
\end{align*}
Note that (c1) holds since we suppose that~\eqref{eq:chosen_determine1} is met when $n=k$, and (c2) holds with $ w'=\lambda'_{H-k-1}(x,\hat{x},q,\\\hat{u})$ according to~\eqref{eq:worst_case_couple}.
Thus, we have~\eqref{eq:chosen_determine1} also holds for $n=k+1$, which completes the proof for~\eqref{eq:chosen_determine1}.
The proof of~\eqref{eq:chosen_determine2} can be proceeded similar to~\eqref{eq:chosen_determine1}, and is omitted here for the sake of brevity.$\hfill\blacksquare$

So far, we are ready to prove the results in Section 5.

\subsection{Required Lemmas, Definitions, and the proof for Theorem~\ref{thm:gua_prmax}}\label{proof2.1}
To show Theorem~\ref{thm:gua_prmax}, we need Lemma~\ref{lem:co-safe_sadv}, Lemma~\ref{prop:MarkovtoC}, and some additional definitions as well.
\begin{lemma}\label{lem:co-safe_sadv}
	Consider a gDTSG $\mathfrak{D} =(X,U,W,X_0,T,$ $Y,h)$ and its finite abstraction $\widehat{\mathfrak{D}}= (\hat X,\hat U,\hat{W},\hat{X}_0,\hat T,Y,$ $\hat{h})$ with $\widehat{\mathfrak{D}}\preceq^{\delta}_{\epsilon}\mathfrak{D}$, and a DFA $\mathcal{A}= (Q, q_0, \Pi, \tau, F)$ modeling the desired property.
	Given a Markov policy $\rho$ for Player~\uppercase\expandafter{\romannumeral1} of the gDTSG $\widehat{\mathfrak{D}}\otimes \mathcal{A}$ over time horizon $[0,H-1]$, 
	we construct a Markov policy $\tilde{\rho}$ for  Player~\uppercase\expandafter{\romannumeral1} of the gDTSG $(\mathfrak{D}||_{\mathscr{R}}\widehat{\mathfrak{D}})\otimes \mathcal{A}$ such that $\forall k\in[0,H-1]$,
	$\tilde{\rho}_k(x,\hat{x},q)=\rho_k(\hat{x},q)$.
	Then, for any Markov policy $\tilde{\lambda}$ for Player~\uppercase\expandafter{\romannumeral2} of $(\mathfrak{D}||_{\mathscr{R}}\widehat{\mathfrak{D}})\otimes \mathcal{A}$, one has
	\begin{equation}\label{eq:relation_operator_cosafe}
	\bar{V}^{\rho,\lambda_*(\rho)}_n(\hat{x},q)\leq \tilde{V}_{n}^{\tilde{\rho},\tilde{\lambda}}(x,\hat{x},q),
	\end{equation}
	for all $n\in[0,H]$ and $(x,\hat{x})\in\mathscr{R}$ as in~\eqref{eq:Rx}, with $\lambda_*(\rho)$ as in~\eqref{eq:policy_max_worst_case}, $\bar{V}_{n}^{\rho,\lambda_*(\rho)}(\hat{x},q)$ computed as in~\eqref{eq:P_general}, and $\tilde{V}_{n}^{\tilde{\rho},\tilde{\lambda}}(x,\hat{x},q)$ as in~\eqref{eq:coupling_inter}.
\end{lemma}

{\bf Proof:}
The proof of Lemma~\ref{lem:co-safe_sadv} is performed by induction. 
We use $\lambda_*$ to denote $\lambda_*(\rho)$ in the following.
Additionally, we only focus on the cases in which $q\notin F$ since~\eqref{eq:relation_operator_cosafe} holds trivially for all $n\in\mathbb{N}$ when $q\in F$.  
According to the initialization of $\bar{V}^{\rho,\lambda_*}_0(\hat{x},q)$ and $\tilde{V}_{0}^{\tilde{\rho},\tilde{\lambda}}(x,\hat{x},q)$, we have $\bar{V}^{\rho,\lambda_*}_0(\hat{x},q)= \tilde{V}_{0}^{\tilde{\rho},\tilde{\lambda}}(x,\hat{x},q)$.
Therefore,~\eqref{eq:relation_operator_cosafe} holds when $n=0$.
Suppose that~\eqref{eq:relation_operator_cosafe} holds when $n=k$. Then, for $n=k+1$, we have
\begin{align*}
&\bar{V}^{\rho,\lambda_*}_{k+1}(\hat{x},q)\\
&=\ (1-\delta)\sum_{\hat{x}'\in\hat{X}}\bar{V}^{\rho,\lambda_*}_{k}(\hat{x}',\underline{q}(\hat{x}',q))\hat{T}(\hat{x}'|\hat{x},\hat{u},\hat{w})\text{ with }\hat{u}=\rho_{H-k-1}(\hat{x},q)\text{ and }\hat{w}=\lambda_{*_{H-k-1}}(\hat{x},q,\hat{u})\\
&\leq (1-\delta)\sum_{\hat{x}'\in\hat{X}}\bar{V}^{\rho,\lambda_*}_{k}(\hat{x}',\underline{q}(\hat{x}',q))\hat{T}(\hat{x}'|\hat{x},\hat{u},f_{\hat{W}})\tag{c1}\\
&\leq (1-\delta)\!\!\sum_{\hat{x}'\in\hat{X}}\!\!\bar{V}^{\rho,\lambda_*}_{k}(\hat{x}',\underline{q}(\hat{x}',q))\big(\frac{1}{1-\delta}\!\!\int_{x'\in \bar{\mathscr{R}}_{\hat{x}'}}\!\!\!\!\!\!\!\!\!\!\!\!\mathscr{L}_{T}(\mathsf dx'|x,\hat{x},\hat{x}',\hat{u},w)\big)\hat{T}(\hat{x}'|\hat{x},\hat{u},\Pi_w(w)),\tag{c2}\\
&= \int_{\mathscr{R}}\bar{V}^{\rho,\lambda_*}_{k}(\hat{x}',\underline{q}(\hat{x}',q))\mathscr{L}_{T}(\mathsf dx'|x,\hat{x},\hat{x}',\hat{u},w)\hat{T}(\hat{x}'|\hat{x},\hat{u},\Pi_w(w))\\
&= \int_{\mathscr{R}}\!\bar{V}^{\rho,\lambda_*}_{k}(\hat{x}',\underline{q}(\hat{x}',q))\mathscr{L}_{T}(\mathsf dx'\!\times\! d\hat{x}'|x,\hat{x},\hat{u},w)\tag{c3}\\
\end{align*}
\begin{align*}
&\leq \int_{\mathscr{R}}\bar{V}^{\rho,\lambda_*}_{k}(\hat{x}',q')\mathscr{L}_{T}(\mathsf dx'\!\times\! d\hat{x}'|\hat{x},x,\hat{u},w),\tag{c4}\\
&\leq \int_{\mathscr{R}}\!\tilde{V}^{\tilde{\rho},\tilde{\lambda}}_k(x',\hat{x}',q')\mathscr{L}_{T}(\mathsf dx'\times d\hat{x}'|x,\hat{x},\hat{u},w)\\
&\leq \int_{X\times\hat{X}}\tilde{V}^{\tilde{\rho},\tilde{\lambda}}_k(x',\hat{x}',q')\mathscr{L}_{T}(\mathsf dx'\times d\hat{x}'|x,\hat{x},\hat{u},w)= \tilde{V}^{\tilde{\rho},\tilde{\lambda}}_{k+1}(x,\hat{x},q),
\end{align*}
where $f_{\hat{W}}$ is a functions that assigns a probability measure over $(\hat{W},\mathcal{B}(\hat{W}))$, $\bar{\mathscr{R}}_{\hat{x}'} = \{x' \in X | (x',\hat{x}')\in \mathscr{R}\}$, 
$\Pi_w(w)$ is as in~\eqref{eq:pi_w}, 
and $\mathscr{L}_{T}(\mathsf dx'|x,\hat{x},\hat{x}',\hat{u},w)$ is the conditional probability of $x'$ as in~\eqref{eq:condition_kernel}.
In the chain of equations above, (c1) holds due to the computation of $\lambda_*$ as in~\eqref{eq:policy_max_worst_case},
(c2) holds with $w=\tilde{\lambda}_{H-k-1}(x,\hat{x},q,\hat{u})$ according to Assumption~\ref{asp1},
(c3) holds according to~\eqref{eq:condition_kernel},
and (c4) holds with $ q'=\tau(q,L\circ h(x'))$, since $\bar{V}^{\rho,\lambda_*}_{k}(\hat{x}',q')\geq \bar{V}^{\rho,\lambda_*}_{k}(\hat{x}',$ $\underline{q}(\hat{x}',q))$ according to the definition of $\underline{q}$ as in~\eqref{eq:underline_p}.
Thus,~\eqref{eq:relation_operator_cosafe} also holds when $n=k+1$, which completes the proof.$\hfill\blacksquare$

Before showing Lemma~\ref{prop:MarkovtoC}, we define how to construct a control strategy $\mathbf{C}_{\rho}$ for Player~\uppercase\expandafter{\romannumeral1} of the gDTSG $\mathfrak{D}||_{\mathscr{R}}\widehat{\mathfrak{D}}$ given a Markov policy $\tilde{\rho}$ for Player~\uppercase\expandafter{\romannumeral1} of $(\mathfrak{D}||_{\mathscr{R}}\widehat{\mathfrak{D}})\otimes \mathcal{A}$.
\begin{definition}\label{def:DAtoD}
	\emph{(Construction of $\mathbf{C}_{\rho}$)} Consider a gDTSG $\mathfrak{D}||_{\mathscr{R}}\widehat{\mathfrak{D}}\!=\!(X\!\times\!\hat{X},\hat{U},W,X_{0||},\mathscr{L}_{T},Y,h_{||})$, a DFA $\mathcal{A} = (Q, q_0, \Pi, \tau, F)$, and a Markov policy $\tilde{\rho}=(\tilde{\rho}_0,\tilde{\rho}_1,\ldots,\tilde{\rho}_{H-1})$ for Player~\uppercase\expandafter{\romannumeral1} of $(\mathfrak{D}||_{\mathscr{R}}\widehat{\mathfrak{D}})\otimes\mathcal{A}=\{\bar{X},\bar{U},\bar{W},\bar{X}_0,\bar{T},\bar{Y},\bar{h}\}$.
	We construct a control strategy $\mathbf{C}_{\rho}=(\mathsf{M},\mathsf{U},\mathsf{Y},\mathsf{H},\mathsf{M}_{0},\pi_{\mathsf{M}},\pi_{\mathsf{Y}})$ for Player~\uppercase\expandafter{\romannumeral1} of $\mathfrak{D}||_{\mathscr{R}}\widehat{\mathfrak{D}}$ with $\mathsf{M}=X\times\hat{X}\times Q$; 
	$\mathsf{U}=X\times \hat{X}$; 
	$\mathsf{Y}=\hat{U}$; 
	$\mathsf{H}=[0,H-1]$; 
	and $\mathsf{M}_{0}=\bar{X}_0$.
	Furthermore, $\pi_{\mathsf{M}}$ updates $\mathsf{m}(k)=(\mathsf{m}_X(k),\mathsf{m}_{\hat{X}}(k),\mathsf{m}_Q(k))\in\mathsf{M}$ at the time instant $k\in \mathsf{H}\backslash\{0\}$ with $(\mathsf{m}_X(k),\mathsf{m}_{\hat{X}}(k))=(x(k),\hat{x}(k))$, where $x(k)\in X$, $\hat{x}(k)\in\hat{X}$, and $\mathsf{m}_Q(k)=\tau\big(\mathsf{m}_Q(k-1),L\circ h(\mathsf{m}_X(k))\big)$;
	$\pi_{\mathsf{Y}}$ updates $\mathsf{y}(k)\in\mathsf{Y}$ at the time instant $k\in \mathsf{H}$ with $\mathsf{y}(k)=\tilde{\rho}_k(\mathsf{m}_X(k),\mathsf{m}_{\hat{X}}(k),\mathsf{m}_Q(k))$.
\end{definition}
In brief, $\mathbf{C}_{\rho}$ takes the state $(x(k),\hat{x}(k))$ of $\mathfrak{D}||_{\mathscr{R}}\widehat{\mathfrak{D}}$ and the state $q(k)$ of $\mathcal{A}$ as its memory state at the time instant $k$.
At runtime, it provides input $\hat{u}(k)$ to $\mathfrak{D}||_{\mathscr{R}}\widehat{\mathfrak{D}}$ according to the Markov policy $\tilde{\rho}_k$ based on its memory state.

\begin{lemma}\label{prop:MarkovtoC}
	Consider a gDTSG $\mathfrak{D}||_{\mathscr{R}}\widehat{\mathfrak{D}} =(X\times\hat{X},\hat{U},W,X_{0||},\mathscr{L}_{T},Y,h_{||})$, a DFA $\mathcal{A}= (Q, q_0, \Pi, \tau, F)$, and their product gDTSG $(\mathfrak{D}||_{\mathscr{R}}\widehat{\mathfrak{D}})\otimes\mathcal{A}$.
	Given a Markov policy $\tilde{\rho}$ for Player~\uppercase\expandafter{\romannumeral1} of $(\mathfrak{D}||_{\mathscr{R}}\widehat{\mathfrak{D}})\otimes \mathcal{A}$, for any control strategy $\mathbf{C}_{\lambda}$ for Player~\uppercase\expandafter{\romannumeral2} of $\mathfrak{D}||_{\mathscr{R}}\widehat{\mathfrak{D}}$, one has
	\begin{align*}
	&\ \mathbb{P}_{(\tilde{\rho},\lambda')\times(\mathfrak{D}||_{\mathscr{R}}\widehat{\mathfrak{D}})\otimes\mathcal{A}}\{\exists k\leq H, q(k)\in F\}\leq\ 
	\mathbb{P}_{(\mathbf{C}_{\rho},\mathbf{C}_{\lambda})\times\mathfrak{D}||_{\mathscr{R}}\widehat{\mathfrak{D}}}\{\exists k\leq H,y_{\omega k}\models\mathcal{A}\},
	\end{align*}
	with $\lambda'$ as in~\eqref{eq:worst_case_couple}, and $\mathbf{C}_{\rho}$ being a control strategy for Player~\uppercase\expandafter{\romannumeral1} of $\mathfrak{D}||_{\mathscr{R}}\widehat{\mathfrak{D}}$ constructed based on $\tilde{\rho}$ as in Definition~\ref{def:DAtoD}.
\end{lemma}

{\bf Proof:}
Given a path $\omega_k\in\Omega$ of $\mathfrak{D}$, the memory state of $\mathbf{C}_{\rho}$ is the same as the state of $(\mathfrak{D}||_{\mathscr{R}}\widehat{\mathfrak{D}})\otimes\mathcal{A}$ according to the construction of $\mathbf{C}_{\rho}$ as in Definition~\ref{def:DAtoD}.
Therefore, the same input $u(k)\in U$ is provided by $\tilde{\rho}$ and $\mathbf{C}_{\rho}$ given the same path $\omega_k$.
Moreover, given $(\omega_k,u(k))$,  we consider, without loss of generality, that $\mathbf{C}_{\lambda}$ chooses its adversarial input $w\in W$ according to a measurable stochastic kernel $T_W(W|\omega_k,u(k))$ over $(W,\mathcal{B}(W))$.
This kernel corresponds to a randomized Markov policy $\lambda_r$ for Player~\uppercase\expandafter{\romannumeral2} of $(\mathfrak{D}||_{\mathscr{R}}\widehat{\mathfrak{D}})\otimes\mathcal{A}$ to select $w(k)$ given the same $\omega_k$ and $u(k)$, such that
\begin{align}
&\ \mathbb{P}_{(\tilde{\rho},\lambda_r)\times(\mathfrak{D}||_{\mathscr{R}}\widehat{\mathfrak{D}})\otimes\mathcal{A}}\big\{\exists k\leq H, q(k)\in F\big\}=\ \mathbb{P}_{(\mathbf{C}_{\rho},\mathbf{C}_{\lambda})\times\mathfrak{D}||_{\mathscr{R}}\widehat{\mathfrak{D}}}\big\{\exists k\leq H,y_{\omega k}\models\mathcal{A}\big\}.\label{eq:step2eq1}
\end{align}
According to~\eqref{eq:chosen_determine1} and~\eqref{eq:proba_coupleV}, we have 
\begin{align}
&\ \mathbb{P}_{(\tilde{\rho},\lambda')\times(\mathfrak{D}||_{\mathscr{R}}\widehat{\mathfrak{D}})\otimes\mathcal{A}}\big\{\exists k\leq H, q(k)\in F\big\}\leq \ \mathbb{P}_{(\tilde{\rho},\lambda_r)\times(\mathfrak{D}||_{\mathscr{R}}\widehat{\mathfrak{D}})\otimes\mathcal{A}}\big\{\exists k\leq H, q(k)\in F\big\},\label{eq:step2eq2}
\end{align}
with synthesized $\lambda'$ based on $\tilde{\rho}$ as in~\eqref{eq:worst_case_couple}.
The proof is then completed by combining~\eqref{eq:step2eq1} and~\eqref{eq:step2eq2}.$\hfill\blacksquare$

Before showing the proof for Theorem~\ref{thm:gua_prmax}, we present how to construct the control strategy $\tilde{\mathbf{C}}_{\rho}$ for Player~\uppercase\expandafter{\romannumeral1} of $\mathfrak{D}$ given the control strategy $\mathbf{C}_{\rho}$ for Player~\uppercase\expandafter{\romannumeral1} of $\mathfrak{D}||_{\mathscr{R}}\widehat{\mathfrak{D}}$.

\begin{definition}\label{def:refine_construction}
	\emph{(Construction of $\tilde{\mathbf{C}}_{\rho}$)}
	Consider a gDTSG $\mathfrak{D} =(X,U,W,X_0,T,$ $Y,h)$ and its finite abstraction $\widehat{\mathfrak{D}}= (\hat X,\hat U,\hat{W},\hat{X}_0,\hat T,Y,\hat{h})$ with $\widehat{\mathfrak{D}}\preceq^{\delta}_{\epsilon}\mathfrak{D}$.
	Given a control strategy $\mathbf{C}_{\rho}=(\mathsf{M},\mathsf{U},\mathsf{Y},\mathsf{H},\mathsf{M}_{0},\pi_{\mathsf{M}},\pi_{\mathsf{Y}})$ for Player~\uppercase\expandafter{\romannumeral1} of $\mathfrak{D}||_{\mathscr{R}}\widehat{\mathfrak{D}}$ that is constructed based on $\tilde{\rho}$ as proposed in Definition~\ref{def:DAtoD}, 
	we construct a control strategy $\tilde{\mathbf{C}}_{\rho}=(\tilde{\mathsf{M}},\tilde{\mathsf{U}},\tilde{\mathsf{Y}},\tilde{\mathsf{H}},\tilde{\mathsf{M}}_{0},\tilde{\pi}_{\mathsf{M}},\tilde{\pi}_{\mathsf{Y}})$ for Player~\uppercase\expandafter{\romannumeral1} of $\mathfrak{D}$, in which
	\begin{itemize}
		\item $\tilde{\mathsf{M}}:=\mathsf{M}\times W\times\hat{W}=X\times\hat{X}\times Q\times W\times\hat{W}$;
		\item $\tilde{\mathsf{U}}:=\mathsf{U}_X\times W =X\times W$;
		\item $\tilde{\mathsf{Y}}:=U$;
		\item $\tilde{\mathsf{H}}:=\mathsf{H}$;
		\item $\tilde{\mathsf{m}}_0\!=\!\big(\tilde{\mathsf{m}}_{X}(0),\tilde{\mathsf{m}}_{\hat{X}}(0),\tilde{\mathsf{m}}_{Q}(0),\tilde{\mathsf{m}}_{W}(0),\tilde{\mathsf{m}}_{\hat{W}}(0)\big)\!\!\in\!\tilde{\mathsf{M}}_0$, with
		$\tilde{\mathsf{m}}_{X}(0)=x_0$, where $x_0\in X_0$; $\tilde{\mathsf{m}}_{\hat{X}}(0)=\hat{x}_0$ such that $(x_0,\hat{x}_0)\in \mathscr{R}$, where $\mathscr{R}$ is as in~\eqref{eq:Rx}; $\tilde{\mathsf{m}}_{Q}(0)=\tau\big(q_0,L\circ h(\tilde{\mathsf{m}}_{X}(0))\big)$; 
		$\tilde{\mathsf{m}}_{W}(0)$ is initialized as $\tilde{\mathsf{m}}_{W}(0)=w(0)$ after Player~\uppercase\expandafter{\romannumeral2} of $\mathfrak{D}$ has chosen $w(0)$, and $\tilde{\mathsf{m}}_{\hat{W}}(0)$ is initialized as $\tilde{\mathsf{m}}_{\hat{W}}(0)=\Pi_w(w(0))$ with $\Pi_w$ as in~\eqref{eq:pi_w};
		\item $\tilde{\pi}_{\mathsf{M}}$ updates $\big(\tilde{\mathsf{m}}_{X}(k),\tilde{\mathsf{m}}_{\hat{X}}(k),\tilde{\mathsf{m}}_{Q}(k),\tilde{\mathsf{m}}_{W}(k),\tilde{\mathsf{m}}_{\hat{W}}(k)\big)$ $\in\tilde{\mathsf{M}}$ at all time instant $k\in\mathsf{H}\backslash\{0\}$, with the following steps:
		\begin{enumerate}[(i)]
			\item update $\tilde{\mathsf{m}}_{\hat{X}}(k)$ with the conditional kernel
			\begin{align*}
			\mathscr{L}_{T}\big(d\hat{x}|\tilde{\mathsf{m}}_{\hat X}(k-1),&\tilde{\mathsf{m}}_{X}(k-1),x(k),\hat{u}(k-1),\tilde{\mathsf{m}}_{W}(k-1)\big)
			\end{align*}
			as in~\eqref{eq:condition_kernel}, with $x(k)$ the state of $\mathfrak{D}$ and $\hat{u}(k-1)=\tilde{\rho}_{k-1}(\tilde{\mathsf{m}}_X(k-1),\tilde{\mathsf{m}}_{\hat X}(k-1),\tilde{\mathsf{m}}_{Q}(k-1))$;
			\item update $\tilde{\mathsf{m}}_{X}(k)$ with $\tilde{\mathsf{m}}_{X}(k)=x(k)$;
			\item update $\tilde{\mathsf{m}}_{Q}(k)$ with $\tilde{\mathsf{m}}_{Q}(k)=\tau\big(\tilde{\mathsf{m}}_{Q}(k-1),L\circ h(\tilde{\mathsf{m}}_{X}(k))\big)$;
			\item update $\tilde{\mathsf{m}}_W(k)$ with $\tilde{\mathsf{m}}_W(k)=w(k)$ after Player~\uppercase\expandafter{\romannumeral2} of $\mathfrak{D}$ has selected $w(k)$ and accordingly update $\tilde{\mathsf{m}}_{\hat{W}}(k)$ as $\tilde{\mathsf{m}}_{\hat{W}}(k)=\Pi_w(w(k))$ with $\Pi_w$ as in~\eqref{eq:pi_w};
		\end{enumerate}
		\item $\tilde{\pi}_{\mathsf{Y}}$ updates $\mathsf{y}(k)\in\mathsf{Y}$ at the time instant $k\in\mathsf{H}$ with $\mathsf{y}(k)=\nu\big(\tilde{\mathsf{m}}_X(k),$ $\tilde{\mathsf{m}}_{\hat X}(k),\tilde{\rho}_k(\tilde{\mathsf{m}}_X(k),\tilde{\mathsf{m}}_{\hat X}(k),$ $\tilde{\mathsf{m}}_{Q}(k))\big)$, with $\nu$ being the interface function associated with the approximate probabilistic relation.	
	\end{itemize}
\end{definition}

\begin{figure*}[htbp]
	\centering
	\subfigure{
		\includegraphics[width=0.45\textwidth]{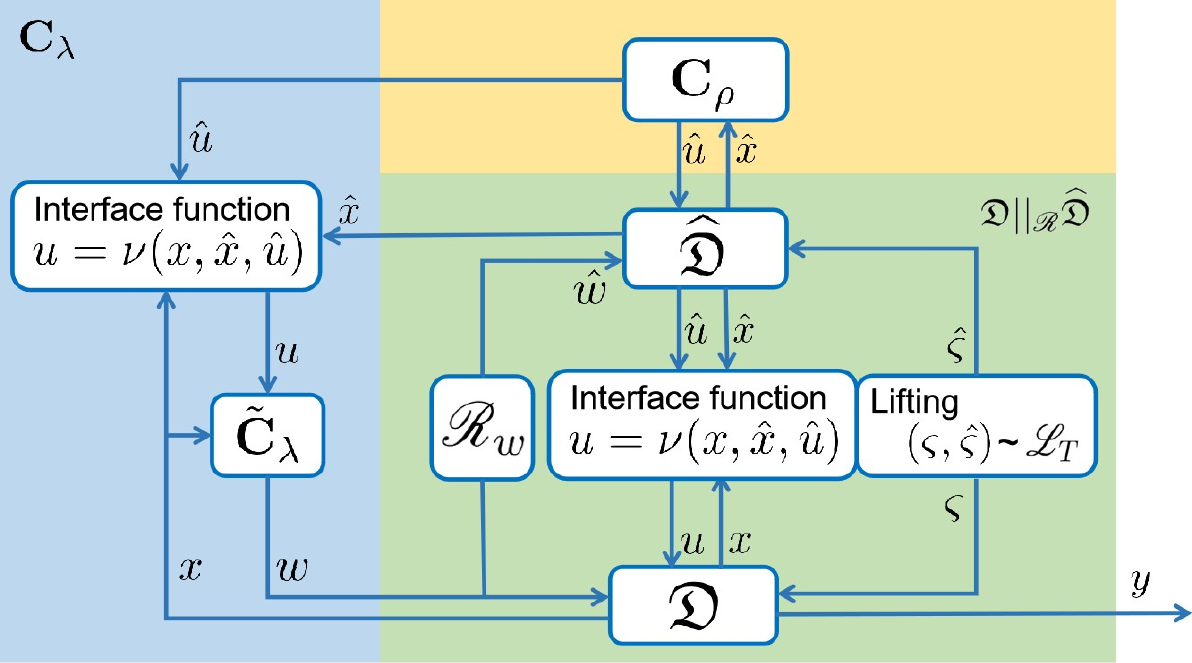}
	}
	\quad
	\subfigure{
		\includegraphics[width=0.45\textwidth]{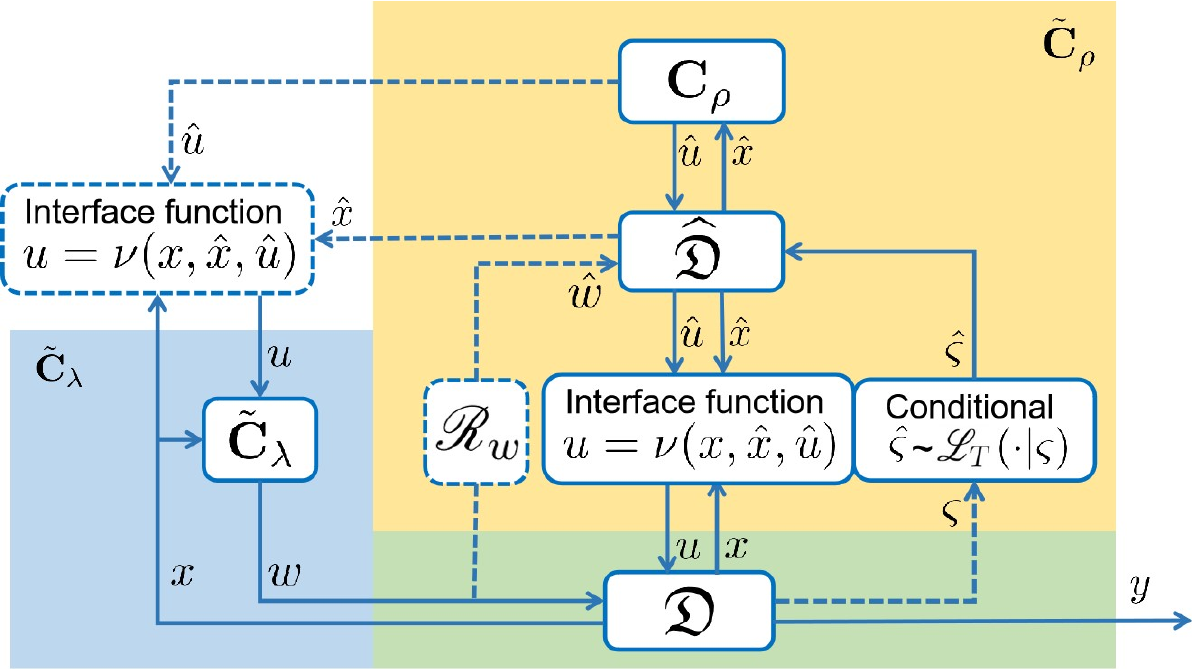}
	}
	\caption{\textbf{Left:} Coupling gDTSG $\mathfrak{D}||_{\mathscr{R}}\widehat{\mathfrak{D}}$ (green region) controlled by $\mathbf{C}_{\rho}$ (yellow region) and $\mathbf{C}_{\lambda}$ (blue region).~\textbf{Right:} A gDTSG $\mathfrak{D}$ (green region) controlled by $\tilde{\mathbf{C}}_{\rho}$ (yellow region) and $\tilde{\mathbf{C}}_{\lambda}$ (blue region).}
	\label{fig:controller_equivalent_proof}
\end{figure*}
Employing Definition~\ref{def:refine_construction}, we construct a control strategy $\tilde{\mathbf{C}}_{\rho}$ for Player~\uppercase\expandafter{\romannumeral1} of the gDTSG $\mathfrak{D}$ given a control strategy $\mathbf{C}_{\rho}$ for Player~\uppercase\expandafter{\romannumeral1} of $\mathfrak{D}||_{\mathscr{R}}\widehat{\mathfrak{D}}$.
Then, given any control strategy $\tilde{\mathbf{C}}_{\lambda}$ for Player~\uppercase\expandafter{\romannumeral2} of $\mathfrak{D}$, the controlled gDTSG $(\tilde{\mathbf{C}}_{\rho},\tilde{\mathbf{C}}_{\lambda})\times\mathfrak{D}$ can be written as a controlled gDTSG $(\mathbf{C}_{\rho},\mathbf{C}_{\lambda})\times\mathfrak{D}||_{\mathscr{R}}\widehat{\mathfrak{D}}$  as depicted in Fig.~\ref{fig:controller_equivalent_proof}, where $\mathbf{C}_{\lambda}$ is constructed by combining $\tilde{\mathbf{C}}_{\lambda}$ with the interface function $\nu(x,\hat{x},\hat{u})$.
Accordingly, we have
\begin{align}
&\ \mathbb{P}_{(\tilde{\mathbf{C}}_{\rho},\tilde{\mathbf{C}}_{\lambda})\times\mathfrak{D}}\{\exists k\leq H, y_{\omega k}\models\mathcal{A}\}=\ \mathbb{P}_{(\mathbf{C}_{\rho},\mathbf{C}_{\lambda})\times\mathfrak{D}||_{\mathscr{R}}\widehat{\mathfrak{D}}}\{\exists k\leq H,y_{\omega k}\models\mathcal{A}\}.\label{eq:equ_control}
\end{align}
Now, we are ready to show the results of Theorem~\ref{thm:gua_prmax}.

{\bf Proof of Theorem~\ref{thm:gua_prmax}:}
Consider $x_0\in X_0$ and $\hat{x}_0\in \hat{X}_0$ with $(x_0,\hat{x}_0)\in\mathscr{R}$ and $\mathscr{R}$ as in~\eqref{eq:Rx}.
According to~\eqref{eq:proba_coupleV} and Lemma~\ref{lem:co-safe_sadv}, for any Markov policy $\tilde{\lambda}$ for Player~\uppercase\expandafter{\romannumeral2} of the gDTSG $(\mathfrak{D}||_{\mathscr{R}}\widehat{\mathfrak{D}})\otimes \mathcal{A}$, we have
\begin{equation}\label{eq:1}
\bar{V}_H^{\rho,\lambda_*(\rho)}(\hat{x}_0,\bar{q}_0)\leq \mathbb{P}_{(\tilde{\rho},\tilde{\lambda})\times (\mathfrak{D}||_{\mathscr{R}}\widehat{\mathfrak{D}})\otimes\mathcal{A}}\{\exists k\leq H, q(k)\in F\},
\end{equation}
with $\tilde{\rho}$ being a Markov policy for Player~\uppercase\expandafter{\romannumeral1} of the gDTSG $(\mathfrak{D}||_{\mathscr{R}}\widehat{\mathfrak{D}})\otimes \mathcal{A}$ that is constructed based on $\rho$ as discussed in Lemma~\ref{lem:co-safe_sadv}.
Moreover, Lemma~\ref{prop:MarkovtoC} indicates that given a Markov policy $\tilde{\rho}$ for Player~\uppercase\expandafter{\romannumeral1} of $(\mathfrak{D}||_{\mathscr{R}}\widehat{\mathfrak{D}})\otimes \mathcal{A}$ and a control strategy $\mathbf{C}_{\rho}$ for Player~\uppercase\expandafter{\romannumeral1} of $\mathfrak{D}||_{\mathscr{R}}\widehat{\mathfrak{D}}$ that is constructed based on $\tilde{\rho}$ as in Definition~\ref{def:DAtoD}, for any control strategy $\mathbf{C}_{\lambda}$ for Player~\uppercase\expandafter{\romannumeral2} of $\mathfrak{D}||_{\mathscr{R}}\widehat{\mathfrak{D}}$, one has
\begin{align}\label{eq:2}
&\ \mathbb{P}_{(\tilde{\rho},\lambda')\times(\mathfrak{D}||_{\mathscr{R}}\widehat{\mathfrak{D}})\otimes\mathcal{A}}\{\exists k\leq H, q(k)\in F\}\leq\ \mathbb{P}_{(\mathbf{C}_{\rho},\mathbf{C}_{\lambda})\times\mathfrak{D}||_{\mathscr{R}}\widehat{\mathfrak{D}}}\{\exists k\leq H,y_{\omega k}\models\mathcal{A}\},
\end{align}
where $\lambda'$ is a Markov policy for Player~\uppercase\expandafter{\romannumeral2} of $(\mathfrak{D}||_{\mathscr{R}}\widehat{\mathfrak{D}})\otimes \mathcal{A}$ computed as in~\eqref{eq:worst_case_couple}. 
Since~\eqref{eq:1} holds for any Markov policy for Player~\uppercase\expandafter{\romannumeral2} of $(\mathfrak{D}||_{\mathscr{R}}\widehat{\mathfrak{D}})\otimes \mathcal{A}$, by combining~\eqref{eq:1} and~\eqref{eq:2}, we have 
\begin{equation}\label{eq:2.5}
\bar{V}_H^{\rho,\lambda_*(\rho)}(\hat{x}_0,\bar{q}_0)\leq \mathbb{P}_{(\mathbf{C}_{\rho},\mathbf{C}_{\lambda})\times\mathfrak{D}||_{\mathscr{R}}\widehat{\mathfrak{D}}}\{\exists k\leq H,y_{\omega k}\models\mathcal{A}\}. 
\end{equation}
Finally, considering~\eqref{eq:2.5} and~\eqref{eq:equ_control}, we have
\begin{equation}\label{eq:3}
\bar{V}_H^{\rho,\lambda_*(\rho)}(\hat{x}_0,\bar{q}_0)\leq \mathbb{P}_{(\tilde{\mathbf{C}}_{\rho},\tilde{\mathbf{C}}_{\lambda})\times\mathfrak{D}}\{\exists k\leq H, y_{\omega k}\models\mathcal{A}\}, 
\end{equation}
where $\tilde{\mathbf{C}}_{\rho}$ is a control strategy for Player~\uppercase\expandafter{\romannumeral1} of $\mathfrak{D}$ that is constructed based on $\mathbf{C}_{\rho}$ as in Definition~\ref{def:refine_construction}.
Considering the construction of $\tilde{\rho}$ as in Lemma~\ref{lem:co-safe_sadv} based on $\rho$, $\mathbf{C}_{\rho}$ as in Definition~\ref{def:DAtoD} based on $\tilde{\rho}$, and $\tilde{\mathbf{C}}_{\rho}$ as in Definition~\ref{def:refine_construction} based on $\mathbf{C}_{\rho}$, 
$\tilde{\mathbf{C}}_{\rho}$ in~\eqref{eq:3} can be constructed as in Definition~\ref{def:C_rho} directly based on a Markov policy $\rho$ for Player~\uppercase\expandafter{\romannumeral1} of $\widehat{\mathfrak{D}}\otimes\mathcal{A}$, which completes the proof.$\hfill\blacksquare$

\subsection{Results for Lemma~\ref{lem:less_con}}\label{proof2.2}
In this subsection, we denote by $\widehat{\mathfrak{D}}= (\hat X,\hat U,\hat{X}_0,\hat T,Y,\hat{h})$ the finite abstraction for the stochastic systems without rational adversarial input, and by $\mathcal{A}= (Q, q_0, \Pi, \tau, F)$ a DFA modeling the desired property.
Additionally, we use $\bar{V}_{n}^{\rho}(\hat{x},q)$ to replace $\bar{V}_{n}^{\rho,\lambda}(\hat{x},q)$ as in~\eqref{eq:P_general}, since $\lambda$ does not play a role in stochastic systems of interest here.
Accordingly, initializing $\bar{V}_{n}^{\rho}(\hat{x},q)$ with $\bar{V}_0^{\rho}(\hat{x},q)=1$ when $q\in F$ and $\bar{V}_0^{\rho}(\hat{x},q)=0$, otherwise, $\bar{V}_{n+1}^{\rho}(\hat{x},q)$ is then recursively computed as
\begin{align*}
\bar{V}_{n+1}^{\rho}(\hat{x},q):= (1-\delta)\sum_{\hat{x}'\in \hat{X}}\bar{V}^{\rho}_{n}(\hat{x}',\underline{q}(\hat{x}',q))\hat{T}(\hat{x}'|\hat{x},\hat{u}),
\end{align*}
when $q\notin F$, and $\bar{V}_{n+1}^{\rho}(\hat{x},q):=1$ otherwise.
Furthermore, $\underline{q}$ as in~\eqref{eq:underline_p} should accordingly be modified as
\begin{equation}\label{eq:underline_p_nw}
\underline{q}(\hat{x}',q) = \mathop{\arg\min}_{q'\in Q'_{\epsilon}(\hat{x}')}\bar{V}^{\rho}_{n}(\hat{x}',q'),
\end{equation}
where  $Q'_{\epsilon}(\hat{x}')$ is the set as in~\eqref{eq:Q_epsilon}.
Before showing the results for Lemma~\ref{lem:less_con}, we briefly introduce some results in~\cite{Haesaert2020Robust} for the sake of completeness.
Considering a Markov policy $\rho=(\rho_{0},\rho_{1},$ $\ldots,\rho_{H-1})$ over time horizon $[0,H-1]$, a value function $V^{\rho}_n:\hat{X}\times Q \rightarrow [0,1]$ is defined in\cite{Haesaert2020Robust}.
Initialized with $V_0^{\rho}(\hat{x},q) = 0$, $V^{\rho}_n(\hat{x},q)$ is then recursively computed as~\cite[equation (41)]{Haesaert2020Robust}:
\begin{align}
V_{k+1}^{\rho}(\hat{x},q) : = \mathbf{L}&\Big(\sum_{\hat{x}'\in \hat{X}} \min_{q'\in\bar{\tau}(q,\hat{x}')}\max \{\mathbf{1}_F(q'),V_k^{\rho}(\hat{x}',q')\}\hat{T}(d\hat{x}'|\hat{x},\hat{u})-\delta \Big),\label{sof_op}
\end{align}
with $\mathbf{L}:\mathbb{R}\rightarrow[0,1]$ being the truncation function $\mathbf{L}(\cdot):=\min(1,\max(0,\cdot))$; 
$\bar{\tau}(q,\hat{x}'):=\{ \tau(q,\alpha)\text{ with }\alpha\in L(\mathcal{N}_{\epsilon}(\hat{h}(\hat{x}')))\}$, where $\mathcal{N}_{\epsilon}(\hat{y}):=\{y\in Y\,|\, \lVert y-\hat{y}\rVert\leq \epsilon \}$;
$\mathbf{1}_F(\cdot)$ being an indicator function for the set $F$, i.e., if $q'\in F$ then $\mathbf{1}_F(q')=1$, otherwise $\mathbf{1}_F(q')=0$; 
and $\hat{u} = \rho_{H-k-1}(\hat{x})$.
With these notations, $\mathcal{S}(\hat{x}_0)$ as in~\eqref{eq:less_con} can be computed as~\cite[equation (43)]{Haesaert2020Robust}:
\begin{align}
\mathcal{S}(\hat{x}_0) := \!\!\!\min_{\bar{q}_0\in\bar{\tau}(q_0,\hat{x}_0)}\!\!\max (\mathbf{1}_F(\bar{q}_0),V_H^{\rho}(\hat{x}_0,\bar{q}_0)).\label{eq:sxq}
\end{align}
Moreover, Lemma~\ref{trick} is required for proving Lemma~\ref{lem:less_con}.
\begin{lemma}\label{trick}
	If we have 
	\begin{equation}
	V_{n}^{\rho}(\hat{x}',q)\leq\bar{V}_{n}^{\rho}(\hat{x}',q),\label{eq:ass1}
	\end{equation}
	for all $\hat{x}'\in \hat{X}$ and $q\in Q$, with $n\in\mathbb{N}$, then we have
	\begin{equation}
	\!\!\!\!\!\!\min_{q'\in\bar{\tau}(q,\hat{x}')}\!\!\!\!\!\max \{\mathbf{1}_F(q'),V_n^{\rho}(\hat{x}',q')\}\!\leq\!\bar{V}^{\rho}_{n}(\hat{x}',\underline{q}(\hat{x}',q)).\label{eq:ass2}
	\end{equation}
\end{lemma}
{\bf Proof of Lemma~\ref{trick}:}
We prove Lemma~\ref{trick} by showing two cases:
\begin{itemize}
	\item (Case 1) If $\exists \hat{x}\in Q'_{\epsilon}(\hat{x}')$ such that $\tau(q,\hat{x})\notin F$, then we have 
	\begin{align}
	&\min_{q'\in\bar{\tau}(q,\hat{x}')}\max \{\mathbf{1}_F(q'),V_n^{\rho}(\hat{x}',q')\}=\!\!\!\min_{q'\in\bar{\tau}(q,\hat{x}')}V_n^{\rho}(\hat{x}',q') = \!\!\!\min_{q'\in Q'_{\epsilon}(\hat{x}')}V_n^{\rho}(\hat{x}',q').\label{p1}
	\end{align}
	Meanwhile, according to the definition of $\underline{q}$ as in~\eqref{eq:underline_p_nw}, we have 
	\begin{equation}
	\bar{V}^{\rho}_{n}(\hat{x}',\underline{q}(\hat{x}',q)) = \min_{q'\in Q'_{\epsilon}(\hat{x}')}\bar{V}^{\rho}_{n}(\hat{x}',q')\label{p2}
	\end{equation}
	Then, with~\eqref{p1},~\eqref{p2} and~\eqref{eq:ass1}, one can readily verify that~\eqref{eq:ass2} holds in Case 1.
	\item (Case 2) If $\forall \hat{x}\in Q'_{\epsilon}(\hat{x}')$ such that $\tau(q,\hat{x})\in F$, we have $\bar{V}^{\rho}_{n}(\hat{x}',\underline{q}(\hat{x}',q))=1$.
	Therefore,~\eqref{eq:ass2} holds trivially in Case 2.
\end{itemize}
Then, we complete the proof for Lemma~\ref{trick} by combining Case 1 and Case 2.
$\hfill\blacksquare$

Now, we are ready to show the results for Lemma~\ref{lem:less_con}.

{\bf Proof of Lemma~\ref{lem:less_con}:}
First, we show that
\begin{align}\label{key}
V_{n}^{\rho}(\hat{x},q)\leq\bar{V}_{n}^{\rho}(\hat{x},q)
\end{align}
holds for all $n\in\mathbb{N}$ by induction. 
Note that we only focus on the cases in which $q\notin F$ since we have $\bar{V}_{n}^{\rho}(\hat{x},q)=1$ when $q\in F$ so that \eqref{key} holds trivially.
According to the initialization of $\bar{V}^{\rho}_0(\hat{x},q)$ and $V_0(\hat{x},q)$, we have $\bar{V}^{\rho}_0(\hat{x},q)\geq V_0(\hat{x},q)$.
Therefore,~\eqref{key} holds when $n=0$.
Suppose that~\eqref{key} is met when $n=k$. 
Then, when $n=k+1$, we only focus on the case in which 
\begin{align}
\sum_{\hat{x}'\in \hat{X}} \min_{q'\in\bar{\tau}(q,\hat{x}')}\max &\{\mathbf{1}_F(q'),V_k^{\rho}(\hat{x}',q')\}\hat{T}(d\hat{x}'|\hat{x},\hat{u})-\delta\geq 0. \label{anyway}
\end{align}
Otherwise, $V_{k+1}^{\rho}(\hat{x},q)\leq\bar{V}_{k+1}^{\rho}(\hat{x},q)$ holds trivially since one has $V_{k+1}^{\rho}(\hat{x},q)=0$ according to the definition of function $\mathbf{L}(\cdot)$ as in~\eqref{sof_op}.
When~\eqref{anyway} holds, we have
\begin{align}
V_{k+1}^{\rho}(\hat{x},q)= &\sum_{\hat{x}'\in \hat{X}} \min_{q'\in\bar{\tau}(q,\hat{x}')}\max \{\mathbf{1}_F(q'),V_k^{\rho}(\hat{x}',q')\}\hat{T}(d\hat{x}'|\hat{x},\hat{u})-\delta,\nonumber\\
\leq & \sum_{\hat{x}'\in \hat{X}} \bar{V}^{\rho}_{k}(\hat{x}',\underline{q}(\hat{x}',q))\hat{T}(d\hat{x}'|\hat{x},\hat{u})-\delta\label{eq:trick}\\
\leq & \sum_{\hat{x}'\in \hat{X}}\!\!\bar{V}^{\rho}_{k}(\hat{x}',\underline{q}(\hat{x}',q))\hat{T}(d\hat{x}'|\hat{x},\hat{u})-\delta\Big(\!\sum_{\hat{x}'\in \hat{X}}\!\!\bar{V}^{\rho}_{k}(\hat{x}',\underline{q}(\hat{x}',q))\hat{T}(d\hat{x}'|\hat{x},\hat{u})\Big)\nonumber\\
= & (1-\delta)\sum_{\hat{x}'\in \hat{X}}\bar{V}^{\rho}_{k}(\hat{x}',\underline{q}(\hat{x}',q))\hat{T}(\hat{x}'|\hat{x},\hat{u})= \bar{V}_{k+1}^{\rho}(\hat{x},q).\nonumber
\end{align}
Note that~\eqref{eq:trick} holds according to Lemma~\ref{trick}.
Therefore, we have~\eqref{key} also hold for $n=k+1$, so that~\eqref{key} holds for all $n\in\mathbb{N}$.
Then, one can readily verify 
\begin{equation*}
\bar{V}_H^{\rho}(\hat{x}_0,\bar{q}_0)\geq\mathcal{S}(\hat{x}_0)
\end{equation*}
by considering~\eqref{eq:underline_p_nw},~\eqref{eq:sxq},~\eqref{key}, and Lemma~\ref{trick}, which completes the proof. $\hfill\blacksquare$

\subsection{Required Lemmas and the proof for Theorem~\ref{thm:gua_prmin}}\label{proof2.3}
In order to show the results of Theorem~\ref{thm:gua_prmin}, we need the Lemma~\ref{lem:safe_sadv} and Lemma~\ref{prop:MarkovtoC2}.

\begin{lemma}\label{lem:safe_sadv}
	Consider a gDTSG $\mathfrak{D} =(X,U,W,X_0,T,Y,h)$ and its finite abstraction $\widehat{\mathfrak{D}}= (\hat X,\hat U,\hat{W},\hat{X}_0,\hat T,Y,$ $\hat h)$ with $\widehat{\mathfrak{D}}\preceq^{\delta}_{\epsilon}\mathfrak{D}$, and a DFA $\mathcal{A}= (Q, q_0, \Pi, \tau, F)$ characterizing the desired property.
	Given a Markov policy $\rho$ for Player~\uppercase\expandafter{\romannumeral1} of the gDTSG $\widehat{\mathfrak{D}}\otimes \mathcal{A}$ over the time horizon $[0,H-1]$, construct a Markov policy $\tilde{\rho}$ for  Player~\uppercase\expandafter{\romannumeral1} of the gDTSG $(\mathfrak{D}||_{\mathscr{R}}\widehat{\mathfrak{D}})\otimes \mathcal{A}$ such that $\forall k\in[0,H-1]$,
	$\tilde{\rho}_k(x,\hat{x},q)=\rho_k(\hat{x},q)$.
	Then, for any Markov policy $\tilde{\lambda}$ for Player~\uppercase\expandafter{\romannumeral2} of $(\mathfrak{D}||_{\mathscr{R}}\widehat{\mathfrak{D}})\otimes \mathcal{A}$, one has
	\begin{equation}\label{eq:relation_operator_safe}
	\ul{V}^{\rho,\lambda^*(\rho)}_n(\hat{x},q)\geq \tilde{V}_{n}^{\tilde{\rho},\tilde{\lambda}}(x,\hat{x},q),
	\end{equation}
	for all $n\in[0,H]$, $(x,\hat{x})\in\mathscr{R}$ as in~\eqref{eq:Rx}, with $\lambda^*(\rho)$ computed as in~\eqref{eq:policy_min_worst_case}, $\ul{V}_{n}^{\rho,\lambda^*(\rho)}(\hat{x},q)$ as in~\eqref{eq:T_general} and $\tilde{V}_{n}^{\tilde{\rho},\tilde{\lambda}}(x,\hat{x},q)$ as in~\eqref{eq:coupling_inter}.
\end{lemma}

{\bf Proof:}
The proof is followed by induction.
We denote $\lambda^*(\rho)$ by $\lambda^*$ for the sake of clarity.
Moreover, we only focus on the cases in which $q\notin F$ since~\eqref{eq:relation_operator_safe} holds trivially for all $n\in\mathbb{N}$ when $q\in F$.
For $n=0$, one can readily verify that $\ul{V}^{\rho,\lambda^*}_0(\hat{x},q)= \tilde{V}_{0}^{\tilde{\rho},\tilde{\lambda}}(x,\hat{x},q)$
according to the initialization of $\ul{V}^{\rho,\lambda^*}_0(\hat{x},q)$ and $\tilde{V}_{0}^{\tilde{\rho},\tilde{\lambda}}(x,\hat{x},q)$.
Thus,~\eqref{eq:relation_operator_safe} holds for $n=0$.
Suppose that~\eqref{eq:relation_operator_safe} holds for $n=k$. Then, for $n=k+1$, one has
\begin{align*}
&1- \ul{V}^{\rho,\lambda^*}_{k+1}(\hat{x},q)\\
&= (1-\delta)-(1-\delta)\!\!\sum_{\hat{x}'\in\hat{X}}\!\!\ul{V}^{\rho,\lambda^*}_{k}(\hat{x}',\bar{q}(\hat{x}',q))\hat{T}(\hat{x}'|\hat{x},\hat{u},\hat{w})\text{ with }\hat{u}=\rho_{H-k-1}(\hat{x})\text{ and } \hat{w}=\lambda^*_{H-k-1}(\hat{x},\hat{u})\\
&\leq (1-\delta)\!-\!(1-\delta)\!\!\sum_{\hat{x}'\in\hat{X}}\!\!\ul{V}^{\rho,\lambda^*}_{k}\!\!(\hat{x}',\bar{q}(\hat{x}',q))\hat{T}(\hat{x}'|\hat{x},\hat{u},f_{\hat{W}})\\
&= (1-\delta)\!\!\sum_{\hat{x}'\in\hat{X}}\!\!\Big(1-\ul{V}^{\rho,\lambda^*}_{k}(\hat{x}',\bar{q}(\hat{x}',q))\Big)\hat{T}(\hat{x}'|\hat{x},\hat{u},f_{\hat{W}})\\
&\leq (1-\delta)\!\!\sum_{\hat{x}'\in\hat{X}}\!\!\Big(1-\ul{V}^{\rho,\lambda^*}_{k}(\hat{x}',\bar{q}(\hat{x}',q))\Big)\Big(\frac{1}{1-\delta}\int_{x'\in \bar{\mathscr{R}}_{\hat{x}'}}\mathscr{L}_{T}(\mathsf dx'|x,\hat{x},\hat{x}',\hat{u},w)\Big)\hat{T}(\hat{x}'|\hat{x},\hat{u},\Pi_w(w)),\\
&\text{ with }w = \tilde{\lambda}_{H-k-1}(x,\hat{x},q,\hat{u})\\
&= \int_{\mathscr{R}}\Big(1-\ul{V}^{\rho,\lambda^*}_{k}(\hat{x}',\bar{q}(\hat{x}',q))\Big)\mathscr{L}_{T}(\mathsf dx'|x,\hat{x},\hat{x}',\hat{u},w)\hat{T}(\hat{x}'|\hat{x},\hat{u},\Pi_w(w))\\
&= \int_{\mathscr{R}}\!\Big(1-\ul{V}^{\rho,\lambda^*}_{k}(\hat{x}',\bar{q}(\hat{x}',q))\Big)\mathscr{L}_{T}(\mathsf dx'\!\times\! d\hat{x}'|x,\hat{x},\hat{u},w)\\
&\leq \int_{\mathscr{R}}\Big(1-\ul{V}^{\rho,\lambda^*}_{k}(\hat{x}',q')\Big)\mathscr{L}_{T}(\mathsf dx'\!\times\! d\hat{x}'|\hat{x},x,\hat{u},w),\text{ with }\ q'=\tau(q,L\circ h(x'))\\
&\leq \int_{\mathscr{R}}\!\Big(1-\tilde{V}^{\tilde{\rho},\tilde{\lambda}}_k(x',\hat{x}',q')\Big)\mathscr{L}_{T}(\mathsf dx'\times d\hat{x}'|x,\hat{x},\hat{u},w)\\
&\leq \int_{X\times\hat{X}}\Big(1-\tilde{V}^{\tilde{\rho},\tilde{\lambda}}_k(x',\hat{x}',q')\Big)\mathscr{L}_{T}(\mathsf dx'\times d\hat{x}'|x,\hat{x},\hat{u},w)= 1-\tilde{V}^{\tilde{\rho},\tilde{\lambda}}_{k+1}(x,\hat{x},q),
\end{align*}
where $f_{\hat{W}}$ is a functions that assigns a probability measure over $(\hat{W},\mathcal{B}(\hat{W}))$, $\bar{\mathscr{R}}_{\hat{x}'} = \{x' \in X | (x',\hat{x}')\in \mathscr{R}\}$, and $\mathscr{L}_{T}(\mathsf dx'|x,\hat{x},\hat{x}',\hat{u},w)$ is the conditional probability of $x'$ as in~\eqref{eq:condition_kernel}.
Note that the chain of equations above hold similarly to those in the proof of Lemma~\ref{lem:co-safe_sadv}.
Thus, we have $\ul{V}^{\rho,\lambda^*}_{k+1}(\hat{x},q)\geq\tilde{V}^{\tilde{\rho},\tilde{\lambda}}_{k+1}(x,\hat{x},q)$ so that~\eqref{eq:relation_operator_cosafe} also holds for $n=k+1$, which concludes the proof.$\hfill\blacksquare$

\begin{lemma}\label{prop:MarkovtoC2}
	Consider a gDTSG $\mathfrak{D}||_{\mathscr{R}}\widehat{\mathfrak{D}} =(X\times\hat{X},\hat{U},W,X_{0||},\mathscr{L}_{T},Y,h_{||})$, a DFA $\mathcal{A}= (Q, q_0, \Pi, \tau, F)$, and their product gDTSG $(\mathfrak{D}||_{\mathscr{R}}\widehat{\mathfrak{D}})\otimes\mathcal{A}$.
	Given a Markov policy $\tilde{\rho}$ for Player~\uppercase\expandafter{\romannumeral1} of $(\mathfrak{D}||_{\mathscr{R}}\widehat{\mathfrak{D}})\otimes \mathcal{A}$, for any control strategy $\mathbf{C}_{\lambda}$ for Player~\uppercase\expandafter{\romannumeral2} of $\mathfrak{D}||_{\mathscr{R}}\widehat{\mathfrak{D}}$, one has
	\begin{align*}
	&\ \mathbb{P}_{(\tilde{\rho},\lambda'')\times(\mathfrak{D}||_{\mathscr{R}}\widehat{\mathfrak{D}})\otimes\mathcal{A}}\{\exists k\leq H, q(k)\in F\}\geq\ 
	\mathbb{P}_{(\mathbf{C}_{\rho},\mathbf{C}_{\lambda})\times\mathfrak{D}||_{\mathscr{R}}\widehat{\mathfrak{D}}}\{\exists k\leq H,y_{\omega k}\models\mathcal{A}\},
	\end{align*}
	with $\lambda''$ as in~\eqref{eq:worst_case_couple2}, and $\mathbf{C}_{\rho}$ being a control strategy for Player~\uppercase\expandafter{\romannumeral1} of $\mathfrak{D}||_{\mathscr{R}}\widehat{\mathfrak{D}}$ constructed based on $\tilde{\rho}$ as in Definition~\ref{def:DAtoD}.
\end{lemma}
Lemma~\ref{prop:MarkovtoC2} can be proved similar to that of Lemma~\ref{prop:MarkovtoC} with the help of~\eqref{eq:chosen_determine2} and~\eqref{eq:worst_case_couple2}.
Employing Lemmas~\ref{lem:safe_sadv} and~\ref{prop:MarkovtoC2}, we show the results of Theorem~\ref{thm:gua_prmin} as follows.

{\bf Proof of Theorem~\ref{thm:gua_prmin}:} 
Consider $x_0\in X_0$ and $\hat{x}_0\in \hat{X}_0$ with $(x_0,\hat{x}_0)\in\mathscr{R}$ and $\mathscr{R}$ as in~\eqref{eq:Rx}.
According to~\eqref{eq:proba_coupleV} and Lemma~\ref{lem:safe_sadv}, for any Markov policy $\tilde{\lambda}$ for Player~\uppercase\expandafter{\romannumeral2} of the gDTSG $(\mathfrak{D}||_{\mathscr{R}}\widehat{\mathfrak{D}})\otimes \mathcal{A}$, we have
\begin{equation}\label{eq:1t}
\ul{V}_H^{\rho,\lambda^*(\rho)}(\hat{x}_0,\bar{q}_0)\geq \mathbb{P}_{(\tilde{\rho},\tilde{\lambda})\times (\mathfrak{D}||_{\mathscr{R}}\widehat{\mathfrak{D}})\otimes\mathcal{A}}\{\exists k\leq H, q(k)\in F\},
\end{equation}
with $\tilde{\rho}$ being a Markov policy for Player~\uppercase\expandafter{\romannumeral1} of the gDTSG $(\mathfrak{D}||_{\mathscr{R}}\widehat{\mathfrak{D}})\otimes \mathcal{A}$ that is constructed based on $\rho$ as in Lemma~\ref{lem:safe_sadv}.
Furthermore, according to Lemma~\ref{prop:MarkovtoC2}, given a Markov policy $\tilde{\rho}$ for Player~\uppercase\expandafter{\romannumeral1} of $(\mathfrak{D}||_{\mathscr{R}}\widehat{\mathfrak{D}})\otimes \mathcal{A}$ and a control strategy $\mathbf{C}_{\rho}$ for Player~\uppercase\expandafter{\romannumeral1} of $\mathfrak{D}||_{\mathscr{R}}\widehat{\mathfrak{D}}$ constructed as in Definition~\ref{def:DAtoD} based on $\tilde{\rho}$, for any control strategy $\mathbf{C}_{\lambda}$ for Player~\uppercase\expandafter{\romannumeral2} of $\mathfrak{D}||_{\mathscr{R}}\widehat{\mathfrak{D}}$, we have
\begin{align}\label{eq:2t}
&\ \mathbb{P}_{(\tilde{\rho},\lambda'')\times(\mathfrak{D}||_{\mathscr{R}}\widehat{\mathfrak{D}})\otimes\mathcal{A}}\{\exists k\leq H, q(k)\in F\}\geq\ \mathbb{P}_{(\mathbf{C}_{\rho},\mathbf{C}_{\lambda})\times\mathfrak{D}||_{\mathscr{R}}\widehat{\mathfrak{D}}}\{\exists k\leq H,y_{\omega k}\models\mathcal{A}\},
\end{align}
where $\lambda''$ is a Markov policy for Player~\uppercase\expandafter{\romannumeral2} of $(\mathfrak{D}||_{\mathscr{R}}\widehat{\mathfrak{D}})\otimes \mathcal{A}$ computed as in~\eqref{eq:worst_case_couple2}. 
Note that~\eqref{eq:1t} holds for any arbitrary Markov policy for Player~\uppercase\expandafter{\romannumeral2} of $(\mathfrak{D}||_{\mathscr{R}}\widehat{\mathfrak{D}})\otimes \mathcal{A}$.
By combining~\eqref{eq:1t} and~\eqref{eq:2t}, one has
\begin{equation}\label{eq:2.5t}
\ul{V}_H^{\rho,\lambda^*(\rho)}(\hat{x}_0,\bar{q}_0)\geq \mathbb{P}_{(\mathbf{C}_{\rho},\mathbf{C}_{\lambda})\times\mathfrak{D}||_{\mathscr{R}}\widehat{\mathfrak{D}}}\{\exists k\leq H,y_{\omega k}\models\mathcal{A}\}. 
\end{equation}
Then, similar to the proof of Theorem~\ref{thm:gua_prmax}, one can readily verify~\eqref{eq:thm_sad_pmin} considering~\eqref{eq:2.5t} and~\eqref{eq:equ_control}, which completes the proof.$\hfill\blacksquare$

\end{document}